\documentclass[12pt]{article}

\usepackage[sorting=none, style=phys]{biblatex}{\small}
\let\cite\supercite

\usepackage[a4paper, margin=1.5cm]{geometry}
\usepackage{newtxtext, newtxmath}
\usepackage{setspace}
\usepackage{float}
\usepackage{indentfirst}
\usepackage{braket}
\usepackage{tabularx}
\usepackage{longtable}
\usepackage{makecell}
\usepackage{multirow}
\usepackage{booktabs}
\usepackage{caption}
\usepackage{setspace}
\usepackage{url}
\renewcommand{\arraystretch}{1.4}
\usepackage{graphicx}
\usepackage{kbordermatrix}
\renewcommand{\kbldelim}{(}
\renewcommand{\kbrdelim}{)}
\usepackage{amsmath}

\begin{document}

\begin{titlepage}
    \centering
    \vspace*{1cm}

    {\LARGE \textbf{Developing fast, accurate and spin pure calculations of organic diradical electronic structure}}\\[2cm]

    {\large \textbf{Author:} Joseph Kielty, Department of Chemistry, UCL}\\[0.5cm]
    {\large \texttt{joey.kielty.22@ucl.ac.uk}}\\[1cm]

    {\large \textbf{Primary Supervisor:} Dr.~Tim Hele, Department of Chemistry, UCL}\\[0.5cm]
    {\large \texttt{t.hele@ucl.ac.uk}}\\[1cm]

    {\large \textbf{Secondary Supervisor:} Dr.~Hugh Burton, Department of Chemistry, UCL}\\[0.5cm]
    {\large \texttt{h.burton@ucl.ac.uk}}\\[1cm]

    \vspace{20pt}
    \begin{figure}[H]
    \centering
    \includegraphics[width=1\linewidth, height=0.44\linewidth]{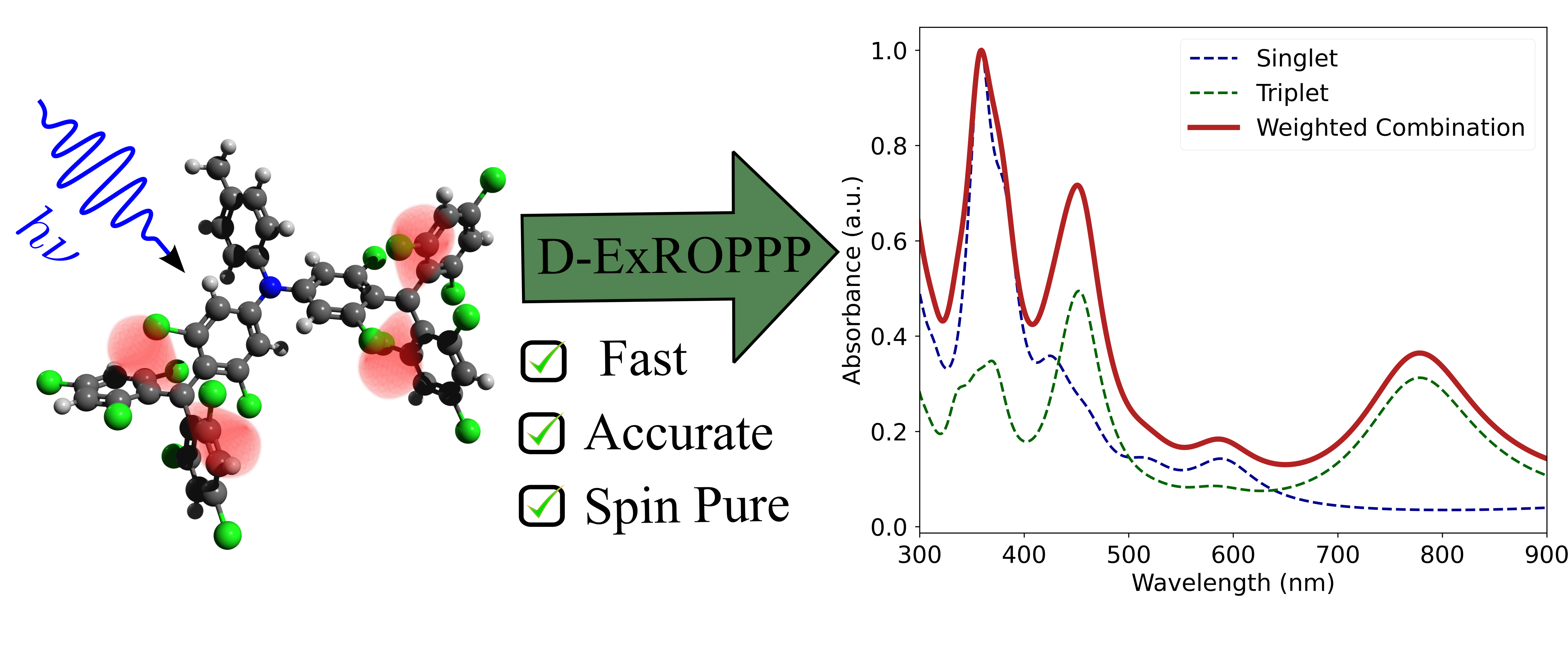}
    \end{figure}

    \vfill
\end{titlepage}

\section*{Abstract}

Organic diradical molecules have emerged in recent years as highly promising candidates for next-generation optoelectronic and quantum information technologies. Hosting two unpaired electrons in nearly degenerate molecular orbitals, diradicals display unusual photophysical properties which are largely driven by the near-degeneracy of their low-lying singlet and triplet states. As efforts continue to synthesise diradicals with tailored photophysical properties, developing our theoretical understanding of their complex electronic structure is crucial to enable rational design strategies. However, existing computational tools struggle significantly with this task. Density functional theory, which is typically the workhorse of computational materials research, is largely ill-suited for describing the multi-reference character of diradical electronic states, making it prone to produce unphysical results that artificially mix states of different spin multiplicity. While a variety of high-level wavefunction-based methods do exist that offer a rigorous treatment of spin statistics, their high computational cost makes them inefficient for simulating the majority of emissive diradicals, which are typically composed of between 75 and 150 atoms.

This work presents a new computational method developed to address this growing demand for accurate yet affordable simulations of diradical molecules. We show that this method, termed Diradical Extended Restricted Open-shell PPP (D-ExROPPP) theory, yields accurate and spin-pure electronic states at a very low computational cost relative to existing approaches. This is achieved by implementing a novel, spin-adapted configuration interaction approach within the semi-empirical Pariser-Parr-Pople framework. D-ExROPPP is then used to study the electronic structure of three prototypical organic diradicals, and benchmarked against results from computationally intensive, post-Hartree-Fock methods. Excitation energies for these molecules are predicted that show a similar level of accuracy as high-level \textit{ab-initio} calculations, while requiring up to five orders of magnitude less computational time. We further apply D-ExROPPP to a number of recently reported stable, emissive, organic diradicals. In the majority of cases, electronic transitions measured in experimental UV-vis spectra are reproduced with high qualitative accuracy and the spin states involved can be reliably assigned. Together, these results establish D-ExROPPP as a promising new tool for efficiently predicting and interpreting the photophysics of organic diradical molecules.

\newpage

\tableofcontents

\newpage

\section{Introduction}

Organic diradical molecules have gained increasing relevance in the past several years, emerging as a promising class of molecules for a variety of applications. While the first stable organic diradical was discovered over 100 years ago \cite{Thiele1904}, interest in the field has exploded since the discovery of synthetic pathways to highly stable, luminescent, organic diradicals \cite{Hattori2019, Feng2021, Mizuno2024, Chowdhury2025}. Hosting two unpaired electrons, these large organic diradicals often have unique spin properties that allow for highly unconventional photophysical properties, including high internal photoluminescence quantum efficiencies and the control of emission properties by temperature and magnetic fields \cite{Zhu2025}.  This has drawn interest towards the development of diradical OLED technologies \cite{Murto2022}, and in particular molecular qubits that make use of organic diradical molecules \cite{Kopp2024, Poh2024}.

Despite this rapidly growing interest, a significant gap remains between experimental advances and our theoretical understanding of diradical molecules and their emission properties. Existing computational methods face major challenges in modelling the highly-correlated electronic structure of these molecules, which typically are made up of 80 - 150 atoms. State-of-the-art multireference methods such as CASSCF and CASPT2 do well to capture electron correlation, but their poor scaling can make them prohibitively expensive for large molecules. Conversely, density-based methods such as time dependent density functional theory (TDDFT) are much more efficient but results often feature a high degree of spin contamination, making it difficult to assign spin states and obtain accurate energies. As research continues into the synthesis and characterisation of organic diradicals, there is a growing need for computational methods that can efficiently and reliably provide insight into their photophysical properties and so help to guide rational design efforts.

This report introduces a novel computational method capable of predicting spin-pure excited electronic states of diradical molecules. We show that this method, termed Diradical Extended Restricted Open-Shell PPP theory (D-ExROPPP) can achieve an accuracy comparable to high-level multireference calculations in some cases, but at a fraction of the computational cost. We ensures spin-purity by constructing configuration state functions (CSFs) which form a basis for an extended configuration interaction approach that includes all single excitations with select double excitations across the $\pi$ orbitals (XCIS-D). The use of machine-learned Pariser-Parr-Pople (PPP) parameters further ensures the method maintains a low computational cost relative to ab-initio methods, while implicitly capturing features of electronic correlation. After establishing the methodology, D-ExROPPP is tested on a number of organic diradical molecules. We first run calculations on several small diradicals and compare results against those obtained with Complete Active Space Self-Consistent Field (CASSCF) and Generalised Multiconfigurational Quasi-Degenerate Perturbation Theory (GMC-QDPT). We then study a set of 14 recently presented large emissive diradicals, computing UV-vis absorption spectra and comparing predictions against experimental data and results from Spin-Flip Time-Dependent Density Functional Theory (SF-TDDFT).

\newpage

\section{Previously Published Work}

\subsection{Electronic Properties of Diradicals}

\subsubsection{Chemical Structure and Bonding}

Diradical molecules are widely defined as open-shell molecules, possessing two degenerate or nearly degenerate singly occupied molecular orbitals (SOMOs) \cite{Salem1972, Stuyver2019}. The presence of two unpaired electrons in diradical molecules lends itself to more complex spin states in both ground and excited electronic states. Diradical electronic structure has long been understood through the two-orbital two-electron model (TOTEM), which looks at the electronic configurations that arise from distributing of two electrons among two spatial orbitals (or four spin-orbitals) \cite{Salem1972, Stuyver2019}. This model yields six possible electronic configurations, of which four have spin magnetic quantum number ($M_s$) of 0 (Fig. 1). Of these four, two are zwitterionic configurations and behave like closed-shell singlets, while the remaining two open-shell configurations are not eigenstates of the $S^2$ operator. Spin-adapted linear combinations of these four configurations yield the zwitterionic $\ket{^1\text{ZW}^\pm}$ states and the open-shell singlet $\ket{^1\text{OS}}$ and triplet $\ket{^3\text{OS}}$ states shown in Fig. 1.

\vspace{1cm}

\begin{figure}[H]
\centering
\includegraphics[width=0.87\linewidth, height=0.6\linewidth]{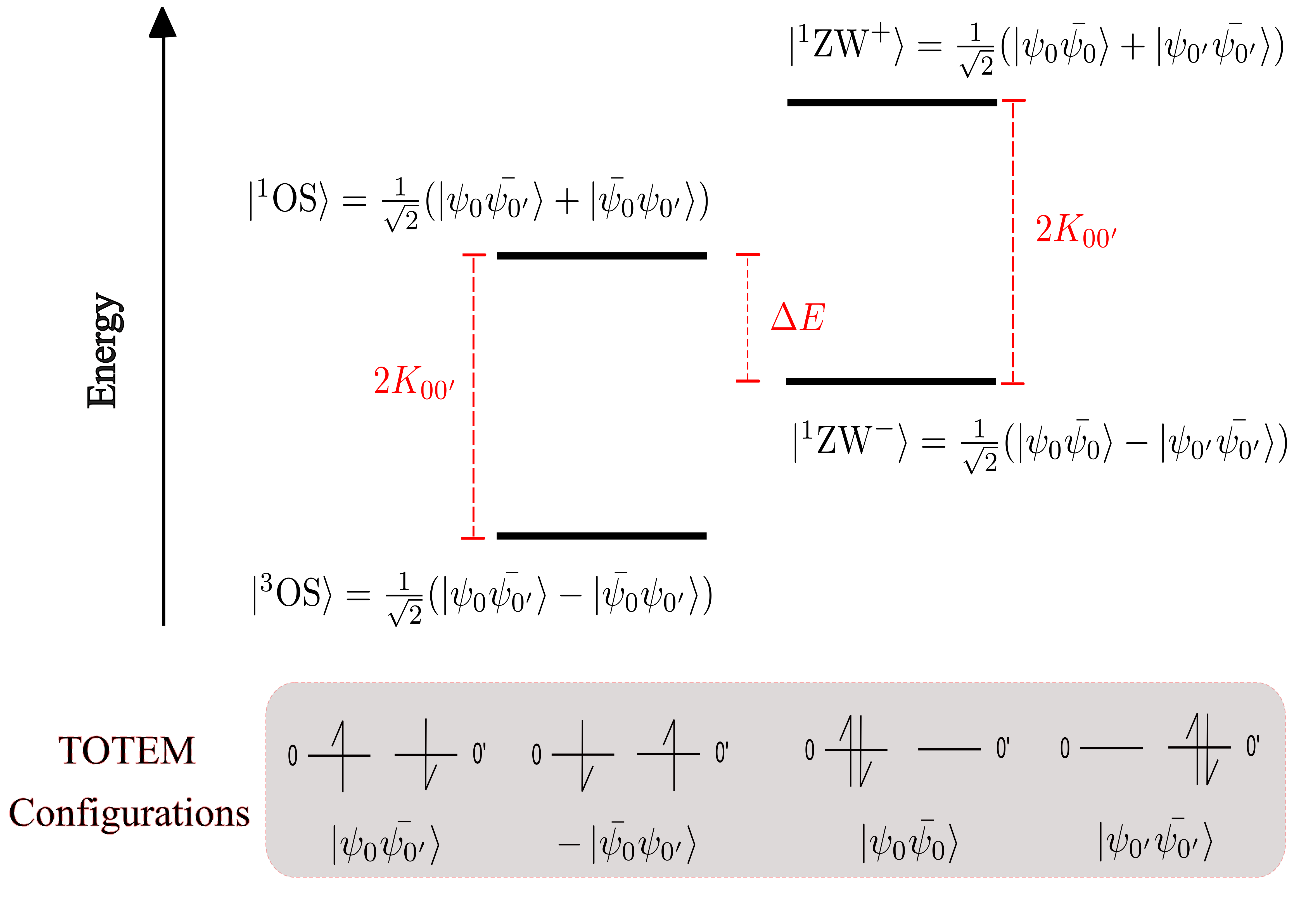}
\caption{Representations of the electronic configurations and spin-adapted electronic states that arise from the two-electron two-orbital model for describing diradical electronic structure. $K_{00'}$ represents the exchange energy between the two SOMOs, $0$ and $0'$. The energy difference between the $\ket{^1\text{OS}}$ and $\ket{^1\text{ZW}^-}$ states is given by $\Delta E = E_\text{OS1} - E_\text{ZW-} = J_{00'} + 2K_{00'} - \frac{1}{2}(J_{00} + J_{0'0'})$, where $J_{ab}$ is the Coulomb energy between the MOs $a$ and $b$. This figure shows the case where $\Delta E>0$, however, this depends on the molecule and it is possible to have $\Delta E<0$ or $\Delta E=0$.}
\end{figure}

It is also important to consider the nature of the SOMOs themselves. In many cases, SOMOs are delocalised across the diradical molecule, meaning that the electron density associated with each unpaired electron is distributed over many atoms in the conjugated $\pi$-system. However, it is possible to rotate SOMO coefficients such that each unpaired electron is mostly or fully localised on different atoms in the molecule. SOMOs that are localised in this way are often called generalised valence bond (GVB) orbitals \cite{CarlLineberger2011, Borden2015}. Molecules in which GVB orbitals do not share electron density between any of the same atoms can be classified as disjoint, wheras molecules without this property are classed as non-disjoint \cite{Borden1977}. Salem and Rowland showed in 1972 that the choice of delocalised versus localised SOMOs alters the description of the low-lying TOTEM electronic states \cite{Salem1972}. For a given molecule, the $\ket{^1\text{OS}}$ state formed from delocalised orbitals maps exactly to the zwitterionic state $\ket{^1\text{ZW}^-}$ when formed from localised orbitals.  Fig. 2 highlights this effect in the case of the widely studied cyclobutadiene (CBD) molecule. The exchange interaction $K_{00'}$ will typically be very small between localised SOMOs, meaning that $\ket{^1\text{OS}}$ and $\ket{^3\text{OS}}$ will be close in energy. However, when SOMOs are delocalised $K_{00'}$ is expected to be much greater, meaning that $\ket{^1\text{ZW}^-}$ will typically be the lowest energy singlet state.

\begin{figure}[H]
\centering
\includegraphics[width=0.92\linewidth, height=0.6\linewidth]{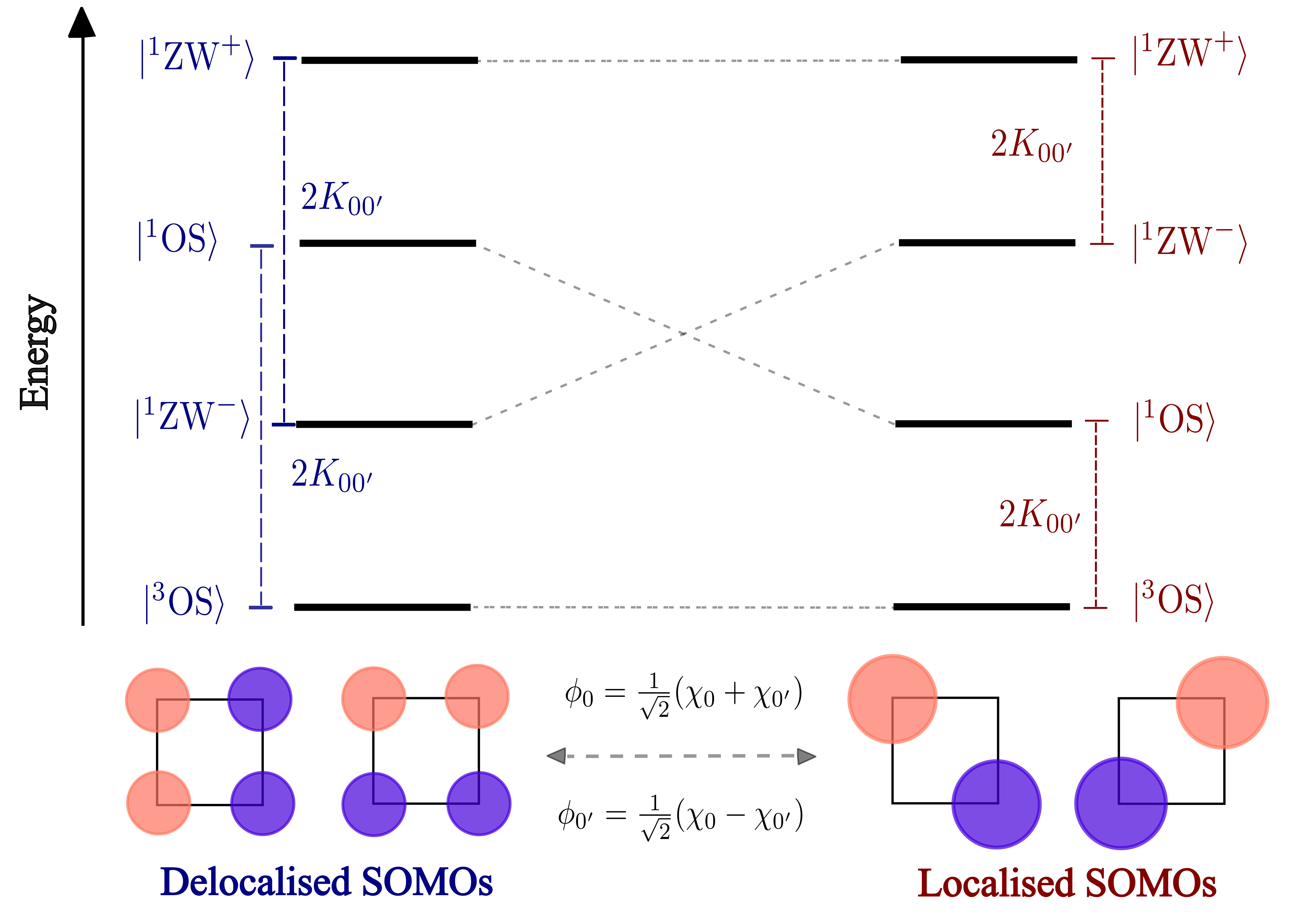}
\caption{Energy level diagram highlighting the transformation between $\ket{^1\text{ZW}^-}$ and $\ket{^1\text{OS}}$ that occurs when transforming from the delocalised to the localised SOMOs of cylobutadiene. $\phi$ and $\chi$ represent localised and delocalised molecular orbitals respectively, and the algebraic transformation between the two sets is shown in grey. As in Fig. 1, the $\ket{^1\text{ZW}^-}$ state is depicted as being lower in energy than $\ket{^1\text{OS}}$ in the delocalised case, however, the relative energies of these states varies depending on the diradical molecule.}
\end{figure}

A closely related class of molecules to diradicals are diradicaloids, which are loosely defined as molecules with an electronic structure that is intermediate between pure open-shell and closed-shell molecules \cite{Stuyver2019}. This arises when a relatively small energy gap exists between the two molecular orbitals, meaning they behave both like the degenerate SOMOs of a pure diradical and the HOMO and LUMO of a pure closed-shell molecule.

\begin{figure}[H]
\centering
\includegraphics[width=0.81\linewidth, height=0.6\linewidth]{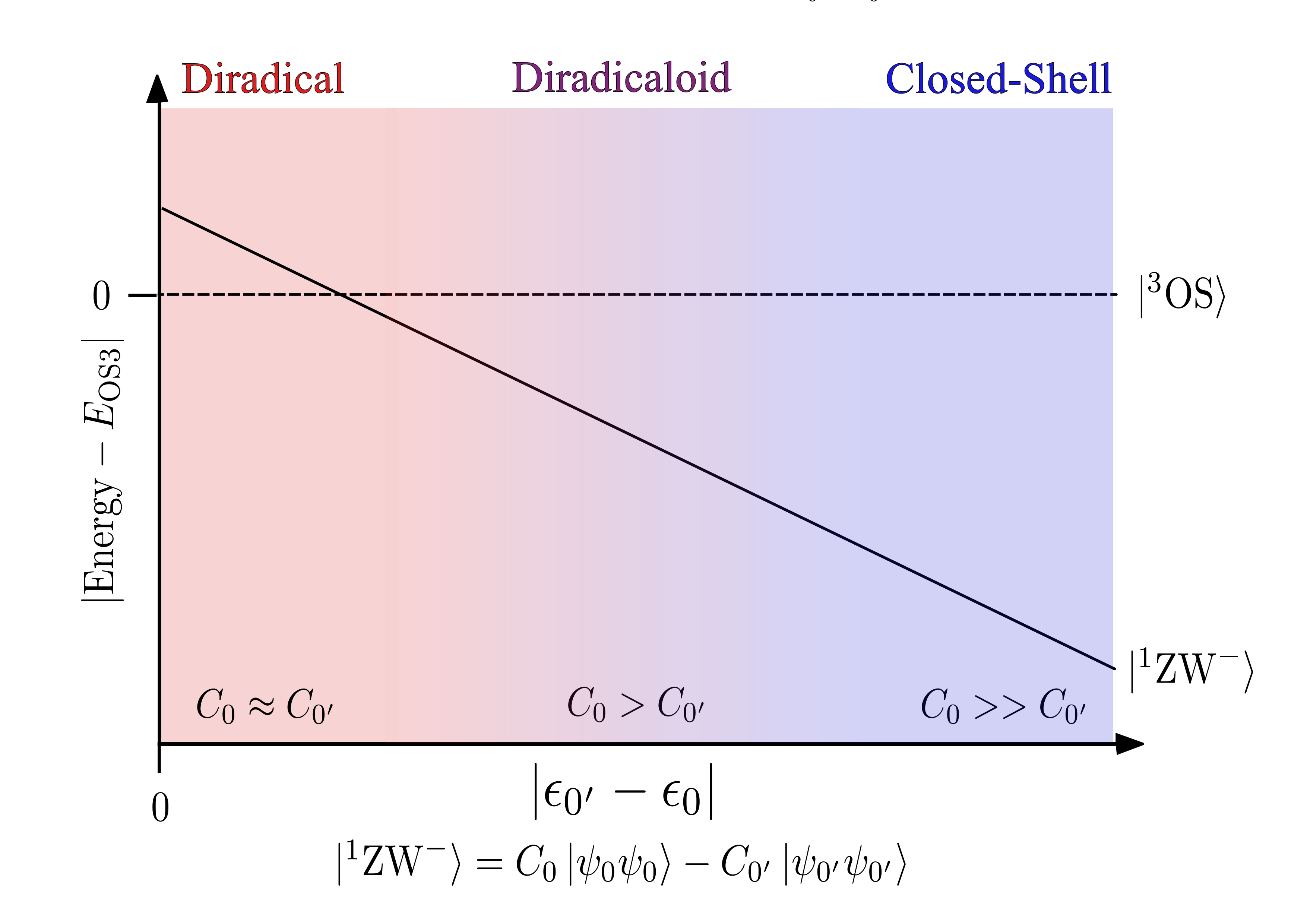}
\caption{Schematic to represent the transition from diradical to diradicaloid to closed shell molecule, inspired by that presented by Stuyver et al. \cite{Stuyver2019}. $\epsilon_0$ and $\epsilon_{0'}$ represent the energies of orbitals 0 and 0' respectively, where orbital 0 is lower in energy and orbital 0' is higher in energy. Energies are shown relative to that of the open-shell triplet in order to highlight how the singlet energy becomes much lower than the triplet energy.}
\end{figure}

Diradicaloids typically possess a singlet ground state which is somewhere in-between the $\ket{^1\text{ZW}^-}$ state of a pure-diradical and the doubly-occupied $S_0$ state of a closed-shell molecule (Fig. 3). In a pure diradical, 0 and 0' are degenerate meaning $\ket{\psi_0\bar{\psi_0}}$ and $\ket{\psi_{0'}\bar{\psi_{0'}}}$ configurations have equal weighting in $\ket{^1\text{ZW}^-}$. As the 0' orbital rises in energy relative to the 0 orbital, its weighting ($C_{0'}$) slowly decreases until it eventually contributes very little to the $\ket{^1\text{ZW}^-}$ state, which is now effectively described by the single $\ket{\psi_0\bar{\psi_0}}$ configuration. We find $\frac{C_{0'}}{C_{0}}$ indirectly characterises the degree of open- vs. closed-shell character, where $\frac{C_{0'}}{C_{0}} = 1$ indicates a pure-diradical and $\frac{C_{0'}}{C_{0}} = 0$ suggests that a molecule can be described as fully closed-shell.

Since the 1970s, there have been many attempts to define a formal metric for a molecules 'diradical character', which is typically expressed using the index $y\in [0,1]$, where $y = 0$ indicates purely closed-shell character and $y=1$ means that a molecule is a pure-diradical \cite{Hayes1971, Takatsuka1978, Nakano2016, Nakano2017, Bachler2002, HeadGordon2003}. One of the most widely used formulations is given by Yamaguchi's equation, presented in Eq. 1 where $n_\text{HONO}$ and $n_\text{LUNO}$ are the occupation numbers of the highest occupied and lowest unoccupied natural orbitals respectively calculated with a broken-symmetry unrestricted Hartree-Fock approach \cite{Yamaguchi1975}. 

\begin{subequations}
\begin{align}
    y_0 &= 1 - \frac{2T_0}{1 + T_0^2}\\
    T_0 &= \frac{n_\text{HONO} - n_\text{LUNO}}{2}
\end{align}
\end{subequations}

Despite the number of different approaches developed, estimating a molecules diradical character remains a theoretical challenge, meanwhile obtaining estimates from experiment is harder still. This may in part explain why the distinction between diradical and diradicaloid molecules is largely ill-defined across the literature, with the two terms sometimes being used interchangeably. 

\vspace{1cm}

\subsubsection{Photophysical Properties}

To date, emissive diradical molecules are most commonly synthesised by coupling two monoradical molecules together with an aromatic linker \cite{Zhu2025}. The most widely used radical precursor is the tris(2,4,6-trichlorophenyl)methyl (TTM) radical, from which a number of diradicals have been designed using a wide variety of linkers \cite{Abdurahman2023, Abdurahman2023-1, Kopp2024, Gorgon2023, Chang2024, Mesto2025, Arnold2025, Feng2023, Hattori2026, Schneburg2026}. Closely related radicals M$_2$TTM as well as the PyBTM (pyridyl bis(2,4,6-trichlorophenyl)methyl) have also seen widespread use \cite{Chowdhury-arxiv, Chowdhury2025, Yu2024, Mizuno2024, Hattori2019, Matsuoka2023, Hattori2024}.

The open-shell electronic structure of organic diradicals typically gives these molecules unique photophysical properties that have the potential to be harnessed in a number of novel technologies. One avenue that has garnered attention is the use of organic diradicals in a new generation of OLED devices \cite{Zhu2025, Abdurahman2023, Hattori2019}. In conventional closed-shell molecules, the excitation of a single electron will yield singlet and triplet excited states in a 1:3 ratio, meaning radiative decay (a.k.a. fluorescence) back to the ground singlet state is typically limited to a maximum internal quantum efficiency (IQE) of 25$\%$ \cite{Abdurahman2020}. In contrast, the near-degeneracy of low-lying singlet and triplet electronic states in diradicals means that there exist many more spin-allowed, radiative pathways back to both singlet and triplet states. Research is ongoing to develop diradical OLEDs with much improved IQEs, and more widely to develop optoelectronic technologies that take advantage of favourable spin-statistics. However, most organic diradicals synthesised to date have shown modest photoluminescence quantum yields of around $1-2\%$ \cite{Abdurahman2023, Chang2024, Yu2024, Kopp2024, Arikawa2023}, highlighting the need for design strategies that are guided by an understanding of diradical electronic structure.

The property that has perhaps attracted the most interest in organic luminescent diradicals is their capacity for magnetoluminescence and thermally activated emission \cite{Zhu2025}. Many emissive diradicals feature very small singlet-triplet energy gaps ($|\Delta E_\text{S-T}| \approx 1 - 10~ \text{meV}$), which is comparable to or smaller than $k_\text{B}T$ at $298$K. This means that the thermal populations of these states are highly sensitive to environmental conditions and can be manipulated with relative ease \cite{Hattori2019, Hattori2024, Tong2025, Abdurahman2023, Matsuoka2023, Mizuno2024}.

 In the case of magnetoluminescence, external magnetic fields can be applied that couple directly to the open-shell triplet state, lifting the degeneracy between states with different $M_s$ and decreasing the relative energy the $M_s = -1$ state \cite{WANG2026}. The result is an increase in the equilibrium population of the $\ket{^3\text{OS}}$ state, meaning the intensity of emissive transitions arising from triplets is increased compared to those associated with singlet states. In 2023, this was observed by Kusamoto et. al after their synthesis of the THDBA-(PyBTM)$_2$ molecule. By applying a magnetic field, they were able to enhance the intensity of a 645nm emission peak by 22\% while decreasing the intensity of a broad 750nm peak by 40\% \cite{Matsuoka2023}. A similar phenomenon that several diradicals have been found to display is thermally activated emission, whereby temperature can be used to influence the relative populations of $S_0$ and $T_0$ states and thus change the intensities of emissive transitions noticeably \cite{Abdurahman2023-1, Mizuno2024}.

Beyond their potential in optoelectronic devices, organic luminescent diradicals have recently emerged as promising candidates for molecular qubits in quantum information science. Solid-state qubit platforms, such as nitrogen-vacancy centres in diamond, have been heavily researched in recent years, but are limited by the restricted tunability of their optical and spin characteristics \cite{Poh2024}. Organic diradicals have attracted attention as potential alternatives, due to their high tuneability and relatively low cost \cite{Kopp2024}. Near degenerate singlet and triplet low-lying states provide a convenient framework for spin-optical interfaces, enabling optical initialisation and read-out of spin states often through spin-selective intersystem crossing pathways \cite{Poh2024}. In 2023, Evans and co-workers exploited this in the diradical molecule (TTM-1Cz)$_2$-An, where the near-degeneracy of the singlet and triplet manifolds facilitated the generation of a high-spin quintet state upon photoexcitation, which could subsequently be optically read out \cite{Gorgon2023}. Building on this, Wasielewski and co-workers established \textit{meta}-TTM-Ph-TTM as an optically addressable molecular qubit in 2024, demonstrating that the spin-polarised ground state arising from the diradical electronic structure could be coherently manipulated and read out optically \cite{Kopp2024}. Together, these results highlight the growing potential of organic diradicals as a chemically tuneable platform for quantum sensing technologies, where once again functionality relies on the precise characterisation of the energies and multiplicities of electronic states.

\subsection{Existing Computational Methods for Diradicals}

\subsubsection{Density-Based Methods}


Density functional theory (DFT) has long been the most widely used computational method across the chemical sciences. The use of exchange-correlation functionals ($V_{xc}[\rho]$) to describe electron correlation decreases the computational cost and scaling of DFT compared to most other \textit{ab-initio} methods, while still yielding sufficient qualitative accuracy for most applications. Time-dependent density functional theory (TD-DFT) can be seen as an extension of DFT that allows for prediction of excited electronic states through linear response theory. The computational cost of TD-DFT is also relatively low compared to other methods for predicting excitation energies, which largely explains its widespread use in the study of emissive diradical molecules and their photophysical properties \cite{Abdurahman2023-1, Kopp2024, Chowdhury2025, Hattori2024, Abdurahman2023, Schneburg2026, Arnold2025, Mesto2025, Chang2024, Tong2025, Hattori2019, Matsuoka2023, Mizuno2024}. However, the single-reference nature of DFT and TD-DFT limits the ability of these methods to reliably describe the electronic structure of diradical molecules. In almost all cases, DFT and TD-DFT are used with an unrestricted reference, meaning that the molecule's $\alpha$ and $\beta$ electrons are allowed to occupy orbitals with different spatial distributions. While this approach typically yields better energy estimates, the electronic states predicted, and particularly the excited states, are highly susceptible to spin contamination, whereby states of different multiplicity are mixed to yield wavefunctions of ill-defined total spin \cite{Pople1995}. Consequently, even in cases where TD-DFT excitation energies are in good agreement with experimental results, the spin multiplicities of the underlying states often cannot be reliably determined and the interpretive value of results is limited. Additionally, the results of DFT and TD-DFT calculations are known to be sensitive to the choice of exchange-correlation functional, with no universally optimal choice having been established for studying diradicals \cite{Sun2025, RetaMaeru2013}. A vast number of functionals have been developed over recent decades, with UB3LYP being the most widely adopted for use with emissive diradical molecules \cite{Abdurahman2023-1, Kopp2024, Chowdhury2025, Hattori2024, Abdurahman2023, Schneburg2026, Arnold2025, Mesto2025, Chang2024, Tong2025, Hattori2019, Matsuoka2023, Mizuno2024}.

A wide variety of approaches have been developed to add multiconfigurational character into DFT and address the issue of spin-contamination. Created over forty years ago, broken symmetry DFT (BS-DFT) offers a way to mimic the open-shell singlet state in diradicals by localising $\alpha$ and $\beta$ electrons in the HOMO onto opposite sites of the molecules \cite{Noodleman1981, Neese2004}. This creates an artificial 'symmetry broken' singlet state which can approximate the $\ket{^1\text{OS}}$ state but is still spin-contaminated meaning it is coupled to triplet and higher multiplicity states \cite{Ferr2015}. Reasonably accurate $\Delta E_{\text{S-T}}$ estimates can often be obtained regardless using spin-projection corrections such as that presented by Yamaguchi (Eq. 2), where $\langle S^2_\text{T} \rangle, \langle S^2_\text{BS} \rangle$ are the expectation values of the total spin operator for the triplet and broken symmetry singlet states respectively \cite{Yamaguchi1988}. Owing to their relative simplicity and low computational cost, broken-symmetry approaches remain a very popular way of approximating the multireference states of diradicals \cite{Chang2024, Abdurahman2023, Abdurahman2023-1, Kopp2024, Arnold2025, Schneburg2026, Mizuno2024, Matsuoka2023}. However, approximating spin-pure solutions through spin-projections of this kind has been shown to be unreliable in a variety of contexts including for organic molecules with high dynamic correlation \cite{Wittbrodt1996, Costa2015,  Rivero2008, Ferr2015, Trinquier2014}.

\begin{equation}
    \Delta E_\text{S-T} = E_\text{S} - E_\text{T} \approx \frac{\langle S^2_\text{T} \rangle}{\langle S^2_\text{T} \rangle - \langle S^2_\text{BS} \rangle}(E_\text{BS} - E_{T})
\end{equation}

Spin-flip time dependent density functional theory (SD-TDDFT) is another method that aims to approximate multi-reference diradical states within the single-reference formalism of TD-DFT \cite{Shao2003}. SF-TDDFT applies spin flips to a high-spin triplet reference determinant ($\ket{\uparrow\uparrow}$) to generate additional configurations ($\ket{\uparrow\downarrow}, \ket{\downarrow\uparrow}$) that are then used in a typical TD-DFT treatment. This approach can recover a decent amount of the mulit-configurational character of low-lying diradical electronic states (often referred to as static correlation), and can yield better estimates of $\Delta E_\text{S-T}$ than broken symmetry approaches \cite{Slipchenko2002, Valero2011}. However, the addition of spin-flips increases the computational cost of the method and typically struggles to produce spin-purity in the higher-lying electronic states of diradicals, meaning spin-contamination and accuracy remain an issue \cite{Sun2025}. This was addressed later by Ho Choi and colleagues with their creation of mixed-reference spin-flip TDDFT (MRSF-TDDFT), which includes a second reference determinant $\ket{\downarrow\downarrow}$ from which spin-flipped excitations can be generated \cite{Lee2018, Komarov2024}. This has been shown to generally improve the accuracy of $\Delta E_\text{S-T}$ predictions for diradical molecules compared to SF-TDDFT and also mostly removes spin-contamination from low-lying electronic states \cite{Horbatenko2021}. The MRSF-TDDFT method in particular shows promise for characterising the excited states of large emissive diradicals, but has not yet widely been adopted, perhaps due in part to its increased computational cost compared to conventional TD-DFT or its broken symmetry equivalent.

\subsubsection{Wave function-Based Methods}


Looking beyond density-based methods, there exist a number of mature wavefunction-based methods which extend the single-reference Hartree-Fock formalism. Perhaps the most widely used of these are multiconfigurational self-consistent field (MCSCF) approaches, which express electronic states as a linear combination of many configurations, which are formed by distributing a defined number of electrons across a defined set of active MOs (Eq. 3). The coefficients of the linear expansion ($c_I$) and the molecular orbital coefficients are simultaneously optimised to minimise the energy of one or more electronic states, depending on whether a state-specific or state-averaged approach is used \cite{Wahl1977}. The most widely used MCSCF method is the complete active space self consistent field (CASSCF) method \cite{Roos1980}, although others have been developed such as RASSCF (restricted active space SCF) which offer slightly more efficient computational scaling \cite{Olsen1988}. MCSCF methods are ideal for capturing multireference character, which has led to their frequent adoption for studying the electronic structure of small and medium-sized diradical and diradicaloid molecules \cite{Matasovi2024, RetaMaeru2013, Poh2024, Starikov2025}. 

\begin{equation}
    \ket{\Psi_\text{MCSCF}} = \sum_{I\in \text{active}}c_I\ket{\Phi_I}
\end{equation}

However, there are several drawbacks to MCSCF methods. Firstly, the accuracy of MCSCF results is highly dependent on an appropriate selection of active space \cite{Moriarty1998, Sun2025, RetaMaeru2013}. For organic diradicals, the frontier $\pi$ orbitals are an obvious choice, however, for large diradicals it is typically unfeasible to include the all $\pi$ MOs in the active space. Additionally, the computational cost of MCSCF methods typically scale very poorly as the size of the active space and the molecule increase, making them much more expensive than density-based alternatives like TD-DFT for simulating large molecules. Finally, while MCSCF methods are excellent for recovering static correlation, they do very little to address short-range repulsive interactions between electrons known as dynamic correlation. It has been known since the 1980s that dynamic correlation is crucial to obtain a quantitatively and sometimes qualitatively accurate description of electronic states even in small diradical molecules like cyclobutadiene \cite{Borden1996, Borden1982, CarlLineberger2011}.

There exist several methods that address the lack of dynamic correlation in MCSCF methods, which are broadly referred to as multi-reference perturbation theory (MRPT) methods. Examples of MRPT methods include CASPT2, which adds dynamic correlation into a CASSCF reference using second-order perturbation theory \cite{Andersson1992}, as well as N-Electron Valence State Pertubation Theory (NEVPT) and Generalised Multiconfigurational Quasi-Degenerate Perturbation Theory (GMC-QDPT) \cite{Angeli2002, Nakano2002}. These methods can typically capture a significant amount of the static and dynamic correlation present in diradicals, and often yield accurate predictions of singlet-triplet energy gaps \cite{Moriarty1998, Sun2025}. However, adding perturbation theory on top of an already expensive MCSCF treatment often makes MRPT calculations prohibitively expensive for larger molecules including emissive diradicals. Alongside this these methods notoriously suffer from intruder state problems that can lead to divergent perturbation expansions \cite{Roos1995}. A promising alternative to conventional MRPT methods, called multiconfigurational pair-density functional theory (MC-PDFT), was presented by Gagliardi and coworkers in 2014 \cite{LiManni2014}. MC-PDFT calculates the total electronic density from an MCSCF reference and then uses adapted versions of density functionals to add in dynamic correlation. This approach has been shown to estimate singlet-triplet gaps in organic diradicals with a similar accuracy to CASPT2, while being less computationally demanding \cite{Stoneburner2018, Sun2025}.

\subsubsection{Semi-Empirical Methods}


A final class of methods to consider are semi-empirical methods, which use parameters derived from experimental data to approximate the one- and two-electron integrals needed in \textit{ab-initio} methods. Of these semi-empirical formalisms, the most widely adopted for the description of organic diradicals is Parisier-Parr-Pople (PPP) theory \cite{Pariser1953, Pariser1953-II, Pople1953}, which has been used since the 1950s to describe the electrons in the conjugated $\pi$ systems of aromatic hydrocarbons. The main advantage of PPP theory is its computational efficiency, largely arising from simplifications made in the calculation of two electron integrals, which is otherwise of the most time consuming steps of a conventional SCF procedure \cite{PopleBeveridge1970}.

Several methods have been developed that combine PPP theory with configuration interaction (CI) and MCSCF methods in order to compute excited states for radical molecules. Like MCSCF methods, CI methods express electronic states as a linear combination of ground and excited electronic configurations, but skip the self-consistent optimisation of underlying MOs so are less computationally demanding. In 2024, Green and Hele created the Extended Restricted Open-shell PPP (ExROPPP) method for monoradicals by combining Maurice and Head-Gordon's extended configuration interaction singles (XCIS) method with a PPP Hamiltonian, enabling UV-vis absorption energies and intensities to be predicted accuracy comparable to GMC-QDPT but at a fraction of the cost \cite{Green2024, Maurice1996}. A similar approach was used by Di Maiolo et al. in their XCISDT-PPP method, which extended the CI basis to include double and triple excitations and again achieved qualitative agreement with CASPT2 for excitation energies of the monoradicals studied \cite{Dubbini2024}.

For diradicals specifically, Yuen-Zhou and colleagues presented a framework in 2024 for describing excited states of diradicals formed by coupling two monoradical 'monomers', using a PPP Hamiltonian for each monomer and treating the linker as a perturbation \cite{Poh2024}. Configuration interaction was then used with a spin-pure basis of local (intra-monomer) excitations and charge transfer (inter-monomer) excitations. The resulting energies showed good agreement with CASSCF calculations across the states considered, although the framework is by construction restricted to diradicals that decompose cleanly into two separated monomers. More recently, Di Maiolo and coworkers developed a PPP-RASCI method which combines PPP theory with the restricted active space configuration interaction (RASCI) method \cite{Olsen1988, Savi2025}. This has been applied to provide insight into the electronic structure of organic diradicals proposed as spin-optical interfaces, and once again results were in good agreement with those from a full CASSCF/QD-NEVPT2 treatment \cite{Barreca2026, Savi2025}.

\subsection{Summary}

In summary, organic diradical molecules possess a unique electronic structure that has motivated extensive research into their applications in optoelectronic and quantum sensing. However, computational methods have so far struggled to reliably provide insight into the photophysics of these molecules, largely due to the substantial static and dynamic correlation inherent in their electronic states. While accurate methods such as CASPT2 and MC-PDFT can rigorously treat this correlation, their high computational cost and reliance on a carefully chosen active space have severely limited their application to emissive diradicals. As a result, TD-DFT and its broken-symmetry variant remain the most popular choices for simulating the excited states of these molecules, owing largely to their relative simplicity and low computational cost. However, spin contamination is an unavoidable pitfall of these approaches, limiting the reliable characterisation spin states and electronic transitions. It remains to establish a method for calculating diradical electronic structure that can efficiently yield accurate and spin-pure results, while avoiding the complications of active space selection or fragment-based decomposition. The development of such a method is presented in this report, providing an efficient and accessible new tool for the interpretation of diradical absorption spectra.

\newpage

\section{Methodology}

Addressing the need for an efficient computational method capable of producing spin-pure, qualitatively accurate excited states for diradical molecules, a novel theoretical approach is presented that combines a multi-reference configuration interaction approach with Pariser-Parr-Pople (PPP) theory. This method builds on the extended restricted open-shell PPP (ExROPPP) approach developed by Green and Hele in 2024 \cite{Green2024}. While ExROPPP was designed for the calculation of excited states in monoradical molecules, this work applies a similar framework to the more complex case of diradical electronic structure, giving rise to a method hereafter referred to as Diradical-ExROPPP (D-ExROPPP). This section will present the theoretical results derived in this project to construct D-ExROPPP, alongside relevant details of its implementation into a custom Python program.

\subsection{Diradical Molecular Orbitals}

\subsubsection{Open-Shell SCF Procedure}

Restricted Hartree-Fock theory is inadequate for describing molecular orbitals in diradicals and similar open-shell systems, since it assumes that all occupied orbitals are doubly-occupied. Here, we introduce a self-consistent field (SCF) procedure for optimising the molecular orbitals (MOs) of diradicals, extending the presented by Longuet-Higgins and Pople \cite{HCLonguetHiggins1955}, and revisited more recently by Hele \cite{Hele2021}, for the treatment of monoradical MOs.

The optimal Hartree-Fock molecular orbitals are typically defined as those that minimise the energy of the ground state $\ket{\Psi_0}$, which for closed shell molecules is simply a single Slater determinant of doubly-occupied orbitals. Since diradicals lack a well-defined single reference ground state, we instead look for orbitals that minimise the energy of the open-shell singlet and triplet ground states, $\ket{^1\text{OS}}$ and $\ket{^3\text{OS}}$ (see Fig. 1). The derivation presented here follows the $\ket{^1\text{OS}}$, however, the same arguments are applied to $\ket{^3\text{OS}}$ in section 1A of the appendix. The first step is to recognise that any infinitesimal variation in MOs is equivalent to mixing small-amounts of singly-excited states into the reference states \cite{HCLonguetHiggins1955}.

\begin{equation}
    \begin{aligned}
    \ket{^1\text{OS}} \rightarrow \ket{^1\text{OS}} + \lambda_{-}\ket{^1\text{ZW}^{-}} + \lambda_{+}\ket{^1\text{ZW}^{-}} + \sum_i^k\lambda_{i0}\ket{^1\text{CS}_i^0} +& \sum_i^k\lambda_{i0'}\ket{^1\text{CS}_i^{0'}} + \sum_p^\infty\lambda_{0p}\ket{^1\text{SV}_0^p} \\ + \sum_p^\infty\lambda_{0'p}\ket{^1\text{SV}_{0'}^p}
    & + \sum_i^k\sum_p^\infty\lambda_{ip}\ket{^{1S}\text{CV}_i^p} + \sum_i^k\sum_p^\infty\lambda_{ip}\ket{^{1T}\text{CV}_i^p}
\end{aligned}
\end{equation}

\noindent where $i$ and $p$ are indices for doubly-occupied and virtual orbitals respectively, and $k$ is the number of doubly occupied (core) spatial orbitals. 

The kets shown in Eq. 4 ($\ket{^1\text{CS}_i^0}, \ket{^1\text{SV}_0^{j}},\text{etc.}$) are defined fully in the following section. For now it is only important to recognise that they are spin-adapted electronic states in which a single electron has been excited from one of the doubly occupied MOs or SOMOs of the open-shell ground state $\ket{^1\text{OS}}$. Note that mixing with triplet states does not need to be considered since Hamiltonian interactions between states of different multiplicity is always formally forbidden (i.e. $\bra{^1\Psi}\hat{H}\ket{^3\Psi} = 0$). An infinitesimal change in energy caused by an infinitesimal variation in MOs is thus given by Eq. 5.

\begin{equation}
\begin{aligned}
    \delta E = & \lambda_-\bra{^1\text{OS}}\hat{H}\ket{^1\text{ZW}^{-}} + \lambda_+\bra{^1\text{OS}}\hat{H}\ket{^1\text{ZW}^{+}} + \sum_i^k\lambda_{i0}\bra{^1\text{OS}}\hat{H}\ket{^1\text{CS}_i^0} \\ & + \sum_i^k\lambda_{i0'}\bra{^1\text{OS}}\hat{H}\ket{^1\text{CS}_i^{0'}} + \sum_p^\infty\lambda_{0p}\bra{^1\text{OS}}\hat{H}\ket{^1\text{SV}_0^p} + \sum_p^\infty\lambda_{0'p}\bra{^1\text{OS}}\hat{H}\ket{^1\text{SV}_{0'}^p} \\ &+ \sum_i^k\sum_p^\infty\lambda_{ip}\bra{^1\text{OS}}\hat{H}\ket{^{1S}\text{CV}_i^p} + \sum_i^k\sum_p^\infty\lambda_{ip}\bra{^1\text{OS}}\hat{H}\ket{^{1T}\text{CV}_i^p}
\end{aligned}
\end{equation}

By the variational priniciple, the best possible MOs to describe the open-shell singlet state of a diradical molecule are those for which $\delta E = 0$, which in turn requires that all of the above brakets are equal to 0. Using Slater-Condon rules to evaluate the Hamiltonian interactions between the ground and excited states, we obtain the following expressions for satisfying this requirement:

\begin{subequations}
\begin{align}
    \bra{^1\text{OS}}\hat{H}\ket{^1\text{ZW}^-} &= (00'|00) - (00'|0'0') = 0 \\
    \bra{^1\text{OS}}\hat{H}\ket{^1\text{ZW}^+} &= 2[F_{00'} + \frac{1}{2}(00'|00) + \frac{1}{2}(00'|0'0')] = 0 \\
    \bra{^1\text{OS}}\hat{H}\ket{^1\text{CS}_i^0} &= -F_{i0} - (i0|00) - (i0|0'0') + 2(i0'|0'0) = 0  \\
    \bra{^1\text{OS}}\hat{H}\ket{^1\text{CS}_i^{0'}} &= F_{i0'} + (i0'|00) + (i0'|0'0') - 2(i0|00') = 0 \\ 
    \bra{^1\text{OS}}\hat{H}\ket{^1\text{SV}_0^p} &= F_{0p} + (0p|0'0') + (00'|0'p) = 0 \\
    \bra{^1\text{OS}}\hat{H}\ket{^1\text{SV}_{0'}^p} &= F_{0'p} + (0'p|00) + (0'0|0p) = 0  \\
    \bra{^1\text{OS}}\hat{H}\ket{^{1S}\text{CV}_i^p} &= \sqrt{2}[F_{ip} + (ip|00) -\frac{1}{2}(i0|0p) + (ip|0'0') - \frac{1}{2}(i0'|0'p)] = 0 \\
    \bra{^1\text{OS}}\hat{H}\ket{^{1T}\text{CV}_i^p} &= \frac{\sqrt{6}}{2}[(i0'|0'p) - (i0|0p)] = 0 
\end{align}
\end{subequations}

\noindent where $F_{ip}$ is the conventional Fock-operator ($F_{ip} = h_{ip} + \sum_{l=1}^{k}[2(pq|ll) - (pl|lq)]$), $h_{ip}$ is the conventional one-electron integral ($h_{ip} = \bra{\psi_i}\hat{h}(1)\ket{\psi_p}$) and $(ij|kl)$ is a two electron integral written in chemist's notation, as shown in eq 7 \cite{szabo1989}.

\begin{equation}
    (ij|kl) = \int d\mathbf{r}_1 \int d\mathbf{r}_2 \psi_i^*(\mathbf{r}_1)\psi_j(\mathbf{r}_1) \frac{1}{r_{12}} \psi_k^*(\mathbf{r}_2)\psi_l(\mathbf{r}_2)
\end{equation}

For Eq. 6g in particular, we notice that the Fock operator $F_{ij}$ is accompanied by several extra two electron integrals prevent the requirement $\bra{^1\Psi_0}\hat{H}\ket{^{1S}\Psi_i^p} = 0$ from being satisfied. This braket represents the interaction between the open-shell singlet state and singly-excited electronic states where an electron is promoted from a doubly-occupied MO ($i$) to an empty orbital ($p$). In almost all molecules, there are many more states of this kind than excited states that promote electrons to and from the SOMOs ($0, 0'$) since typically there are many more doubly occupied and virtual orbitals than SOMOs. We are thus motivated to use the approximate diradical Fock-like operator ($F^\text{d}$) shown in Eq 8.

\begin{equation}
\begin{split}
F_{ab}^{\text{d}} & = F_{ab} + (ij|00) - \frac{1}{2}(a0|0b) + (ab|0'0') - \frac{1}{2}(a0'|0'b)] \\
& = h_{ab} + \sum_{l=1}^{k}[2(ab|ll) - (al|lb)] + \frac{1}{2}[2(ab|00) - (a0|0b) + 2(ab|0'0') - (a0'|0'b)]
\end{split}
\end{equation}

\noindent Substituting the expression in Eq. 8 back into equations 6a-h, we end up with equations 9a-h

\begin{subequations}
\begin{align}
    (00'|00) - (00'|0'0') &= 0 \\
    2F_{00'}^\text{d} &= 0 \\
    F_{i0'}^\text{d} + \frac{1}{2}(i0'|0'0') - \frac{3}{2}(i0|00') &= 0 \\
    F_{i0}^\text{d} + \frac{1}{2}(i0|00) - \frac{3}{2}(i0'|0'0) &= 0 \\
    F_{0'p}^\text{d} - \frac{1}{2}(0'p|00) + \frac{3}{2}(0'0|0p) &= 0 \\
    F_{0p}^\text{d} - \frac{1}{2}(0p|0'0') + \frac{3}{2}(00'|0'p) &= 0 \\
    \sqrt{2}F_{ip}^\text{d} &= 0 \\
    \frac{\sqrt{6}}{2}[(i0'|0'p) - (i0|0p)] &= 0 
\end{align}
\end{subequations}

Now, the extra two electron terms have either disappeared from equations 9a-9h or are reduced in magnitude. We can now make a similar approximation to Longuet-Higgins and Pople in assuming that the two electron terms ($(i0|00), (i0'|0'0),...$) that persist for brakets with excitations that involve the SOMOs will be small and so can be  ignored \cite{HCLonguetHiggins1955}. This approximation is made to simplify eqs 9a-h such that $F_{pq}^\text{d} = 0$ for all core, virtual and occupied molecular orbitals $p$ and $q$. This result allows us to conveniently adopt the conventional approach of iteratively solving the Roothan equations $\bf{F^\textbf{d}C} = \bf{SC\epsilon}$ to obtain molecular orbital coefficients \cite{szabo1989}, now using the diradical Fock-operator instead of the closed-shell variant. In practice, this is most conveniently implemented by changing the standard expression for the density matrix $P$ to a 'diradical' density matrix $P^\text{d}$ shown in eq 10 (where $k$ is once again an index for all doubly-occupied MOs). Using the diradical density matrix with the standard Fock operator is shown in section I-A of the appendix to be equivalent to using the standard density matrix with the diradical Fock operator.

\begin{equation}
\begin{aligned}
    P_{\mu \nu} &= \sum_{1=1}^k 2C_{\mu l}^{*} C_{\nu l}  \\
    \rightarrow P_{\mu \nu}^{\text{d}} &= \sum_{1=1}^k (2C_{\mu l}^{*} C_{\nu l}) + C_{\mu 0}^* C_{\nu 0} + C_{\mu 0'}^* C_{\nu 0'}
\end{aligned}
\end{equation}

\subsubsection{PPP Theory}

Having introduced the diradical Fock operator, $F^\text{d}$, and density matrix, $P^\text{d}$, we now require a way to calculate the one- and two-electron integrals necessary for the SCF procedure. For this, we turn to the semi-empirical formalism developed by Pariser, Parr and Pople in 1953 to describe the electronic structure of $\pi$-electrons in unsaturated hydrocarbons \cite{Pople1953, Pariser1953, Pariser1953-II}. PPP theory has been proven to accurately reproduce the excited spectra of $\pi$-conjugated closed-shell molecules \cite{Hele2019, Alvertis2019} as well as more recently in open-shell molecules \cite{Green2024, Dubbini2024, Savi2025, Poh2024}. It's main advantage is it's minimal computational cost relative to \textit{ab-initio} methods, which is largely the result of several approximations that we briefly describe below.

The first approximation is use of linear combinations of atomic orbitals (LCAO) to represent molecular orbitals. As opposed to the standard Gaussian basis functions, MOs ($\phi$) are expanded as a weighted sum of atom-centred basis functions,
\begin{equation}
    \phi(\mathbf{r}) = \sum_\mu C_{\mu p} \chi_\mu(\mathbf{r}),
\end{equation}
where $\chi_\mu$ represents a $2p_z$ atomic orbital for an atom $\mu$. This is typically a Carbon atom but can be any atom with one or two electrons in the $2p_z$ AOs that constitute the $\pi$-bonding system (e.g. Nitrogen, Chlorine, etc.). In PPP theory only the $\pi$-electron framework is treated explicitly, with the $\sigma$-bonded core absorbed into an effective potential. A closely related approximation is the neglect of differential overlap (NDO), where the overlap integral between any two atomic orbitals on different atoms is set to zero. This has two key implications, the first being that the overlap matrix $\bf{S}$ reduces to the identity matrix and so does not play a role in the SCF procedure, as it typically would when gaussian basis functions are used. Secondly, the four-index two-electron integrals that are routinely evaluated during the SCF procedure collapse to a single two-index quantity as shown in Eq. 12.

\begin{equation}
    (\mu\nu|\rho\sigma) = \delta_{\mu\nu}\delta_{\rho\sigma}\gamma_{\mu\rho} = \frac{U}{(1 + \frac{r_{\mu\nu}}{r_0})}
\end{equation}

$\gamma_{\mu\rho}$ is a function of the distance between two atoms, here defined with the Mataga-Nishimoto form \cite{Mataga1957} that is used in the D-ExROPPP implementation. U is a semi-empirical parameter that describes the Coulomb repulsion for a given atom. Using Eq. 12 yields a dramatic reduction in computational cost compared to evaluating two-electron integrals from first principles, and lowers the computational scaling from $O(N^4)$ to $O(N^2)$ where N is the number of basis functions \cite{PopleBeveridge1970}. The remaining one- and two-electron integrals are similarly approximated by semi-empirical parameters. The diagonal core integrals $h_{\mu\mu}$ are replaced by valence state ionisation potentials ($W_\mu$) which approximate the energy of an electron in the field of its own core. The off-diagonal core integrals $h_{\mu\nu}$ are replaced with resonance integrals ($\beta_{\mu\nu}$) that are only non-zero only between bonded atoms. The end result, when combined with the diradical density matrix derived previously, is the effective Fock-operator given in eq. 13 \cite{Jug1990}.

\begin{subequations}
\begin{align}
    F_{\mu \mu}^{\text{c}} &= W_{\mu} + \frac{1}{2}P_{\mu \mu} (\mu\mu|\mu\mu) + \sum_{\rho \ne \mu}(P_{\rho \rho} - Z_{\rho})(\mu \mu | \rho \rho) \\
    F_{\mu \nu}^{\text{c}} &= \beta_{\mu \nu} - \frac{1}{2}P_{\mu \nu} (\mu\mu|\nu\nu)
\end{align}
\end{subequations}

As well as offering drastic improvements to computational speed, the fitting of parameters to empirical data allows them to typically capture some of the dynamical correlation between electrons in $\pi$-conjugated systems, which is entirely missing from the \textit{ab-initio} Hartree-Fock model. To enhance this capability, D-ExROPPP makes use of machine-learned parameters that were optimised by Shen \textit{et al.} for the prediction of monoradical excited states \cite{Shen2025}. While no diradicals were included in the training set for this study, a sizeable portion of the data originated from monoradicals based on TTM, PyBTM and similar molecules. Since many emissive diradicals are formed as dimers of TTM and PyBTM monoradicals, we expect these parameters to similarly capture their excited states

\subsection{Spin-Pure Diradical Excited States}

\subsubsection{Constructing Configuration State Functions}

While the two electron two-orbital model (TOTEM) provides a good description of the low-lying electronic states of diradical molecules, it is unable to capture the higher energy excitations that play a key role in their absorption and emission properties. To address this, we begin by extending the TOTEM model to include the highest occupied molecular orbital (HOMO) and the lowest unoccupied molecular orbital (LUMO), thus establishing a four-electron four-orbital model (FOFEM). Figure 4 shows the three general ways that a single electron can be excited from the low-lying Slater determinants of a diradical. The HOMO is more generally referred to as a core orbital, whereas the general name for the LUMO is a virtual orbital. Excitations described with the FOFEM are conveniently generalised to higher-energy excitations involving any core or virtual MO in the $\pi$ system of diradicals. In these cases, core orbitals are represented with indices $i,j,k,l$ and virtual orbitals are given indices $p,q,r,s$.

\begin{figure}[H]
\centering
\includegraphics[width=0.98\linewidth, height=0.69\linewidth]{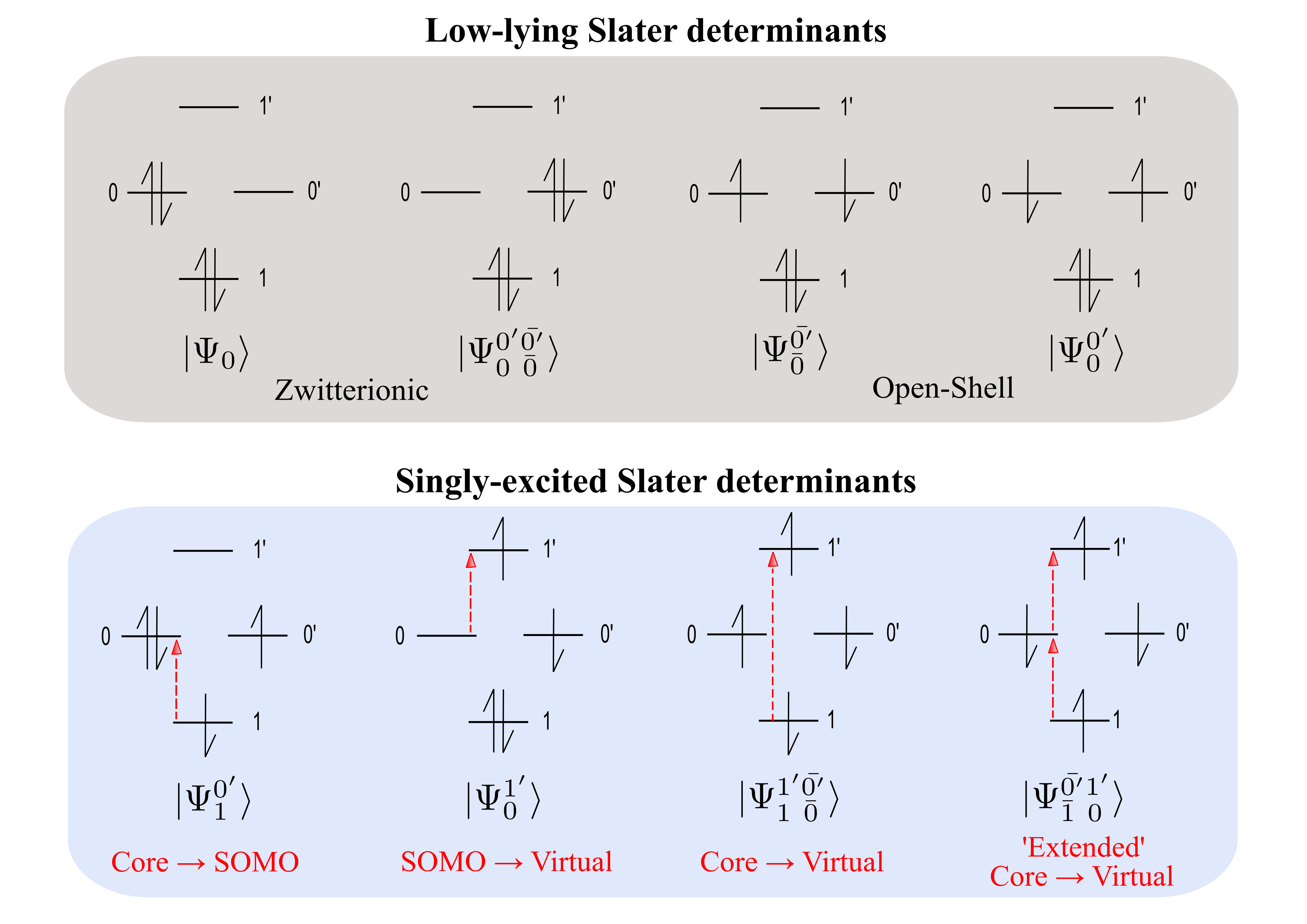}
\caption{Examples of electronic configurations involved in describing diradical electronic structure within the FOFEM model. Indices with a ' refer to virtual orbitals, while no prime suggests a core orbital (not including 0 and 0'). In the absence of well-defined, single-reference ground state, we arbitrarily label $\ket{\Psi_0}$ to be the Slater determinant for the zwitterionic state with the HOMO and SOMO(0) doubly occupied. This choice is made to allow other Slater determinants to be more concisely represented with ket vectors that show which spin-orbitals are occupied.}
\end{figure}

Since none of the excited Slater determinants are eigenstates of the $\hat{S^2}$ operator, spin-adapted linear combinations, known as configuration state functions (CSFs), must be formed to ensure spin-purity \cite{Helgaker2000}. For core $\rightarrow$ SOMO (CS) and SOMO $\rightarrow$ LUMO (SV) excitations, the number of unpaired electrons in the system is conserved which makes it relatively straightforward to form CSFs. Similar to the $\ket{^1\text{OS}}$ and $\ket{^3\text{OS}}$ states, taking symmetric and antisymmetric combinations excited Slater determinants with opposite spin orientations in the unpaired MOs yields spin-pure singlets and triplets respectively (Table 1). Constructing CSFs to describe core $\rightarrow$ virtual (CV) excitations becomes more complicated. Moving an electron from a doubly-occupied core MO increases the number of unpaired electrons in the system from two to four, thus increasing the number of ways that the two $\alpha$ and two $\beta$ electrons can be distributed among the four singly-occupied orbitals. Furthermore, Maurice and Head-Gordon previously showed for monoradicals that CSFs for CV excitations can only be formed with the inclusion of specific doubly-excited configurations that simultaneously excite an electron from a core to a virtual orbital, while also flipping the spin of the singly occupied orbital \cite{Maurice1996}. These 'extended single' excitations (Fig. 4) are the key insight from their extended configuration interaction singles (XCIS) method, which was designed as a spin-pure formulation of standard CIS for monoradical molecules. Building from this, we are left us with a basis of six excited Slater determinants needed to describe the excitation of one electron from a core orbital to a virtual orbital.

\begin{table}[H]
\caption{Singlet and triplet configuration state functions for excitations from an arbitrary core orbital, denoted by the index $i$, to either of the SOMOs ($0$ or $0'$), and from either of the SOMOs ($0$ or $0'$) to an arbitrary virtual orbital, denoted by the index $p$.}
    \centering
    \renewcommand{\arraystretch}{1.2} 
    \begin{tabular}{p{6.5cm}|p{6.5cm}}
         Core $\rightarrow$ SOMO CSFs & SOMO $\rightarrow$ Virtual CSFs \\ \hline
         $\ket{^3\text{CS}_i^{0}} = \frac{1}{\sqrt{2}}
        (\ket{\Psi_{i}^{0'}} - \ket{\Psi_{\bar{i}}^{\bar{0'}}})$ & 
        $\ket{^3\text{SV}_0^p} = \frac{1}{\sqrt{2}}
        (\ket{\Psi_{0 ~\bar{0}}^{p ~\bar{0'}}} - \ket{\Psi_{\bar{0}~0}^{\bar{p} ~0'}})$ \\
        $\ket{^1\text{CS}_i^{0}} = \frac{1}{\sqrt{2}}
        (\ket{\Psi_{i}^{0'}} + \ket{\Psi_{\bar{i}}^{\bar{0'}}})$ & 
        $\ket{^1\text{SV}_0^p} = \frac{1}{\sqrt{2}}
        (\ket{\Psi_{0 ~\bar{0}}^{p ~\bar{0'}}} + \ket{\Psi_{\bar{0}~0}^{\bar{p} ~0'}})$ \\
        
        $\ket{^3\text{CS}_i^{0'}} = \frac{1}{\sqrt{2}}
        (\ket{\Psi_{i~\bar{0}}^{0'\bar{0'}}} - \ket{\Psi_{\bar{i} ~0}^{\bar{0'} 0'}})$ & 
        $\ket{^3\text{SV}_{0'}^p} = \frac{1}{\sqrt{2}}
        (\ket{\Psi_{0}^{p}} - \ket{\Psi_{\bar{0}}^{\bar{p}}})$ \\
        $\ket{^1\text{CS}_i^{0'}} = \frac{1}{\sqrt{2}}
        (\ket{\Psi_{i~\bar{0}}^{0'\bar{0'}}} + \ket{\Psi_{\bar{i} ~0}^{\bar{0'} 0'}})$ & 
        $\ket{^1\text{SV}_{0'}^p} = \frac{1}{\sqrt{2}}
        (\ket{\Psi_{0}^{p}} + \ket{\Psi_{\bar{0}}^{\bar{p}}})$ \\ 

    \end{tabular}
\end{table}

To obtain CSFs, the matrix representation of the total-spin ($S^2$) operator was calculated in the basis of these six determinants, where the $S^2$ operator is defined as $\hat{S}^2 \ket{\Phi} = (\hat{S}_-\hat{S}_+ + \hat{S}_z^2 + \hat{S}_z)\ket{\Phi}$ \cite{szabo1989}. $\hat{S_+}$ and $\hat{S_-}$ are the standard raising and lowering operators for spin, while $\hat{S_z}$ operator is simply the $z$-component of the spin-angular momentum operator which is $0$ for all determinants included.

\begin{equation}
\bf{\hat{S^2}} =
\kbordermatrix{
& \ket{\Psi_{\bar{1} ~\bar{0}}^{\bar{1'} \bar{0'}}} &
\ket{\Psi_{1 ~\bar{0}}^{1' \bar{0'}}} &
\ket{\Psi_{\bar{1} ~0}^{\bar{0'} 1'}}  &
\ket{\Psi_{1 ~\bar{0}}^{0' \bar{1'}}} &
\ket{\Psi_{\bar{1} ~0}^{\bar{1'} 0'}} &
\ket{\Psi_{1 ~0}^{1' 0'}}\\
    & ~~~2 & -1 & ~~~1 & ~~~1 & -1 & ~~~0 \\
    & -1 & ~~~2 & -1 & -1 & ~~~0 & -1 \\
    & ~~~1 & -1 & ~~~2 & ~~~0 & -1 & ~~~1 \\
    & ~~~1 & -1 & ~~~0 & ~~~2 & -1 & ~~~1 \\
    & -1 & ~~~0 & -1 & -1 & ~~~2 & -1 \\
    & ~~~0 & -1 & ~~~1 & ~~~1 & -1 & ~~~2 \\
}
\end{equation}

The eigenvectors and eigenvalues of the matrix shown in Eq 14 were calculated using a standard linear algebra package in Python. This produced two eigenvectors with an eigenvalue of 0, three with an eigenvalue of 2 and one with an eigenvalue of 6. Using the formula to determine the multiplicity of an electronic state ($\hat{S}^2 \ket{\Phi} = S(S+1) \ket{\Phi}$) these CSFs can be recognised as two singlets ($S=0$), three triplets ($S=1$) and one quintet ($S=2$). The CSFs obtained are given in table A1 in the appendix.

Of these, only the quintet CSF is a non-degenerate eigenvector and thus uniquely defined. For the triplets and singlets, it follows from standard linear algebra that any linear combination of CSFs with the same multiplicity will also be a valid CSF of that multiplicity. Recognising this, we sought to further differentiate the singlets and triplets based on their transition dipole moments with the low-lying singlet and triplet open-shell states. The transition dipole moment operator ($\hat{\mu}$) is a one-electron operator, meaning that analytical expressions for transition dipole moments can be computed relatively easily according to the Slater-Condon rules \cite{szabo1989}. By taking linear combinations of the CSFs obtained from the $\hat{S^2}$ matrix, we are able to concentrate transition dipole moment intensity into only of the singlets and triplets (Table 2). These 'bright' singlet and triplet states are denoted as 1S and 3T to highlight that they can be accessed by dipole-allowed electronic transitions from $\ket{^1\text{OS}}$ and $\ket{^3\text{OS}}$ respectively, wheras transitions to and from the 'dark' 1T, 3S and 3X states are formally forbidden. On the whole, while this specific basis of core $\rightarrow$ virtual CSFs is not uniquely defined, it is expected to make interpretation of absorption and emission spectra more convenient.

\begin{table}[H]
\caption{Core $\rightarrow$ virtual CSFs representing excitations from a generic core orbital ($i$) to a generic virtual orbital ($p$) alongside their transition dipole moments with the open-shell CSFs. Transitions between states of different multiplicity are always formally forbidden meaning TDMs between singlets and triplets are all 0.}
    \centering
    \begin{tabular}{|m{8.5cm}|m{5.5cm}|} 
    \hline
    \multicolumn{1}{|c|}{\textbf{Core $\rightarrow$ Virtual CSF}} & \multicolumn{1}{|c|}{\textbf{Transition dipole moment}} \\ \hline
    
    \makecell{$\begin{aligned}
    \rule{0pt}{4ex} \ket{^{5}\text{CV}_i^{p}} = \tfrac{1}{\sqrt{6}}
        (&\ket{\Psi_{i ~\bar{0}}^{p \bar{0'}}} - \ket{\Psi_{\bar{i} ~\bar{0}}^{\bar{p} \bar{0'}}} -
        \ket{\Psi_{\bar{i} ~0}^{\bar{0'} p}} \\  + & 
        \ket{\Psi_{\bar{i} ~0}^{\bar{p} 0'}} -
        \ket{\Psi_{i ~0}^{p 0'}} -
        \ket{\Psi_{i ~\bar{0}}^{0' \bar{p}}}) \rule[-2ex]{0pt}{0pt}
    \end{aligned}$} 
    & \makecell{
    $\begin{aligned}
        &\bra{^3\text{OS}}\mu \ket{^{5}\text{CV}_i^{p}} = 0 \\
        &\bra{^1\text{OS}}\mu \ket{^{5}\text{CV}_i^{p}} = 0
    \end{aligned}$} \\ \hline

    \makecell{$\ket{^{3T}\text{CV}_i^{p}} = \frac{1}{2}(\ket{\Psi_{\bar{i} ~\bar{0}}^{\bar{p} \bar{0'}}} +
        \ket{\Psi_{i ~\bar{0}}^{p \bar{0'}}} -
        \ket{\Psi_{i ~0}^{p 0'}} -
        \ket{\Psi_{\bar{i} ~0}^{\bar{p} 0'}})$} 
    & 
    \makecell{$\begin{aligned}
        &\bra{^3\text{OS}}\mu \ket{^{3T}\text{CV}_i^{p}} = \sqrt{2}\mu_{ip} \rule[-1ex]{0pt}{0pt}
    \end{aligned}$} \\ \hline
    
    \makecell{$\ket{^{3S}\text{CV}_i^{p}} = \frac{1}{2}(\ket{\Psi_{\bar{i} ~\bar{0}}^{\bar{p} \bar{0'}}} +
        \ket{\Psi_{\bar{i} ~0}^{\bar{p} 0'}} -
        \ket{\Psi_{i ~0}^{p 0'}} -
        \ket{\Psi_{i ~\bar{0}}^{p \bar{0'}}})$ }
    & 
    \makecell{$\begin{aligned}
        &\bra{^3\text{OS}}\mu \ket{^{3S}\text{CV}_i^{p}} = 0 \rule[-1ex]{0pt}{0pt}
    \end{aligned}$} \\ \hline

    \makecell{$\ket{^{3X}\text{CV}_i^{p}} = \frac{1}{\sqrt{2}}
        (\ket{\Psi_{i ~\bar{0}}^{0' \bar{p}}} -
        \ket{\Psi_{\bar{i} ~0}^{\bar{0'} p}})$} 
    & 
    \makecell{$\begin{aligned}
        &\bra{^3\text{OS}}\mu \ket{^{3X}\text{CV}_i^{p}} = 0 \rule[-1ex]{0pt}{0pt}
    \end{aligned}$} \\ \hline

    \makecell{$\ket{^{1S}\text{CV}_i^{p}} = \frac{1}{2}(\ket{\Psi_{\bar{i} ~\bar{0}}^{\bar{p} \bar{0'}}} +
        \ket{\Psi_{\bar{i} ~0}^{\bar{p} 0'}} +
        \ket{\Psi_{i ~0}^{p 0'}} +
        \ket{\Psi_{i ~\bar{0}}^{p \bar{0'}}})$}
    & 
    \makecell{$\begin{aligned}
        &\bra{^1\text{OS}}\mu \ket{^{1S}\text{CV}_i^{p}} = \sqrt{2}\mu_{ip} \rule[-1ex]{0pt}{0pt}
    \end{aligned}$} \\ \hline

    \makecell{$\begin{aligned}
    \rule{0pt}{4ex} \ket{^{1T}\text{CV}_i^{p}} = \tfrac{1}{2\sqrt{3}}(&2\ket{\Psi_{\bar{i} ~0}^{\bar{0'} p}} +
        \ket{\Psi_{i ~\bar{0}}^{p \bar{0'}}} -
        \ket{\Psi_{\bar{i} ~\bar{0}}^{\bar{p} \bar{0'}}} \\ + & 
        2\ket{\Psi_{i ~\bar{0}}^{0' \bar{p}}} +
        \ket{\Psi_{\bar{i} ~0}^{\bar{p} 0'}} -
        \ket{\Psi_{i ~0}^{p 0'}}) \rule[-2ex]{0pt}{0pt} 
    \end{aligned}$} 
    & 
    \makecell{$\begin{aligned}
        &\bra{^1\text{OS}}\mu \ket{^{1T}\text{CV}_i^{p}} = 0
    \end{aligned}$} \\ \hline
    
    \end{tabular}
\end{table}

Combining the CV CSFs with those representing CS and SV excitations, we have a complete, spin-pure description of single excitations in diradical molecules. However, many emissive diradicals including many based on the TTM radical have a property known as alternacy symmetry, a consequence of which is that their core and virtual molecular orbitals come in paired sets with energies that are symmetrically distributed about the energy of the SOMOs ($\epsilon_\text{SOMO} - \epsilon_\text{core} = \epsilon_\text{virt} - \epsilon_\text{SOMO}$) \cite{Hele2021, Cho2025, Poh2024}. In diradicals with this property, the energy gap between a core MO and its alternant virtual pair will be very similar or identical to twice the energy gap from that core MO to one of the SOMOs, or twice the energy gap from one of the SOMOs to that virtual MO. This implies that double excitations of this kind will have similar transition energies to those of single core $\rightarrow$ virtual excitations. This motivates the inclusion of these specific double excitations into our existing basis of singly-excited CSFs (Fig. 5). 

\begin{figure}[H]
\centering
\includegraphics[width=0.88\linewidth, height=0.45\linewidth]{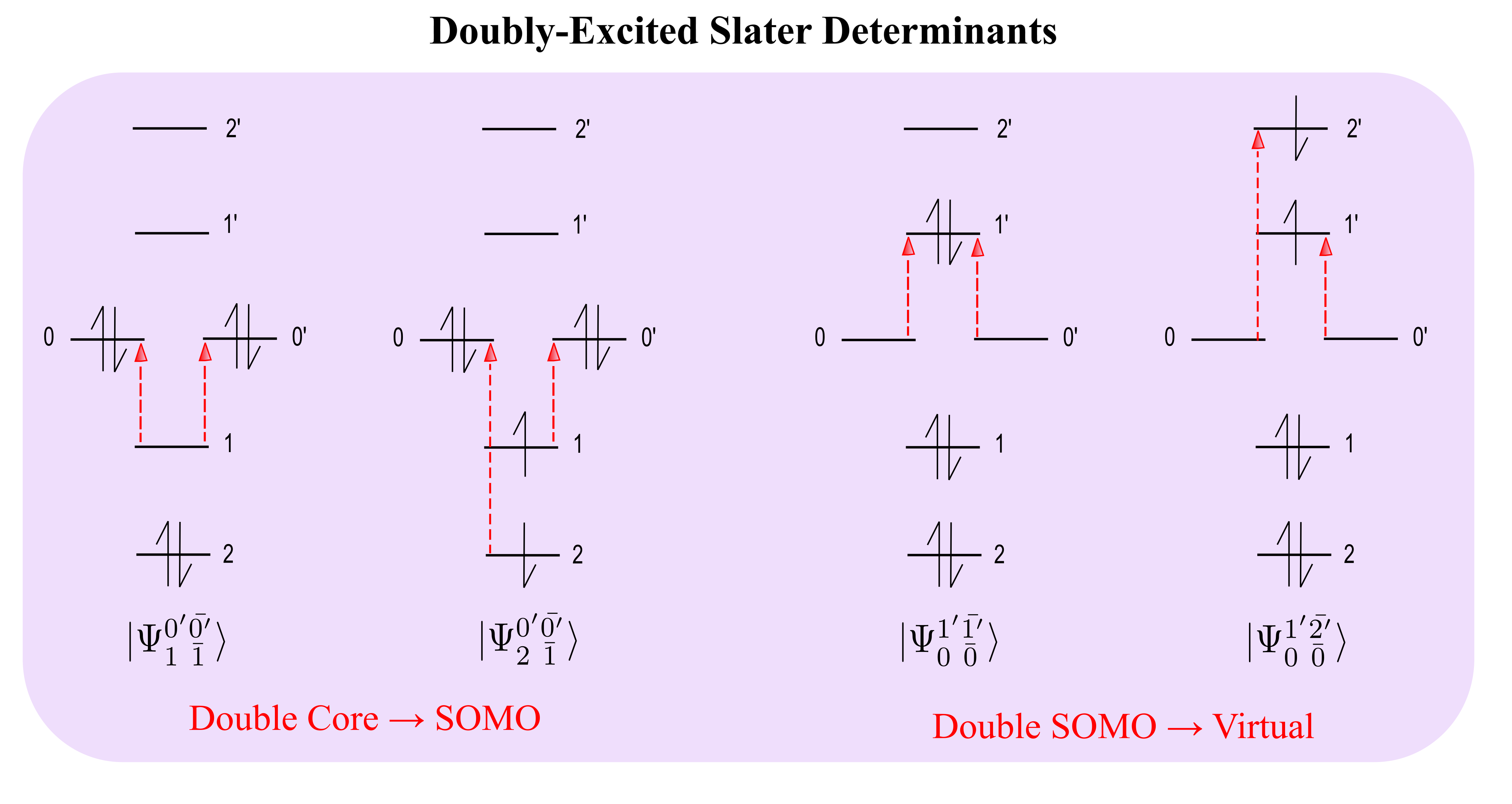}
\caption{Examples of doubly-excited electronic configurations included in the description of diradical excited states.}
\end{figure}

For these core to SOMO double (CSD) excitations and SOMO to virtual double (SVD) excitations, the formation of CSFs is mostly straightforward. As shown in figure 5, in the case that both electrons are excited either from the same core orbital or to the same virtual orbital, all MOs become doubly occupied meaning these configurations can be described straightforwardly with a single Slater determinant of singlet multiplicity. If CSD excitations are from two different MOs, however, we once again have one unpaired $\alpha$ and $\beta$ electron across two MOs so need to form +/- singlet and triplet CSFs. The same applies to SVD excitations that excite electrons to different virtual orbitals.

\begin{table}[h]
\caption{Slater determinants and CSFs describing the excitation of two electrons from core orbitals ($i, j$) $\rightarrow$ SOMOs and from the SOMOs $\rightarrow$ virtual orbitals ($p, q$). In the case that excitations are from/to the same core/virtual orbital, electrons in all occupied orbitals are paired meaning they are spin-pure singlets and don't need to be spin-adapted.}
    \centering
    \renewcommand{\arraystretch}{2} 
    \begin{tabular}{p{6.5cm}|p{6.5cm}}
        Double Core $\rightarrow$ SOMO CSFs & Double SOMO $\rightarrow$ Virtual CSFs \\ \hline
         $\ket{^1\text{CS}_{i ~\bar{i}}^{0' \bar{0'}}} = \ket{\Psi_{i ~ \bar{i}}^{0' \bar{0'}}}$ & $\ket{^1\text{SV}_{0' \bar{0'}}^{p ~\bar{p}}} = \ket{\Psi_{0' \bar{0'}}^{p ~\bar{p}}}$ \\
         
        $\ket{^3\text{CS}_{i ~\bar{j}}^{0' \bar{0'}}} = \frac{1}{\sqrt{2}} (\ket{\Psi_{i~\bar{j}}^{0' \bar{0'}}} - \ket{\Psi_{\bar{i}~j}^{\bar{0'} 0'}})$ & $\ket{^3\text{SV}_{0' \bar{0'}}^{p ~\bar{q}}} = \frac{1}{\sqrt{2}} (\ket{\Psi_{0'\bar{0'}}^{p ~\bar{q}}} - \ket{\Psi_{\bar{0'}0'}^{\bar{p}~ q}})$
          \\

        $\ket{^1\text{CS}_{i ~\bar{j}}^{0' \bar{0'}}} = \frac{1}{\sqrt{2}} (\ket{\Psi_{i~\bar{j}}^{0' \bar{0'}}} + \ket{\Psi_{\bar{i}~j}^{\bar{0'} 0'}})$ & $\ket{^1\text{SV}_{0' \bar{0'}}^{p ~\bar{q}}} = \frac{1}{\sqrt{2}} (\ket{\Psi_{0'\bar{0'}}^{p ~\bar{q}}} + \ket{\Psi_{\bar{0'}0'}^{\bar{p}~ q}})$
        \\

    \end{tabular}
\end{table}

\subsubsection{Configuration Interaction}

Having established a spin-pure basis of configuration state functions to represent all single and low-energy double excitations in diradical molecules, configuration interaction is used to calculate excited states and their energies. We refer to this novel implementation as extended configuration interaction singles with selected doubles (XCIS-D). This approach goes beyond extended configuration interaction singles, with the inclusion of a subset of double excitations, but stops short of XCISD since certain double excitations such as double core $\rightarrow$ virtual excitations are excluded. The Hamiltonian matrix is constructed in the CSF basis using analytical expressions for matrix elements that were derived using a custom Python script to implement Slater-Condon rules. Diagonal elements ($H_{aa} = \bra{\Phi_a}\hat{H}\ket{\Phi_a}$) are the approximate energies of each CSF and their expressions can be found in Table A2 of the Appendix. Off-diagonal elements ($H_{ab} = \bra{\Phi_a}\hat{H}\ket{\Phi_b}$) are energetic couplings between CSFs, where large couplings cause CSFs to mix strongly into a stabilised lower-energy state and a destabilised higher-energy state \cite{Shavitt1977}. Analytical expressions obtained for all couplings in the XCIS-D matrix are given in tables A3-A5.

\begin{figure}[H]
\centering
\includegraphics[width=1\linewidth, height=0.5\linewidth]{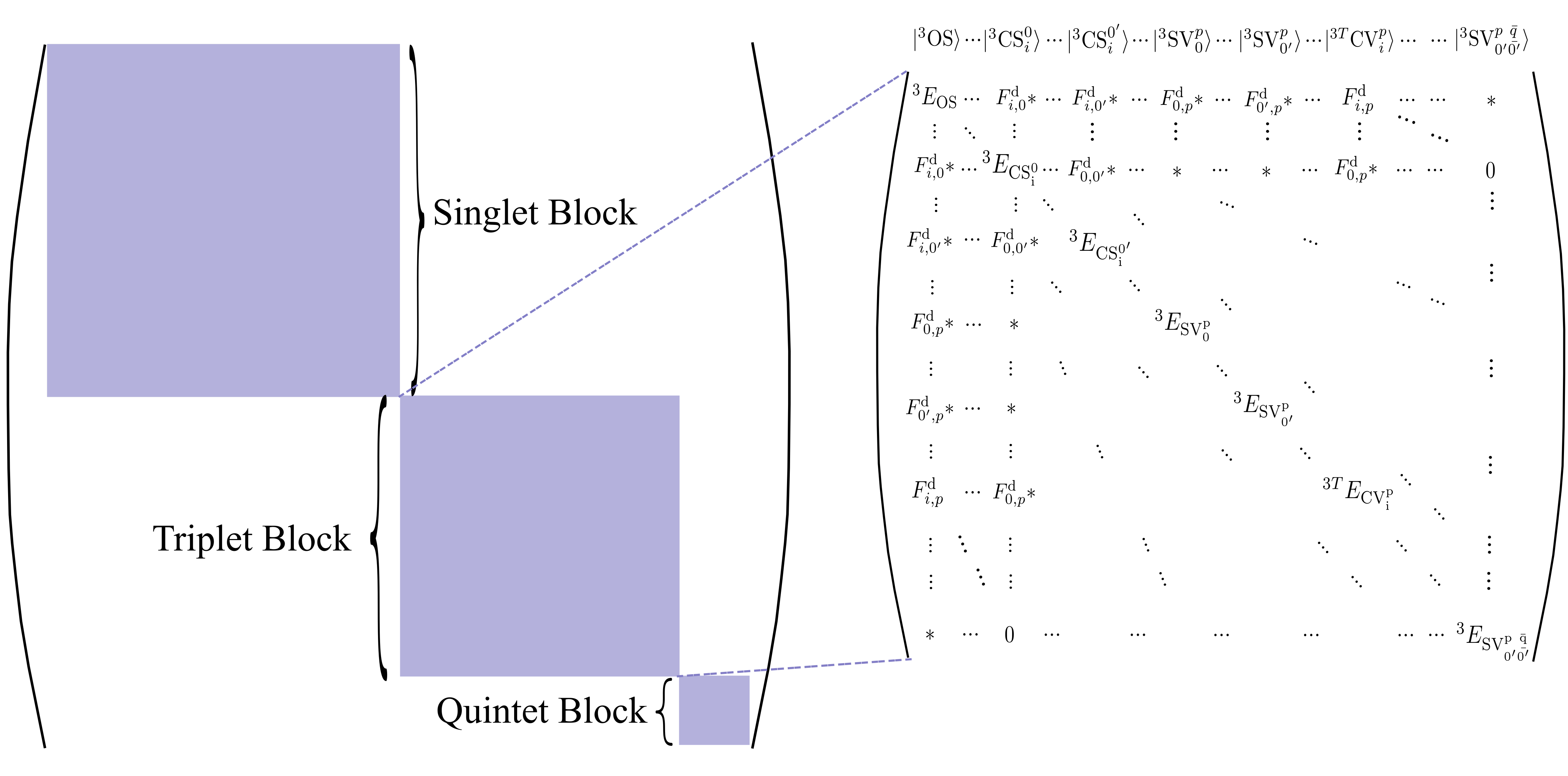}
\caption{A depiction of the structure of the XCIS-D matrix. Hamiltonian matrix elements between CSFs with different multiplicities are always 0, creating a distinct block-diagonal structure. The triplet block is highlighted, showing the diagonal matrix elements which represent the energies of each CSF as well as the off-diagonal couplings. $F_{a,b}^\text{d}$ is the diradical Fock operator defined previously. Stars (*) indicate the presence of mixed two-electron terms e.g. $(ij|kl)$ that are not shown for conciseness.}
\end{figure}

Diagonalisation of the XCIS-D matrix yields the approximate energies of electronic states (eigenvalues) and their configuration coefficients that give the weightings of each CSF to the state (eigenvectors). Diagonalising the full matrix is typically inefficient meaning lowest energy states are instead iteratively obtained using the Lanczos algorithm, implemented in the ARPACK library \cite{Lehoucq1998}. The use of CSFs rather than Slater determinants as the CI basis yields two significant benefits. Firstly, electronic states are guaranteed to be spin-pure by construction \cite{Shavitt1977, Maurice1996}, and secondly, since matrix elements between CSFs of different multiplicities are zero, the XCIS-D matrix has a block-diagonal structure. This enables the use of a more efficient divide-and-conquer approach in which singlet, triplet, and quintet blocks are diagonalised independently. Although not yet implemented in D-ExROPPP, singlet, triplet, and quintet blocks can also be diagonalised in parallel, which could provide further speedups to calculations in the future.

XCIS-D uses an efficient basis for configuration interaction compared to other CI and MCSCF methods that include higher-order excitations. For a $\pi$ system where the number of doubly occupied MOs is equal to the number of virtual MOs ($N_\text{core}=N_\text{virt}$), as is often the case when using an LCAO basis, the XCIS-D matrix contains $(8*N_\text{core}^2 + 8 * N_\text{core} + 4)$ CSFs, representing a $ O(N^2_\text{core})$ scaling in big-O notation. This low polynomial scaling allows the inclusion of excitations across all $\pi$ MOs, removing the need to define an active space to make CI computationally tractable. Including all core $\rightarrow$ virtual double excitations in a full XCISD treatment would raise the scaling to $O(N^4_\text{core})$. By contrast, active-space methods like CASSCF and RASCI feature scale much less favourably, with the number of Slater determinants typically growing combinatorially with the size of the active space. This is the main reason that these methods can be prohibitively expensive for large molecules, even when the chosen active spaces only captures a small subspace of the $\pi$ system.

\subsubsection{Simulating Spectra}

In order to evaluate transition intensities between electronic states obtained from XCIS-D, the dipole moment operator $\bf{\hat{\mu}}$ was calculated in the same basis of CSFs as $\bf{\hat{H}}$. Similarly, a custom python script implementing Slater-Condon rules was used to obtain analytical expressions for each of these matrix elements, which can be found in tables A6 - A8 in the appendix. The point charge approximation is used to calculate the dipole moment between two orbitals $\mu_{ij}$ with Eq. 15 where n is the number of atoms in the molecule and $C_{\nu i}$ is the coefficient for MO $i$ on atom $\nu$.

\begin{equation}
    \mu_{ij} = \sum _\nu^n C_{\nu i} ~\cdot \textbf{r}_{\nu} ~ \cdot~ C_{\nu j}
\end{equation}

Transition dipole interactions between CSFs are then contracted with the CI state coefficients to calculate transition dipole moments between the lowest energy singlet and triplet states ($S_0$ and $T_0$) and excited electronic states. This is then used to calculate an oscillator strength for each transition ($\ket{\Phi} \leftarrow \ket{\Psi}$), as shown in Eq. 16 where $\omega_{\Phi\Psi} = E_\Phi - E_{\Psi}$ is the transition energy in Hartrees.

\begin{equation}
    f_{\Phi\Psi} = \frac{2}{3}\omega_{\Phi\Psi}|\mu_{\Phi\Psi}|^2
\end{equation}

Finally, Lorentzian line broadening is applied to peaks in the simulated spectra using a full width half maximum (FWHM) value of 20 nm for a peak at 300 nm.

\newpage

\section{Results and Discussion}

\subsection{Small and Medium-Sized Diradicals}

The D-ExROPPP method was first tested on a selection of small organic diradical molecules\cite{CarlLineberger2011}. Results are compared with those obtained from multi-reference wavefunction-based methods and experimental data where possible. In each case, the open shell structure was estimated using unrestricted DFT geometry optimisation of the lowest energy triplet state. The methods used for comparison are the state-averaged complete active space self-consistent field (SA-CASSCF) method and generalised multiconfigurational quasi-degenerate perturbation theory (GMC-QDPT), with active spaces that include all $\pi$-orbitals. Further details on the computational procedure are provided in the appendix.

\subsubsection*{Cyclobutadiene}

Square cyclobutadiene (CBD) is perhaps the most structurally simple organic diradical, consisting of only four Carbon and four Hydrogen atoms. Despite this, the attempts of organic chemists to synthesise CBD have been unsuccessful for over a century, and the nature of its electronic structure was a subject of intense scrutiny for much of that time \cite{Maier1974}. The ground state of square CBD was widely believed to be a triplet for many years until it was recognised by Borden in the late 1970s that accounting for correlation between electrons in the HOMO and the SOMO leads to the correct prediction of a singlet ground state instead \cite{Borden1975, Borden1977}.

\begin{table}[H]
\caption{Summary of calculated energies and configurations for the first four electronic states of CBD obtained with various methods. Energies are given relative to that of their respective ground state energy and do not include zero-point energy corrections. The most important configuration(s) in each state with the highest weighting(s) (coefficient given in brackets) is(/are) given.}
    \centering
    \begin{tabular}{|c|l|c|l|c|l|}
         \hline
         \makecell{\textbf{D-ExROPPP} \\ \textbf{Energy}} & \makecell{\textbf{D-ExROPPP} \\ \textbf{Configurations}} & \makecell{\textbf{CASSCF} \\ \textbf{Energy}} & \makecell{\textbf{CASSCF} \\ \textbf{Configurations}} & \makecell{\textbf{GMC-QDPT} \\ \textbf{Energy}} & \makecell{\textbf{GMC-QDPT} \\ \textbf{Configurations}}\\ \hline
         
         0 & $\begin{aligned}
         \rule{0pt}{3ex} &\ket{^1\text{OS}} (0.97), \\ &\ket{^{1T}\text{CV}_1^{1'}} (0.23) \rule[-2ex]{0pt}{0pt} 
         \end{aligned}$ & 0 & $\begin{aligned}
         \rule{0pt}{3ex} &\ket{^1\text{OS}} (0.97), \\ &\ket{^{1T}\text{CV}_1^{1'}} (0.24) \rule[-2ex]{0pt}{0pt}
         \end{aligned}$ & 0 & $\begin{aligned}
         \rule{0pt}{3ex} &\ket{^1\text{ZW}^-} (0.74), \\ &\ket{^{1}\text{OS}} (-0.63) \rule[-2ex]{0pt}{0pt}
         \end{aligned}$ \\ \hline
         
         0.335 & $\begin{aligned}
         \rule{0pt}{3ex} &\ket{^3\text{OS}} (0.97), \\ &\ket{^{3S}\text{CV}_1^{1'}} (0.24) \rule[-2ex]{0pt}{0pt} \end{aligned}$  & 0.172 & $\begin{aligned}
         \rule{0pt}{3ex} &\ket{^3\text{OS}} (0.98), \\ &\ket{^{3S}\text{CV}_1^{1'}} (0.21) \rule[-2ex]{0pt}{0pt} \end{aligned}$ & 0.121 & $\begin{aligned}
         \rule{0pt}{3ex} &\ket{^3\text{OS}} (0.97), \\ &\ket{^{3S}\text{CV}_1^{1'}} (0.25) \rule[-2ex]{0pt}{0pt} \end{aligned}$ \\  \hline
         
         1.938 & $\begin{aligned}
         \rule{0pt}{3ex} &\ket{^1\text{ZW}^+} (0.95), \\ &\ket{^1\text{CS}_{1~\bar{1}}^{0'\bar{0'}}} (-0.21) \rule[-2ex]{0pt}{0pt} \end{aligned}$ & 2.024 & $\begin{aligned}
         \rule{0pt}{3ex} &\ket{^1\text{ZW}^+} (0.94), \\ &\ket{^1\text{CS}_{1~\bar{1}}^{0'\bar{0'}}} (-0.24) \rule[-2ex]{0pt}{0pt} \end{aligned}$ & 1.310 & $\begin{aligned}
         \rule{0pt}{3ex} &\ket{^1\text{ZW}^+} (0.94), \\ &\ket{^1\text{CS}_{1~\bar{1}}^{0'\bar{0'}}} (-0.24) \rule[-2ex]{0pt}{0pt} \end{aligned}$ \\ \hline
          
         2.770 & \makecell{$\ket{^1\text{ZW}^-}$ $(1.0)$} & 2.95 & \makecell{$\ket{^1\text{ZW}^-}$ $(1.0)$} & 1.766 & $\begin{aligned}
         \rule{0pt}{3ex} &\ket{^{1}\text{OS}} (0.76), \\ &\ket{^1\text{ZW}^-} (0.65) \rule[-2ex]{0pt}{0pt}
         \end{aligned}$ \\ \hline
         
    \end{tabular}
\end{table}

Results in table 4 show that D-ExROPPP correctly predicts the open-shell singlet to be the ground state, suggesting it is able to sufficiently describe the aformentioned correlation between electrons in the HOMO and SOMO. While CBD's instability has prevented its electronic structure from being characterised experimentally, benchmark results have been calculated by Loos et al. at a CCSDTQ/aug-cc-pVTZ level of theory \cite{Monino2022}. The CCSDTQ (coupled-cluster singles doubles triples quadruples) method is a highly expensive post-Hartree Fock method that, when used with the large aug-cc-pVTZ basis, is expected to yield highly accurate energy predictions. D-ExROPPP overestimates the energy of the higher energy $\ket{^3\text{OS}}$ state relative to the to the CCSDTQ benchmark of 0.144eV \cite{Monino2022}, wheras CASSCF and GMC-QDPT show relatively good agreement on this front. For the zwitterionic states, CASSCF and D-ExROPPP both appear to overestimate energies relative to the benchmark results of 1.500 eV and 1.849 eV for $\ket{^1\text{ZW}^+}$ and $\ket{^1\text{ZW}^-}$ respectively \cite{Monino2022}. However, compared to CASSCF which does not treat dynamic correlation, results from D-ExROPPP are consistently closer to the benchmark energies. D-ExROPPP is able to predict the correct ordering of CBD's low-lying electronic states and shows relatively good qualitative agreement with the high-level GMC-QDPT and CCSDTQ methods.

\subsubsection*{Trimethylenemethane}

Trimethylenemethane (TMM) is another widely studied molecule that is often used a minimal example of a non-disjoint diradical (Fig. 7) \cite{Borden1977, CarlLineberger2011, Stuyver2019}. The agreement between results obtained from D-ExROPPP and those from CASSCF and GMC-QDPT is slightly worse for TMM compared to CBD (Table 5), noticeable particularly in predictions of the singlet-triplet gap ($\Delta E_\text{S-T}$). This may be partially explained by the presence of core $\rightarrow$ virtual double excitations in CASSCF and GMC-QDPT results, which act to stabilise the $\ket{^3\text{OS}}$ state slighltly relative to $\ket{^1\text{OS}}$. However, D-ExROPPP results actually agree better with photoelectron spectroscopy measurements of the TMM$^-$ anion, which predict $\Delta E_\text{S-T} =0.699 \pm 0.006 ~\text{eV}$ \cite{Wenthold1996}. This experimental result has been further supported by high-level benchmark calculations, including with the highly-correlated equation-of-motion spin-flip coupled-cluster(2,3) (EOM-SF-CC(2,3)) method which predicts $\Delta E_\text{S-T} =0.697 ~\text{eV}$ \cite{Slipchenko2005}.

\begin{figure}[H]
\centering
\includegraphics[width=0.95\linewidth, height=0.33\linewidth]{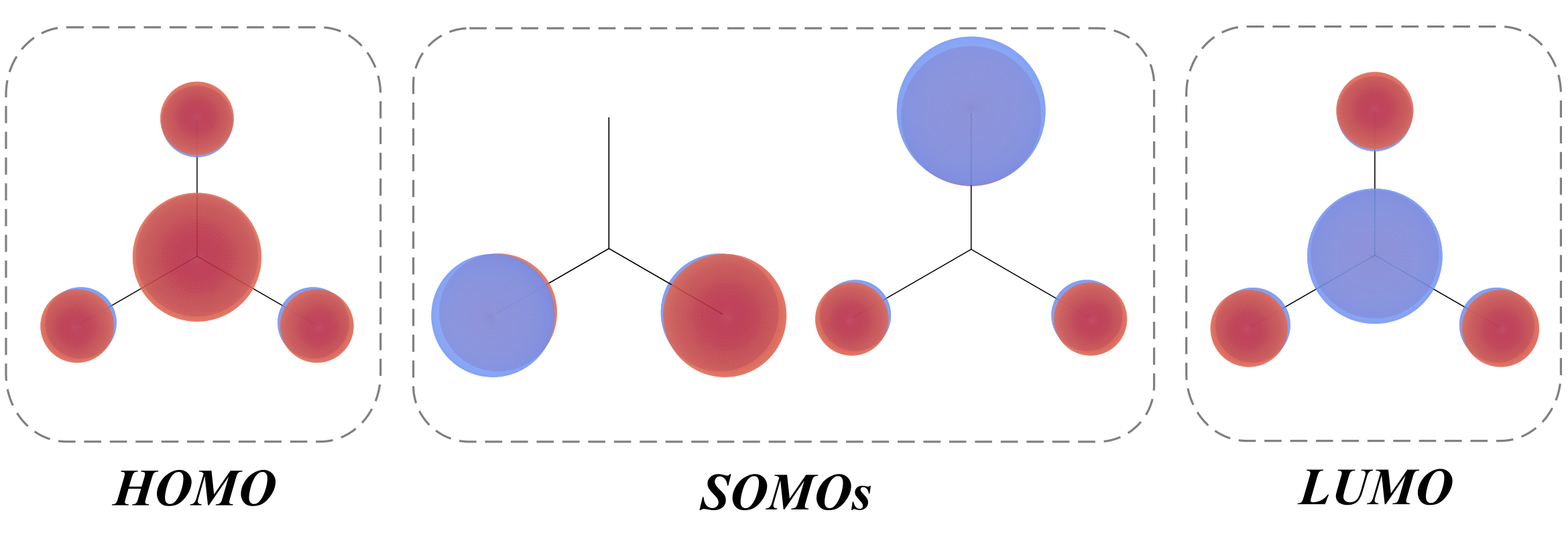}
\caption{$\pi$ molecular orbitals of the trimethylenemethane diradical. This diradical is classed as non-disjoint since both SOMOs have electron density on the bottom two Carbon atoms.}
\end{figure}

\begin{table}[H]
\caption{Summary of calculated energies and configurations for the first five electronic states of TMM obtained with various methods. Energies are given relative to that of their respective ground state energy and do not include zero-point energy corrections. The most important configuration(s) with the highest weighting(s) (coefficient given in brackets) is shown, sometimes alongside lower-weighted configurations that stabilise them. The experimental prediction of the lowest energy singlet's energy relative to $\ket{^3\text{OS}}$ is $0.699 \pm 0.006 ~\text{eV}$ \cite{Wenthold1996}.}
    \centering
    \begin{tabular}{|c|l|c|l|c|l|}
    \hline
         \makecell{\textbf{D-ExROPPP} \\ \textbf{Energy}} & \makecell{\textbf{D-ExROPPP} \\ \textbf{Configurations}} & \makecell{\textbf{CASSCF} \\ \textbf{Energy}} & \makecell{\textbf{CASSCF} \\ \textbf{Configurations}} & \makecell{\textbf{GMC-QDPT} \\ \textbf{Energy}} & \makecell{\textbf{GMC-QDPT} \\ \textbf{Configurations}}\\ \hline

         0 & $\begin{aligned}
         \rule{0pt}{3ex} &\ket{^3\text{OS}} (0.97), \\ &\ket{^{3X}\text{CV}_1^{1'}} (-0.23) \rule[-2ex]{0pt}{0pt} \end{aligned}$ & 0 & $\begin{aligned}
         \rule{0pt}{3ex} &\ket{^3\text{OS}} (0.96), \\ &\ket{^{3X}\text{CV}_1^{1'}} (-0.24) \rule[-2ex]{0pt}{0pt} \end{aligned}$ & 0 & $\begin{aligned}
         \rule{0pt}{3ex} &\ket{^3\text{OS}} (0.96), \\ &\ket{^{3X}\text{CV}_1^{1'}} (-0.24) \rule[-2ex]{0pt}{0pt} \end{aligned}$ \\ \hline
         
         0.893 & $\begin{aligned}
         \rule{0pt}{3ex} &\ket{^1\text{OS}} (0.84) \rule[-2ex]{0pt}{0pt} \end{aligned}$  & 1.119 & $\begin{aligned}
         \rule{0pt}{3ex} &\ket{^1\text{OS}} (0.84) \rule[-2ex]{0pt}{0pt} \end{aligned}$ & 1.145 & \makecell{$\ket{^1\text{OS}}$ $(0.9)$} \\ \hline
         
         0.893 & $\begin{aligned}
         \rule{0pt}{3ex} &\ket{^1\text{ZW}^-} (0.84) \rule[-2ex]{0pt}{0pt} \end{aligned}$ & 1.119 & $\begin{aligned}
         \rule{0pt}{3ex} &\ket{^1\text{ZW}^-} (0.84) \rule[-2ex]{0pt}{0pt} \end{aligned}$ & 1.145 & \makecell{$\ket{^1\text{ZW}^-}$ $(0.9)$} \\ \hline
          
         3.177 & $\begin{aligned}
         \rule{0pt}{3ex} &\ket{^1\text{CS}_{1}^{0'}} (0.57), \\ &\ket{^1\text{SV}_{0}^{1'}} (0.57) \rule[-2ex]{0pt}{0pt} \end{aligned}$ & 4.447 & $\begin{aligned}
         \rule{0pt}{3ex} &\ket{^1\text{CS}_{1}^{0'}} (0.69), \\ &\ket{^1\text{SV}_{0}^{1'}} (0.56) \rule[-2ex]{0pt}{0pt} \end{aligned}$ & 4.508 & $\begin{aligned}
         \rule{0pt}{3ex} &\ket{^1\text{CS}_{1}^{0'}} (0.77), \\ &\ket{^1\text{SV}_{0}^{1'}} (0.46) \rule[-2ex]{0pt}{0pt} \end{aligned}$ \\ \hline
         
         3.736 & \makecell{$\ket{^1\text{ZW}^+}$ $(0.96)$} & 5.866 & \makecell{$\ket{^1\text{ZW}^+}$ $(0.96)$} & 3.621 & \makecell{$\ket{^1\text{ZW}^+}$ $(0.96)$} \\
         \hline
    \end{tabular}
\end{table}

While there are no experimental insights available into the energies of higher lying states, the D-ExROPPP and GMC-QDPT predictions of 3.736 eV and 3.621 eV respectively for the relative energy of the $\ket{^1\text{ZW}^+}$ state are in good agreement with the benchmark EOM-SF-CC(2,3) prediction of 3.551 eV \cite{Slipchenko2005}. On the other hand, the CASSCF energy of 5.866 eV for this state is severely overestimated relative to methods which account for dynamic correlation. While D-ExROPPP does obtain a good prediction of the $\ket{^1\text{ZW}^+}$ state energy, largely thanks to stabilisation from double excitations $\ket{^1\text{CS}_{1~\bar{1}}^{0'\bar{0'}}}$ and $\ket{^1\text{SV}_{0~\bar{0}}^{1'\bar{1'}}}$, the energetic ordering of states is slightly thrown off by the apparent underestimation of $\ket{^1\text{CS}_{1}^{0'}}/\ket{^1\text{SV}_{0}^{1'}}$ state energy. The fact that the weighting of $\ket{^1\text{CS}_{1}^{0'}}$ state increases relative to $\ket{^1\text{SV}_{0}^{1'}}$ for the MCSCF and MCPT methods might suggest that the self-consistent orbital optimisation is driving the energies of these states apart.

\subsubsection*{\textit{meta}-Xylylene}

The \textit{meta}-xylylene (m-XYL) diradical, also known as \textit{meta}-benzoquinodomethane (MBQDM), is a moderately-sized, stable diradical that has been widely studied \cite{RetaMaeru2013, Hrovat1998, CarlLineberger2011, Wang2005}. It is frequently used as a model system for assessing the quality of electronic structure methods for diradicals since it's electronic states are relatively well characterised by experiment \cite{LeJeune1984, Neuhaus2008, Haider1989, Migirdicyan1975, Wright1983, Wenthold1997}.

\begin{figure}[H]
\centering
\includegraphics[width=1\linewidth, height=0.6\linewidth]{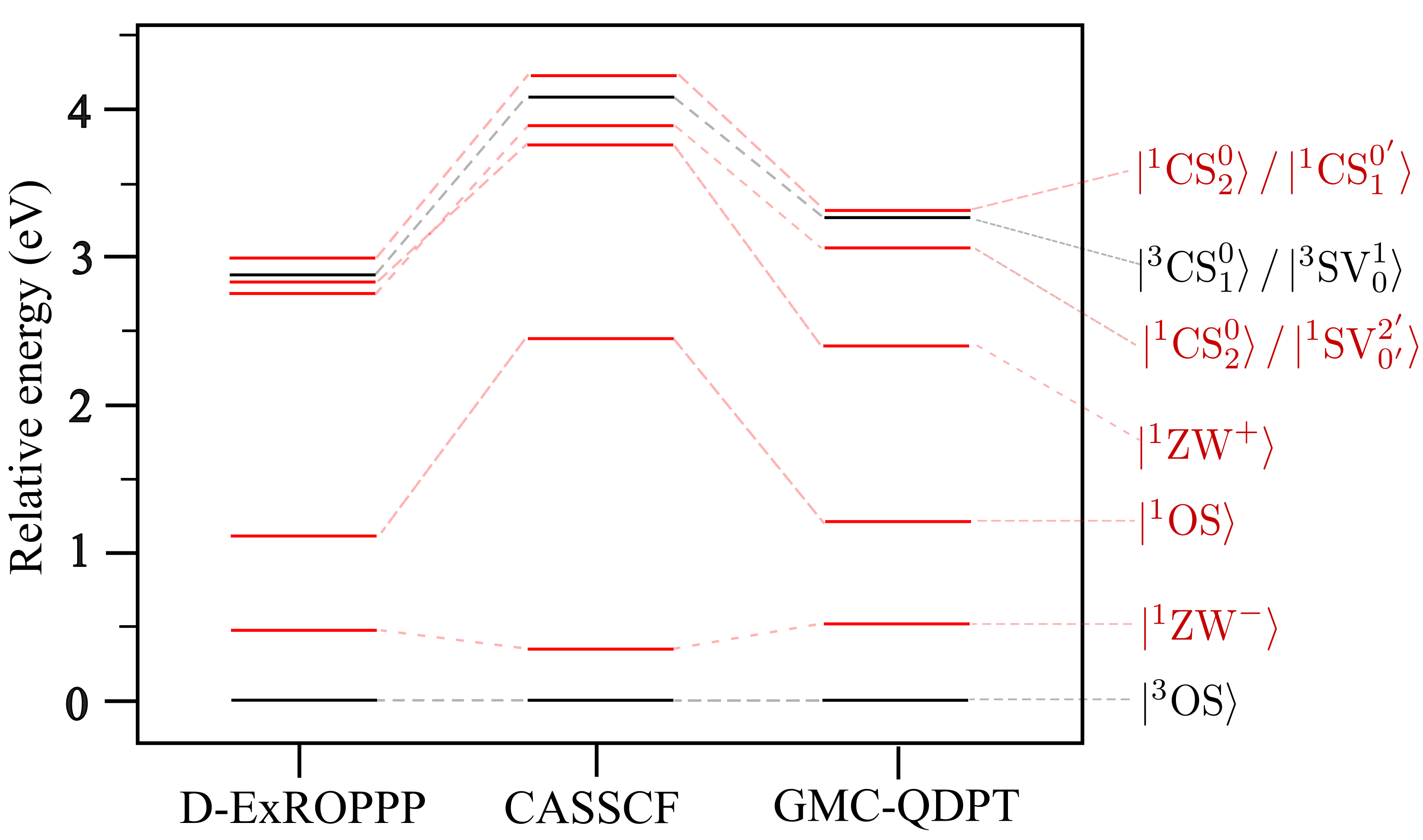}
\caption{Energy level diagram for electronic states of the m-XYL diradical calculated with various methods. Triplet states are shown in black, while singlets are in red. States of the same character are joined by dotted lines with their highest weighted configuration(s) given right. Energy levels for each method are given relative to the energy of the $\ket{^3\text{OS}}$ calculated with that method. Data can be found in table A9 of the appendix.}
\end{figure}

Fig. 8 shows excellent agreement between D-ExROPPP and GMC-QDPT for the energies for the low-lying excited states of m-XYL. Predictions of 0.485 eV and 0.502 eV made by D-ExROPPP and GMC-QDPT for the relative energy of the $\ket{^1\text{ZW}^-}$ state (Table A9) are in good agreement with the experimental measurement of $(0.416 \pm 0.009)~\text{eV}$ obtained by Linenberger and colleagues with photoelectron spectroscopy \cite{Wenthold1997}. Similarly for the $\ket{^1\text{OS}}$ state, D-ExROPPP obtains a relative energy of 1.116 eV which is slightly lower than the GMC-QDPT prediction of 1.235 eV but closer to the experimental value of $\le0.932 ~\text{eV}$ \cite{Wenthold1997}. For the $\ket{^1\text{ZW}^+}$ state, D-ExROPPP prediction of 2.813 eV is approximately 0.45 eV higher than that obtained by GMC-QDPT. For the higher-lying states composed predominantly of core to SOMO and SOMO to virtual single excitations, experimental results are not readily available. However, we note that similar to TMM, estimates obtained by D-ExROPPP are on average 11.8 \% lower than those obtained by GMC-QDPT for the first 3 states of this kind. Regarding CASSCF results, for all but the lowest energy singlet, energies are consistently shifted $\approx 1~\text{eV}$ higher than those obtained with D-ExROPPP and GMC-QDPT, suggesting once again that the lack of dynamic correlation is causing significant issues.

\begin{figure}[H]
\centering
\includegraphics[width=1\linewidth, height=0.71\linewidth]{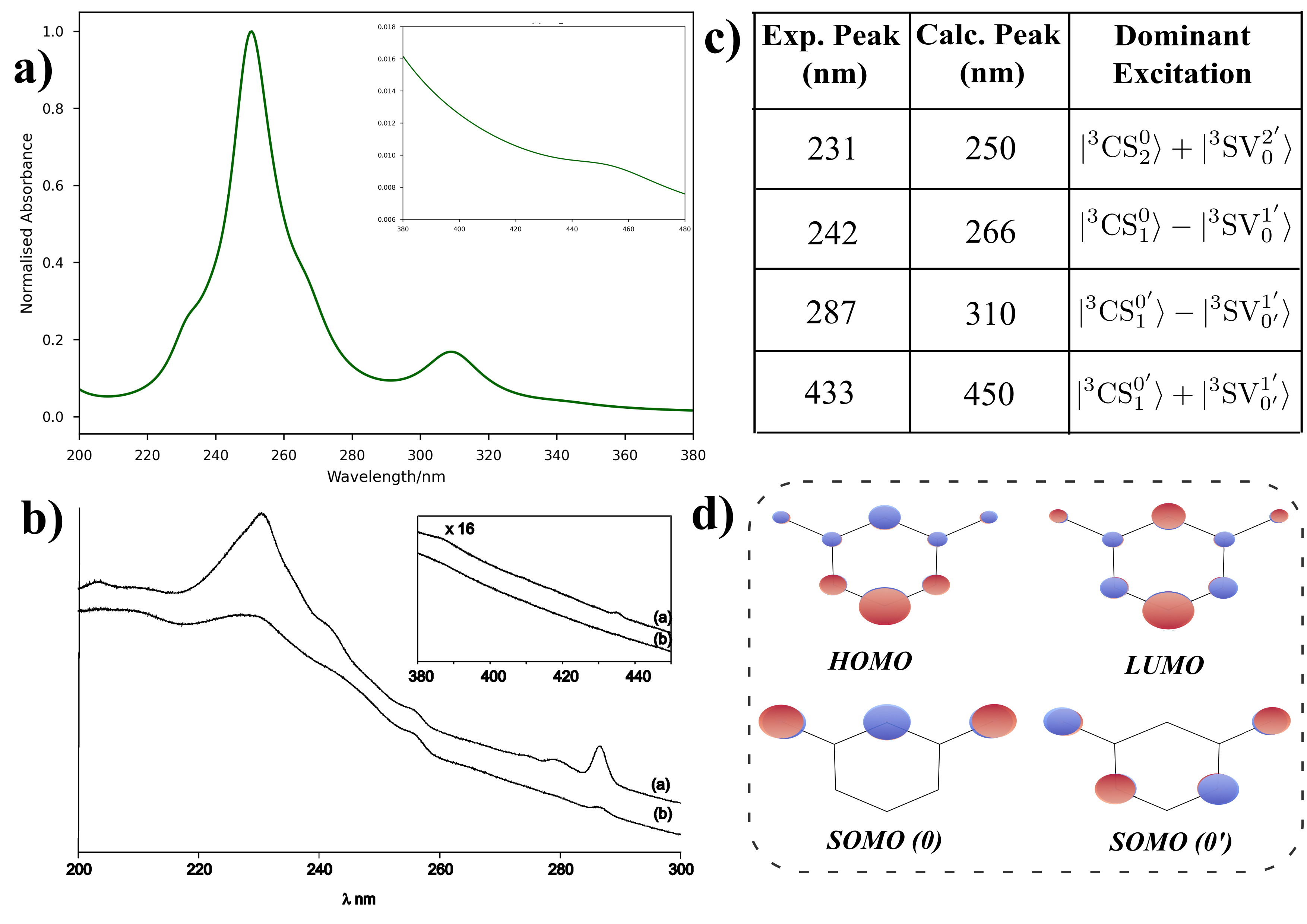}
\caption{a) UV-vis absorption spectrum for the \textit{meta}-Xylylene diradical simulated by D-ExROPPP. b) Experimentally measured UV-vis absorption spectrum for the \textit{meta}-Xylylene diradical obtained by Sander et al. \cite{Neuhaus2008}. c) Comparison of main absorption peaks found in the experimental spectrum and the D-ExROPPP spectrum. The main configurations that contribute to the respective excited states are also given. d) Frontier $\pi$ orbitals for \textit{meta}-Xylylene.}
\end{figure}

The \textit{meta}-xylylene diradical also acts as a useful test case for D-ExROPPP since it features several dipole-allowed electronic transitions that have been measured with UV-vis absorption spectroscopy \cite{Neuhaus2008, Migirdicyan1975}. Since m-XYL features a relatively large singlet-triplet energy gap ($\Delta E_\text{S-T} > 0.4$ eV), we can be confident that the lowest energy singlet state is not significantly populated at room temperature and thus all observed transitions are between triplet states. Comparing the UV-vis spectrum produced by D-ExROPPP to the experimental spectrum measured by Sander et. al (Fig. 9), we find good qualitative agreement. All of the main features of the experimental spectrum are reproduced by D-ExROPPP, although transition energies are underestimated an average of 6.6 \%. Furthermore, to the best of our knowledge, this is the first time that these electronic transitions have been assigned, highlighting the potential that D-ExROPPP holds as a tool for elucidating the photophysical properties of diradical molecules.

\subsubsection*{Influence of SOMO rotation}

In pure diradicals such as CBD, TMM and MXYL, the SOMOs ($\phi_0$, $\phi_{0'}$) obtained in an SCF procedure are eigenfunctions of the Fock-operator with exactly degenerate eigenvalues, meaning any unitary rotation within the degenerate SOMO subspace will yield a valid pair of SOMOs. In an exact treatment, configuration interaction results should be invariant with respect to rotations of this kind. However, Borden showed in 1996 that truncated CI methods can sometimes violate this invariance, yielding excited electronic states that differ slightly depending on the choice of SOMOs \cite{Borden1996}.

To assess the sensitivity of D-ExROPPP to SOMO rotation, we take orthogonal linear combinations of the SOMOs obtained from SCF (Fig. 10), and compare CI results obtained for the standard SOMOs versus the rotated ones. In general, D-ExROPPP excited state energies are found to be sensitive to the choice of SOMOs. This effect is observed for CBD and \textit{m}-XYL (Tables A10 and A12), although notably not for the TMM diradical (Table A11). Inspecting this more closely, CSF energies are found to remain invariant to SOMO rotation while off-diagonal couplings between states in the XCIS-D matrix that are responsible for the observed sensitivity. This is illustrated most clearly in the excited states of CBD (Table A10). With localised SOMOs (MO diagrams shown in Fig. 2), a strong coupling is present between the $\ket{^3\text{OS}}$ and $\ket{^{3S}\text{CV}_1^{1'}}$ states, which reduces to zero when configurations are formed from delocalised SOMOs (Fig. 2). Conversely, coupling between the $\ket{^1\text{ZW}^+}$ and  $\ket{^{1T}\text{CV}_1^{1'}}$ states is present for delocalised but not localised SOMOs. In each case, off-diagonal couplings yield an energy stabilisation of $\approx 0.4$ eV for the overall state compared to states with solely $\ket{^3\text{OS}}$ or $\ket{^1\text{ZW}^+}$ character. Comparison with benchmark CCSDTQ results \cite{Monino2022} does not clearly favour either set of SOMOs, as the relative energy of the $\ket{^3\text{OS}}$ state is better predicted with localised orbitals, wheras a more accurate $\ket{^1\text{ZW}^+}$ energy is obtained with delocalised orbitals.

\begin{figure}[H]
\centering
\includegraphics[width=0.82\linewidth, height=0.4\linewidth]{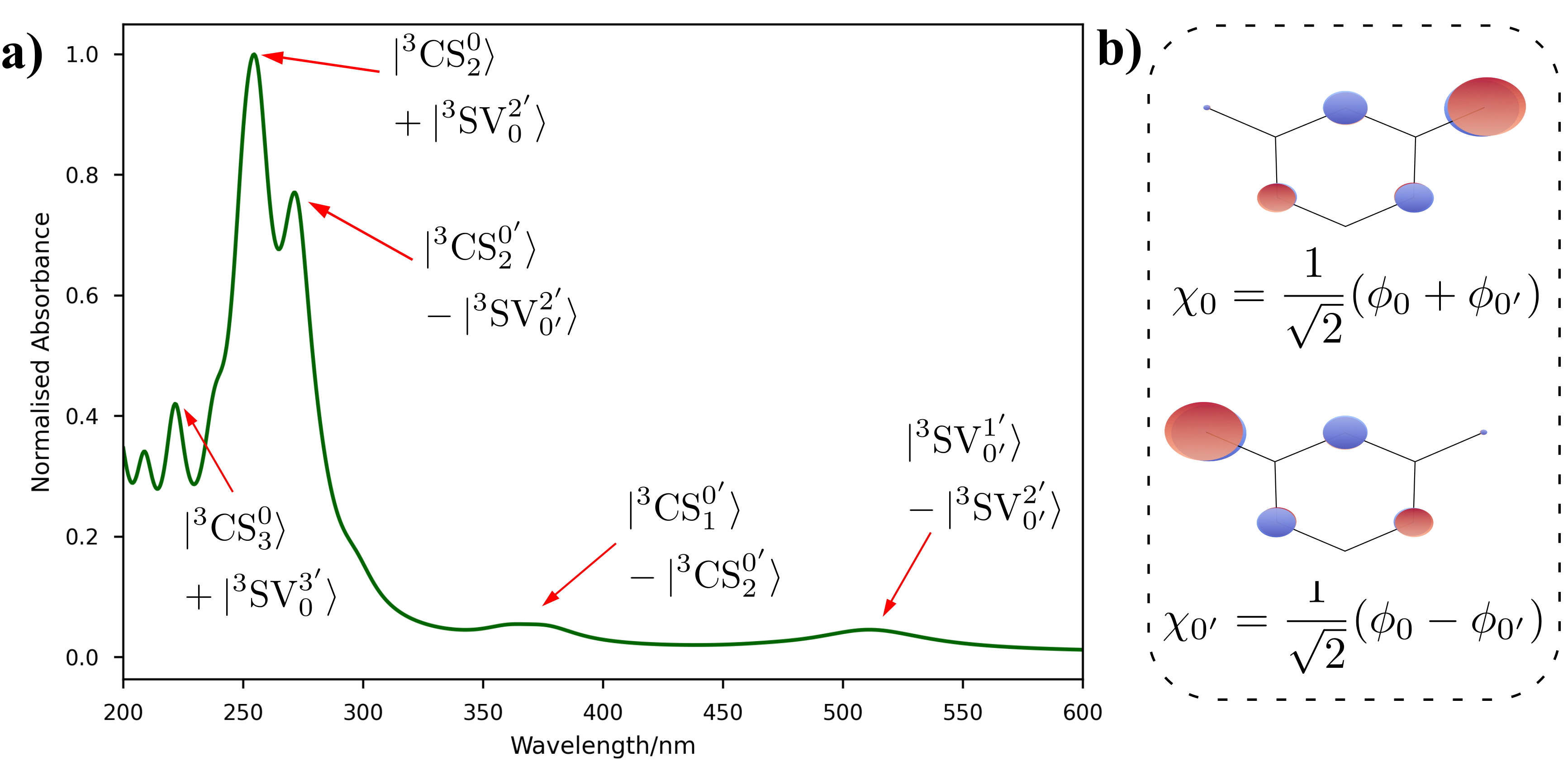}
\caption{a) UV-vis absorption spectrum calculated for the meta-xylylene diradical using the rotated SOMOs shown in b) in the configuration interaction treatment. Peaks are labelled according to their main excitation character. b) also shows the linear transformation used to rotate SOMOs (where $\phi_0$ and $\phi_{0'}$ are the unrotated SOMOs shown in Figure 9).}
\end{figure}

\textit{meta}-Xylylene shows largely the same behaviour as CBD, with couplings between CSFs in the CI matrix changing depending on the choice of SOMOs (Table A12). However, the effect on the excited states appears to be more dispersed for \textit{m}-XYL, likely due to the higher number of excited configurations contributing to each state. Relative energies of all of the first five excited states are affected by SOMO rotation, with a mean absolute percentage change of 7.6\% between results obtained with rotated and unrotated SOMOs. Notably, the sensitivity of D-ExROPPP results is higher when an XCIS basis is used instead of XCIS-D, with the mean absolute \% change rising to 8.8\%. Figure 12 shows the negative effect that SOMO rotation has on the predicted UV-vis absorption spectrum of \textit{m}-XYL, where the unrotated spectrum is in better agreement with experimental measurements (Fig. 9) than that obtained with rotated SOMOs. This is most evident in the splitting of the main 250 nm peak into two peaks with slightly higher absorption energies. This appears to be primarily caused by changes in transition intensities rather than state energies. Dipole moments involving the SOMOs are highly sensitive to their orbital coefficients (eq 15), meaning the rotation of SOMOs can cause transitions that were previously dark ($\mu_{i0} \approx 0$) to gain intensity ($|\mu_{i0}| > 0$). This appears to be the case in Fig. 10, where the previously weak \textit{HOMO - 1 (2)} $\rightarrow$ SOMO(0') and SOMO(0') $\rightarrow$ \textit{LUMO + 1 (2')} transitions gain intensity due to the polarisation of the rotated SOMO(0'). While the rotated absorption spectrum still shows decent qualitative agreement with the experimental spectrum, we recognise this lack of invariance to SOMO rotation as a limitation of D-ExROPPP that will be carefully considered moving forward.

\subsubsection*{Computational Efficiency}

The speed of the excited state calculations using D-ExROPPP is compared with that of the post-Hartree Fock methods. D-ExROPPP calculations were run on a laptop using a single CPU core, meanwhile GMC-QDPT and CASSCF calculations were run on UCL's Myriad HPC cluster with 8 CPU cores. Given this drastic difference in computational hardware, the single-core wall-clock time of the post-Hartree Fock methods is roughly estimated using eq. 17. The multi-core wall clock time and CPU utilisation \% are both obtained from the GAMESS output and $n_\text{cores} = 8$ for all calculations. While parallelisation over multiple CPU cores will not truly yield a $\times8$ speed-up, this overestimation is expected to be more than offset by the significantly higher single-core performance and memory bandwith of the HPC cluster relative to the laptop hardware used.
\vspace{-0.5cm}

\begin{equation}
    \text{Single-core wall-clock time} \approx (\text{Multi-core wall-clock time}) \times n_\text{cores} \times \frac{\text{CPU utilisation \%}}{100}
\end{equation}
\vspace{-0.5cm}

\begin{table}[H]
\caption{Comparison of the wall-clock time taken by various computational methods in calculations on the small diradicals studied.
\\ $^{*}$ Mean wall clock time taken over 5 calculations.\\ $^{**}$ Estimated single-core wall clock time calculated using eq. 17 . \\ $^{***}$ Approximate increase in computational speed when running D-ExROPPP with an XCIS-D basis compared to GMC-QDPT.}
    \centering
    \begin{tabular}{|c|l|l|l|l|l|}
    \hline
         \makecell{\textbf{Molecule} \\ \textbf{Name}} & \makecell{$t_\text{D-ExROPPP}$$^*$ \\ (XCIS) / s} & \makecell{$t_\text{D-ExROPPP}$$^*$ \\ (XCIS-D) / s} & \makecell{$t_\text{CASSCF}$$^{**}$ \\ / s} & \makecell{$t_\text{GMC-QDPT}$$^{**}$ \\ / s} & \makecell{\textbf{Speed-up} \\ \textbf{Factor}$^{***}$} \\ \hline
         
         Cyclobutadiene & 0.008 & 0.009 & 12.08 & 56.12 & $\approx 6000$ \\ \hline
         
         Trimethylenemethane & 0.008 & 0.010 & 18.19 & 77.366 & $\approx 7500$ \\ \hline
         
         meta-Xylylene & 0.051 & 0.101 & 780.21 & 20926.63 & $\approx 200,000$ \\ \hline

         \hline
    \end{tabular}
\end{table}

While the comparisons are approximate, Table 6 highlights the efficiency of D-ExROPPP for calculating diradical electronic structure compared to multireference approaches like CASSCF and GMC-QDPT. Perhaps most importantly, we find that, as expected, D-ExROPPP scales much more favourably with respect to the size of the molecule. MCSCF and MCPT methods typically scale steeply with the number of basis functions, which itself grows linearly with the number of atoms being simulated in a molecule increases. Despite \textit{m}-XYL only containing twice as many atoms as CBD, CASSCF and GMC-QDPT calculations respectively take $\approx$65x and 370x longer. This unfavourable scaling becomes a severe problem for large emissive diradicals, which often exceed 100 atoms and thus require thousands of basis functions. In contrast, the time taken for D-ExROPPP/XCIS-D calculations only increases $\approx$ ten-fold for \textit{m}-XYL compared to CBD, largely owing to the low $O(N_\text{occ}^2)$ scaling of the XCIS-D CSF basis and the use of atomic orbitals instead of basis functions.

\subsection{Large emissive diradicals}

Having verified the speed, accuracy and spin-purity of D-ExROPPP for the calculation of electronic states for small diradicals, we apply it the method to the more challenging task of predicting the excited states of large, emissive diradicals. A survey of the literature since 2019 identified 18 novel organic diradical molecules with publicly available UV-Vis absorption data and optimised molecular coordinates. Molecules including heteroatoms such as boron and fluorine were excluding due to the unavailability of their machine-learned PPP parameters \cite{Shen2025}. While running D-ExROPPP on these 19 molecular geometries, 5 failed to achieve appropriate SCF convergence and had to be discarded. Results for the remaining 14 diradicals, are presented in this section are the subject of this section.


Large emissive diradicals almost all feature a very small or sometimes negligible energy gap between the lowest energy singlet and triplet electronic states ($\Delta E_\text{S-T}$). $\Delta E_\text{S-T}$ can be predicted experimentally by performing electron paramagnetic resonance (EPR) (also referred to as electron spin resonance (ESR)) measurements at variable temperatures, and fitting the results to the Bleaney-Bowers equation \cite{Bleaney1952}. The ground state multiplicity is found to be singlet for the majority of the TTM- and PyBTM-based diradicals studied, which indicates so-called 'anti-ferromagnetic coupling' ($\Delta E_\text{S-T} <0$). As for cyclobutadiene, the presence of a singlet ground state goes against the typical TOTEM energy ordering (Fig. 1), thus suggesting that there is an appreciable degree of correlation between electrons in the core MOs and the SOMOs which needs to be accounted for to obtain a qualitatively correct prediction.

{\renewcommand{\arraystretch}{1.2}
\begin{table}[H]
\caption{Comparison of the predictions of $\Delta E_\text{S-T}$ obtained by different methods for the 12 emissive diradicals that have had $\Delta E_\text{S-T}$ predicted from VT-EPR. Spin-flip time-dependent density functional theory (SF-TD-DFT) is used with three popular exchange correlation functionals (in brackets). Full $\Delta E_\text{S-T}$ data is given in Tables A13 and A14. \\ $^*$ SF calculations severely overestimate $\Delta E_\text{S-T}$ for PCz-(PyBTM')$_2$ by over 3 eV. This is treated as an outlier and is not included in SF results.}
    \centering
    \begin{tabular}{|c|l|l|l|}
    \hline
          \makecell{\textbf{Method}} & \makecell{\textbf{Root Mean} \\ \textbf{Squared Error (eV)}} & \makecell{\textbf{Mean Signed} \\ \textbf{Error (eV)}} & \makecell{\textbf{Correct GS} \\ \textbf{Mult. Prediction}} \\ \hline
         
         \makecell{D-ExROPPP} & \makecell{0.049} & \makecell{-0.034} & \makecell{9/9} \\ \hline
         \makecell{SF-TD-DFT \\ (B3LYP)$^*$} & \makecell{0.029} & \makecell{0.0070} & \makecell{8/9} \\ \hline
         \makecell{SF-TD-DFT \\ (CAM-B3LYP)$^*$} & \makecell{0.027} & \makecell{0.0097} & \makecell{8/9}  \\ \hline
         \makecell{SF-TD-DFT \\ (PEB0)$^*$} & \makecell{0.029} & \makecell{0.0059} & \makecell{8/9} \\ \hline

         \hline
    \end{tabular}
\end{table}
}

Ground state multiplicities and singlet triplet gaps ($\Delta E_\text{S-T}$) calculated by D-ExROPPP are presented alongside experimental results in Table A13 in the appendix. Of the 9 diradicals with well-characterised ground state multiplicities, D-ExROPPP correctly predicts the mulitplicity in every case. In the 8 molecules with singlet ground states, singly excited CSFs couple more strongly to $\ket{^1\text{ZW}^-}$ than to $\ket{^3\text{OS}}$ state, which drives the stabilisation of the singlet. While D-ExROPPP produces good qualitative agreement with experimental measurements of $\Delta E_\text{S-T}$, Table A13 and Fig. A1 in the appendix show that the quantitative accuracy of predictions is fairly limited. To benchmark D-ExROPPP against established methods for determining $\Delta E_\text{S-T}$, spin-flip time-dependent density functional theory (SF-TD-DFT) calculations were performed with three widely-used functionals (Table 7). SF-TD-DFT consistently predicts smaller $|\Delta E_\text{S-T}|$ values than D-ExROPPP, which has a tendency to overestimate the stabilisation of the singlet state relative to the triplet.

\begin{figure}[H]
\centering
\includegraphics[width=1\linewidth, height=0.45\linewidth]{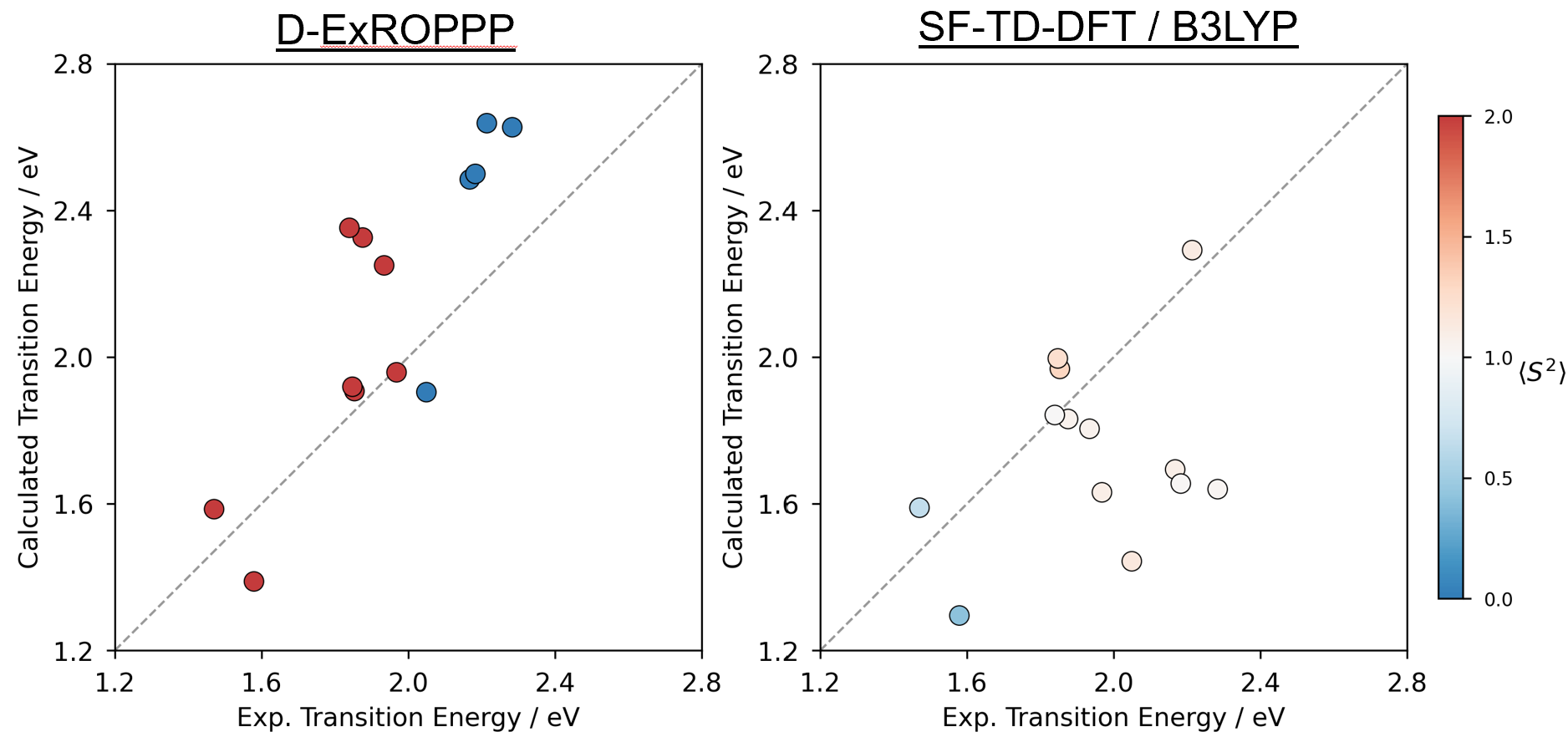}
\caption{Predicted versus experimental $\lambda_\text{abs}^\text{max}$ for the 13 emissive diradicals studied. Right: D-ExROPPP; left: spin-flip time dependent density functional theory with the uB3LYP functional. Dot colour indicates the $\langle \hat{S}^2 \rangle$ value of the final state. The dotted line denotes perfect agreement with experiment. SF-TDDFT significantly overestimates $\lambda_\text{abs}^\text{max}$ for the PCz-(PyBTM')$_2$ diradical meaning this data point is treated as an outlier and is not shown.}
\end{figure}

The advantages of D-ExROPPP over TD-DFT approaches become more apparent when considering higher energy electronic states. Figure 11 compares results obtained from D-ExROPPP and SF-TDDFT/B3LYP for the energies of the highest wavelength UV-vis transitions ($\lambda_\text{abs}^\text{max}$) for each of the large emissive diradical molecules studied. D-ExROPPP achieves a root mean squared error (RMSE) of 0.323 eV with respect to experimental UV-vis data, wheras the RMSE of SF-TDDFT calculations is 0.558 eV. Even if the PCz-(PyBTM')$_2$ outlier is removed, the improved RMSE of SF predictions (0.345 eV) is still higher than that of D-ExROPPP (0.297 eV). This is perhaps attributable in part to the spin-purity of D-ExROPPP's excited states, where in contrast, most SF-TD-DFT states are heavily spin contaminated, with $\langle \hat{S}^2 \rangle \approx 1$ indicating a mixing of singlet and triplet states. Beyond improving the accuracy of D-ExROPPP relative to SF-TD-DFT, the lack of spin contamination also allows the multiplicity of states involved in transitions to be more reliably characterised. This is particularly important for the $\lambda^\text{max}$ absorption peak, since it is typically relevant to many of the molecule's optical properties.

Finally, D-ExROPPP also outperforms SF-TD-DFT in terms of computational efficiency. Eq 17 is used again to estimate the single-core wall-clock time for SF-TD-DFT calculations, which were performed on 24 CPU cores of an HPC cluster. The CPU-utilisation of each calculation was not readily available and was estimated to be $75\%$, in line with the mean observed for GAMESS calculations. The resulting timings (Table A16) estimate that D-ExROPPP on average takes 86$\times$ less time than SF-TD-DFT/B3LYP to calculate the first 30 excited electronic states of TTM- and PyBTM-based diradicals. As noted previously, the superior computing architecture of the HPC cluster relative to the laptop used for D-ExROPPP calculations makes this a relatively conservative estimate, with the true speedup likely exceeding $100\times$. The combination of improved accuracy, spin-purity and computational efficiency make D-ExROPPP a practical and reliable tool for calculating and characterising the excited states of large emissive diradicals.

Looking at UV-Vis spectra more widely, we recognise that the small $\Delta E_\text{S-T}$ of the emissive diradicals studied means that in many cases states of both singlet and triplet multiplicity will be thermally populated at room temperature. To account for this behaviour, D-ExROPPP combines the absorption spectra predicted for each thermally accessible state into a single spectrum, where the intensities of singlet and triplet transitions are weighted according to their thermal population (Eq. 18). Since we find D-ExROPPP has a tendency to overestimate $\Delta E_\text{S-T}$, we use the values obtained from variable-temperature electron paramagnetic resonance measurements.

\begin{equation}
    w_\text{singlet} = \frac{\exp\left(-\Delta E_\text{S-T} / k_\text{B} T\right)}
               { 3 + \exp\left(-\Delta E_\text{S-T} / k_\text{B} T\right)}, 
    \qquad
    w_\text{triplet} = \frac{3}
               { 3 + \exp\left(-\Delta E_\text{S-T} / k_\text{B} T\right)}, 
\end{equation}

\begin{figure}[H]
\centering
\includegraphics[width=1\linewidth, height=0.66\linewidth]{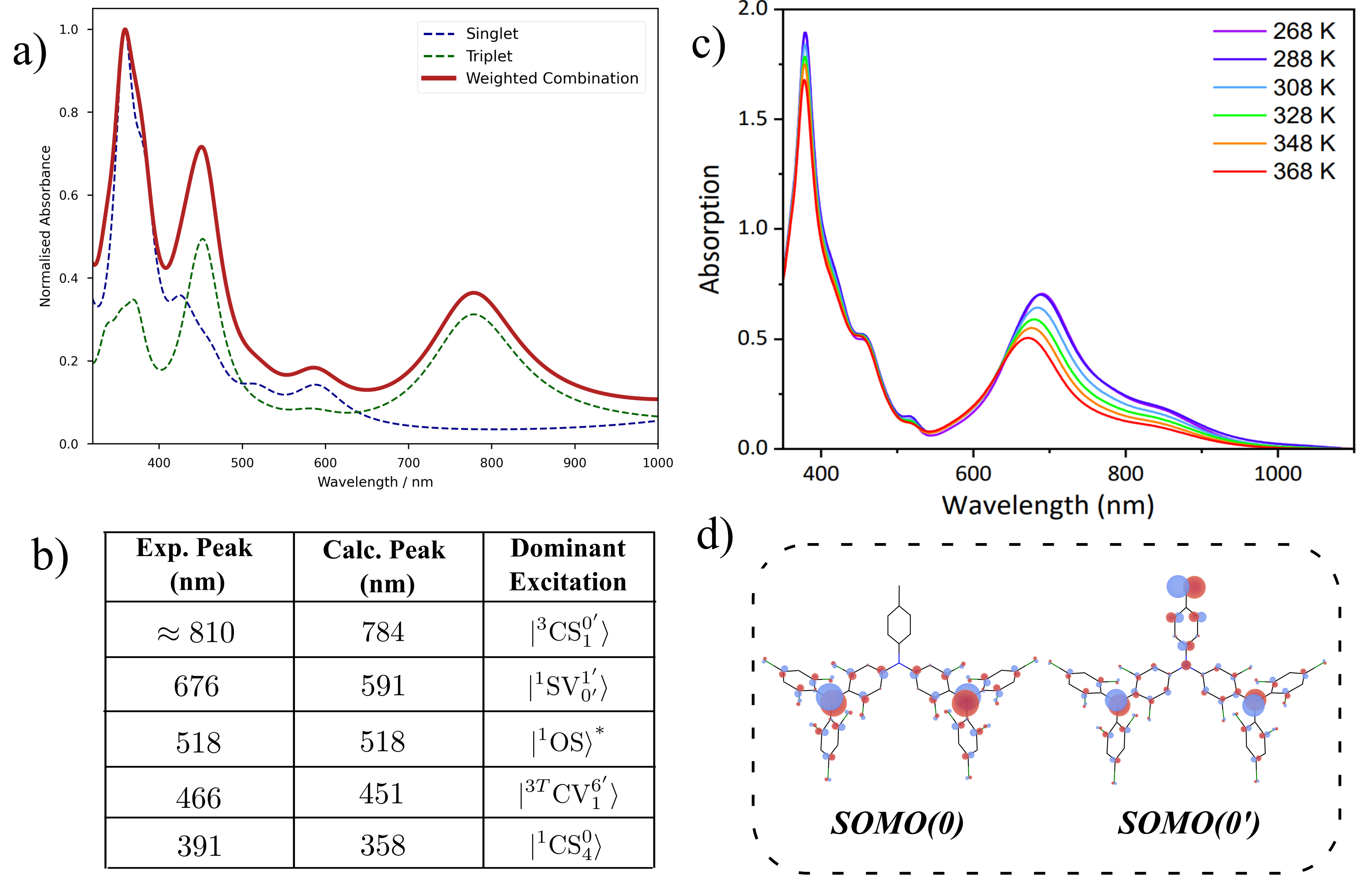}
\caption{a) Visible/Near-IR absorption spectrum of the DTA diradical calculated by D-ExROPPP, where the weighted combination is performed according to Boltzmann weights shown in Eq. 18 with $T = 298~\text{K}$. b) Comparison of experimental peak positions versus those calculated by D-ExROPPP. Typically many CSFs contribute to each of the excited states of large emissive diradicals, but the CSF with the highest weighting is also given. $^*$Note $\ket{^1\text{ZW}^-}$ not $\ket{^1\text{OS}}$ is the lowest energy singlet state predicted for DTA. c) Variable-temperature Vis/Near-IR absorption spectrum of DTA experimentally measured by Dong \textit{et. al} \cite{Tong2025}. d) MO diagrams for the two singly occupied molecular orbitals of DTA calculated by D-ExROPPP.}
\end{figure}


This approach successfully reproduces many of the absorption features observed in experimental UV-vis spectra across the 14 diradicals studied. For the DTA diradical (Fig. 12), all 5 peaks observed in the experimental spectrum are replicated by D-ExROPPP. Although some of peak energies and intensities are not exactly accurate in some cases, this offers an excellent demonstration of the method's ability to offer insight into the multiplicities and configuration character of electronic transitions. Indeed the multiplicity assignments are largely supported by variable temperature absorption data, which shows a decrease in the intensities of singlet peaks as the temperature increases and drives up the relative population of the triplet ground state. ON the other hand, there is a slight increase in intensity for the shoulder peak near 460nm, suggesting it does indeed arise from an electronic transition between triplet states. Many of the peak's in DTA's absorption spectrum are found to be characteristic of TTM- and PyBTM- based diradicals. Figure 13 shows the predicted energies of the triplet shoulder peak often observed between 400 - 500 nm, and the higher energy peak between 360 - 400 nm, which is typically also the most intense peak in emissive diradicals.

\begin{figure}[H]
\centering
\includegraphics[width=1\linewidth, height=0.45\linewidth]{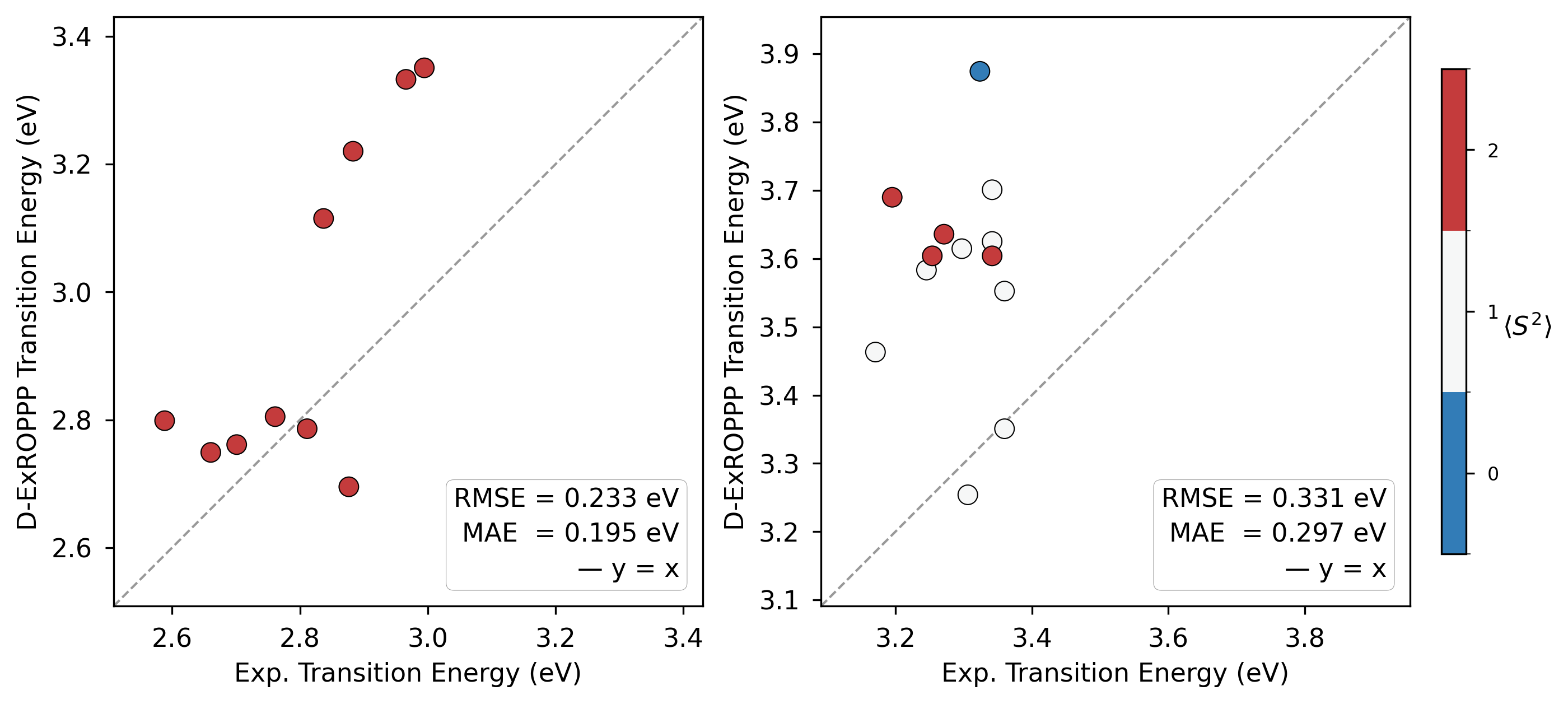}
\caption{Predicted (D-ExROPPP) versus experimental transition energies for high intensity peaks in the spectra of large emissive diradicals. Left: Medium intensity shoulder peaks found between 400 - 500 nm in experimental spectra. Right: High intensity peaks found between 360 - 400 nm in experimental spectra. The dotted line denotes perfect agreement with experiment. White dots do not mean that excited states are spin-contaminated, but instead signifies that the peak could not be definitively assigned as arising from singlet or triplet transitions but instead features contributions from states of both multiplicities.}
\end{figure}

Once again, D-ExROPPP shows relatively good agreement with experimentally measured transition energies for the two higher-energy peaks that are typical in emissive organic diradicals. Energies are mostly overestimated, as was observed for $\lambda_\text{abs}^\text{max}$ peaks, which corresponds to an underestimation of wavelengths. This is unsurprising since configuration interaction singles (CIS) and extended configuration interaction singles (XCIS) are known to typically overestimate excited state energies due to a lack of an insufficient description of electron correlation \cite{Maurice1996}. Combining results for the lower-energy maximum absorption peak, the root mean squared error of all the peaks predictions is 0.301 eV, which falls short of quantitative agreement but is expected to be sufficient for the reliable qualitative assignment of most absorption features.

\section{Conclusion}

In conclusion, this report presents the methodology and initial results for D-ExROPPP, a novel semi-empirical method designed for the calculation of spin-pure ground and excited electronic states of organic diradical molecules. An approximate SCF procedure is used for obtaining open-shell molecular orbitals of any diradical molecule, which is combined with Pariser-Parr-Pople theory to ensure computational efficiency and low-scaling with respect to molecule size. A spin-pure basis of configuration state functions is constructed to describe single excitations within the $\pi$ system of a diradical molecule, with selected doubles incorporated to increase the accuracy of excited state energies while maintaining favourable computational scaling. Hamiltonian interactions between these CSFs are calculated and implemented into a block-diagonal configuration interaction matrix, which is efficiently solved to yield spin-pure excited states and their energies.

The accuracy, efficiency and spin purity of D-ExROPPP were then assessed with calculations on a variety of small and large organic diradical molecules. For the small molecules studied, D-ExROPPP achieves qualitative and in many cases quantitative agreement with the highly-correlated GMC-QDPT method, while offering much improved scaling with respect to system size. For the higher-lying states of \textit{meta-}xylylene, D-ExROPPP faithfully replicates the experimental UV-vis absorption spectrum measured by Sander \textit{et al.} \cite{Neuhaus2008}, with energies deviating only by an average of 6.6\%. We then similarly apply the D-ExROPPP method to study the electronic structure of 14 large emissive diradicals drawn from the recent literature. Ground state multiplicities are accurately predicted in every case, although singlet-triplet energy gaps are systematically overestimated. D-ExROPPP is however, found to reliably reproduce the majority of absorption peaks found in experimental UV-vis spectra, with an RMSE of 0.301 eV across a range of peaks with transition energies between 1.5 and 3.5 eV. We find evidence that D-ExROPPP can predict transition energies in these large diradical molecules with a higher accuracy and lower computational cost than SF-TDDFT, while also offering spin purity. Future work will aim to more rigourously compare the accuracy and efficiency of D-ExROPPP relative to existing methods for computing diradical electronic states.

This work establishes D-ExROPPP as a promising tool for calculating the electronic structure of organic diradicals, and in particular for interpreting their highly complex UV-vis absorption spectra. We anticipate that the method will prove especially valuable for high-throughput screening or data driven approaches to molecular design, where its low computational cost and minimal user-input would be particularly advantageous. However, there remains plenty of scope for improving the method, particularly in regards to the sensitivity of excited state energies to the choice of SOMOs. Additional double excitations could be incorporated into the XCIS-D basis without increasing scaling beyond $O(N_\text{occ}^2)$, while modifications to the SCF procedure, such as the use of natural orbitals or self-consistent MO optimisation, could yield a more physically motivated orbital basis \cite{Suaud2012}. Together, these developements would further strengthen D-ExROPPP as a reliable and accessible tool for the study of organic diradical molecules.

\newpage

\section{Acknowledgements and Declaration of Contributions}

\subsubsection*{Declaration of Contributions}

The Python code for the original EXROPPP program, which provided a useful basis from which the D-ExROPPP code was heavily adapted, was written by Dr. James Green and Kevin Ma. Dr. Prashant Kumar prepared input files and ran TD-DFT calculations on ORCA for large emissive diradicals. D-ExROPPP results for the cyclobutadiene molecule were checked by Dr. Hugh Burton using the QUANTEL package, allowing small bugs in the implementation to be identified and fixed. Finally, Dr. Hele and Dr. Burton provided both helped significantly with conceptualising the project and developing ideas for the methodology.

\subsubsection*{Acknowledgements}

I would like to first and foremost thank my primary supervisor Dr. Hele and secondary supervisor Dr. Burton for their useful guidance, inspiration and support throughout the duration of the project. In addition, I thank all the members of the Hele group for creating a welcoming and stimulating academic environment. In particular, I thank Dr. Prashant Kumar, Dr. James Green and Jingkun Shen for their help in navigating unfamiliar computational methods and software. Finally I thank my friends and family for their encouragement and support throughout the project and throughout the full duration of my undergraduate studies.

\section{Use of Generative AI Declaration}

Generative AI was used in the following capacity in this project:
\begin{itemize}
    \item Anthropic's Claude Sonnet 4.5 (\textit{URL: www.Claude.ai}) was used on several occasions to check for bugs in code files that were otherwise written without any AI assistance. Additionally this model was used at times to assist in the creation and formatting of overleaf tables and figures designed for the project report, where data was then entered individually. On a limited number of occasions, the model was used to provide suggestions for the best way to phrase sentences in drafts of the project report, although no text was copied directly.
    \item Google Scholar labs (\textit{URL: www.scholar.google.com/scholar\_labs}) was used to support literature review by helping to find relevant academic papers relevant in the field.
\end{itemize}

\clearpage

\addcontentsline{toc}{section}{References}

\printbibliography

\clearpage

\section*{Appendix}
\addcontentsline{toc}{section}{Appendix}
\setcounter{figure}{0}
\renewcommand{\thefigure}{A\arabic{figure}}
\setcounter{table}{0}
\renewcommand{\thetable}{A\arabic{table}}

\appendix

\subsection*{I - Further Methodological Details}

\subsubsection*{A - Further derivation of Diradical-Fock Operator}

Here we present the derivation for the diradical Fock-operator but showing that it applies similarly to the description of triplet open-shell states. Once again, the first step is to recognise that any infinitesimal variation in MOs is equivalent to mixing small-amounts of singly-excited states into the reference states. This time, singly excited states have triplet multiplicity and we have an additional core $\rightarrow$ virtual CSF to be concerned with.

\begin{equation}
    \begin{aligned}
    \ket{^3\text{OS}} \rightarrow \ket{^3\text{OS}} + & \sum_i^k\lambda_{i0}\ket{^3\text{CS}_i^0} + \sum_i^k\lambda_{i0'}\ket{^3\text{CS}_i^{0'}} +   \sum_p^\infty\lambda_{0p}\ket{^3\text{SV}_0^p} + \sum_p^\infty\lambda_{0'p}\ket{^3\text{SV}_{0'}^p} 
 \\ + & \sum_i^k\sum_p^\infty\lambda_{ip}\ket{^{3T}\text{CV}_i^p} + \sum_i^k\sum_p^\infty\lambda_{ip}\ket{^{3S}\text{CV}_i^p} + \sum_i^k\sum_p^\infty\lambda_{ip}\ket{^{3X}\text{CV}_i^p}
\end{aligned}
\end{equation}

where once again $i$ and $p$ are indices for doubly-occupied and virtual orbitals respectively, and $k$ is the number of doubly occupied (core) spatial orbitals. An infinitesimal change in energy caused by an infinitesimal variation in MOs is thus given by

\begin{equation}
\begin{aligned}
    \delta E = & \sum_i^k\lambda_{i0}\bra{^3\text{OS}}\hat{H}\ket{^3\text{CS}_i^0} + \sum_i^k\lambda_{i0'}\bra{^3\text{OS}}\hat{H}\ket{^3\text{CS}_i^{0'}} + \sum_p^\infty\lambda_{0p}\bra{^3\text{OS}}\hat{H}\ket{^3\text{SV}_0^p} \\ & + \sum_p^\infty\lambda_{0'p}\bra{^3\text{OS}}\hat{H}\ket{^3\text{SV}_{0'}^p} + \sum_i^k\sum_p^\infty\lambda_{ip}\bra{^3\text{OS}}\hat{H}\ket{^{3T}\text{CV}_i^p} + \sum_i^k\sum_p^\infty\lambda_{ip}\bra{^3\text{OS}}\hat{H}\ket{^{3S}\text{CV}_i^p} \\
    & + \sum_i^k\sum_p^\infty\lambda_{ip}\bra{^3\text{OS}}\hat{H}\ket{^{3X}\Psi_i^p}
\end{aligned}
\end{equation}

By the variational priniciple, the best possible MOs to describe the open-shell triplet state of a diradical molecule are those for which $\delta E = 0$, which in turn requires that all of the above brakets are equal to 0. Using Slater-Condon rules to evaluate the Hamiltonian interactions between the ground and excited states, we obtain the following expressions for satisfying this requirement:

\begin{subequations}
\begin{align}
    \bra{^3\text{OS}}\hat{H}\ket{^3\text{CS}_i^0} &= F_{i0} + (i0|00) + (i0|0'0') = 0  \\
    \bra{^3\text{OS}}\hat{H}\ket{^3\text{CS}_i^{0'}} &= F_{i0'} + (i0'|0'0') + (i0'|00) = 0 \\
    \bra{^3\text{OS}}\hat{H}\ket{^3\text{SV}_0^p} &= F_{0p} + (0p|00) + (00'|0'p) = 0 \\
    \bra{^3\text{OS}}\hat{H}\ket{^3\text{SV}_{0'}^p} &= F_{0'p} + (0'p|0'0') + (0'0|0p) = 0 \\
    \bra{^3\text{OS}}\hat{H}\ket{^{3T}\text{CV}_i^p} &= \sqrt{2}[F_{ip} + (ip|00) -\frac{1}{2}(i0|0p) + (ip|0'0') - \frac{1}{2}(i0'|0'p)] = 0 \\
    \bra{^3\text{OS}}\hat{H}\ket{^{3S}\text{CV}_i^p} &= \frac{\sqrt{2}}{2}[(i0|0p) - (i0'|0'p)] = 0 \\
    \bra{^3\text{OS}}\hat{H}\ket{^{3X}\text{CV}_i^p} &= (i0|0p) + (i0'|0'p) = 0 
\end{align}
\end{subequations}

$F_{ij}$ is the conventional Fock-operator as previously. We notice that the expression in Eq. 21e is identical to that of eq 6g - i.e. the coupling between the open-shell triplet and the bright core $\rightarrow$ virtual triplet CSF is the same as the coupling between the open-shell singlet and the bright core $\rightarrow$ virtual singlet CSF. As such, in the same way we replace $F$ with the diradical Fock-operator $F^\text{d}$ defined in Eq. 8. Once again substituting this same $F^\text{d}$ into equations 21a - 21g yields

\begin{subequations}
\begin{align}
    F_{i0}^\text{d} + \frac{1}{2}(i0|00) + \frac{1}{2}(i0'|0'0) &= 0 \\
    F_{i0'}^\text{d} + \frac{1}{2}(i0'|0'0') + \frac{1}{2}(i0|00') &= 0 \\
    F_{0'p}^\text{d} - \frac{1}{2}(0'p|00) - \frac{1}{2}(0'0|0p) &= 0 \\
    F_{0p}^\text{d} - \frac{1}{2}(0p|0'0') - \frac{1}{2}(00'|0'p) &= 0 \\
    \sqrt{2}F_{ip}^\text{d} &= 0 \\
    \frac{\sqrt{2}}{2}[(i0|0p) - (i0'|0'p)] &= 0 \\
    (i0|0p) + (i0'|0'p) &= 0
\end{align}
\end{subequations}

We now have a similar scenario where the bright core $\rightarrow$ virtual transition has a coupling with the open shell ground state given by $\sqrt{2}F_{ip}^\text{d} = 0$. Once again choosing to ignore mixed two electron terms, we find that couplings between the open shell triplet state and all singly excited CSFs can be approximated as $F_{ab}^\text{d}$ for some MOs a and b. Thus the same SCF procedure of iteratively solving the Roothaan equations with the diradidcal Fock operator yields is found to yield appropriate MOs for representing the open shell triplet state $\ket{^3\text{OS}}$

We can now show the equivalence between the use of the diradical Fock operator $F^\text{d}$ with the standard density matrix $P$, and the use of the standard Fock operator $F$ with the diradical density matrix $P^\text{d}$

\noindent Where $C_{\mu l}$ is the coefficient of a doubly-occupied molecular orbital $l$ on atom $\mu$. Orbital 0 refers to the singly occupied molecular orbital. \\

\noindent $P^\text{d}$ is defined as follows:

$$P_{\mu \nu}^{\text{d}} = \sum_{1=1}^k (2C_{\mu l}^{*} C_{\nu l}) + C_{\mu 0}^* C_{\nu 0} + C_{\mu 0'}^* C_{\nu 0'}$$

\noindent Where $C_{\mu l}$ is the coefficient of a doubly-occupied molecular orbital $l$ on atom $\mu$. Orbitals 0 and 0' are the two SOMOs as usual. This definition is also equivalent to

$$P^{\text{d}} = C\Lambda C^{T}$$

\noindent where $P^{\text{d}}$ and $C$ are the diradical density matrix and matrix of MO coefficients respectively, and

$$\Lambda = \begin{bmatrix}
    2 & 0 & 0 & \dots  & 0 & 0 & 0 \\
    0 & 2 & 0 & \dots  & 0 & 0 & 0 \\
    \vdots & \vdots & \vdots & \ddots & \vdots & \vdots & \vdots\\
    0 & 0 & 0 & \dots  & 2 & 0 & 0 \\
    0 & 0 & 0 & \dots  & 0 & 1 & 0 \\
    0 & 0 & 0 & \dots  & 0 & 0 & 1 \\
\end{bmatrix}$$

\noindent The above definition for the diradical density matrix ($P_{\mu \nu}^{\text{d}}$) can then be substituted into the standard closed-shell form of the Fock operator in the AO basis. As shown below, this yields the same expression as the earlier defined diradical Fock operator ($F_{pq}^{\text{d}}$).

\begin{equation*}
\begin{split}
F_{\mu \nu}^{\text{d}} & = h_{\mu \nu} + \sum_{\lambda \sigma}P_{\lambda \sigma}^{\text{d}}[(\mu \nu| \lambda \sigma) - \frac{1}{2}(\mu \sigma|\lambda \nu)] \\
 & = h_{\mu \nu} +  \sum_{1=1}^k \sum_{\lambda \sigma}(2C_{\lambda l}^{*} C_{\sigma l})[(\mu \nu| \lambda \sigma) - \frac{1}{2}(\mu \sigma|\lambda \nu)] + \sum_{\lambda \sigma}C_{\lambda 0}^* C_{\sigma 0}[(\mu \nu| \lambda \sigma) - \frac{1}{2}(\mu \sigma|\lambda \nu)] + \sum_{\lambda \sigma}C_{\lambda 0'}^* C_{\sigma 0'}[(\mu \nu| \lambda \sigma) - \frac{1}{2}(\mu \sigma|\lambda \nu)] \\ \\
 & \text{ ( Simplifying with }\sum_{\lambda \sigma}C_{\lambda i}^*C_{\sigma i}\phi_{\lambda}^*\phi_{\sigma} = \psi_i^* \psi_i \text{ ) } \\
 & = h_{\mu \nu} + \sum_{l=1}^k[2(\mu \nu| ll) - (\mu l|l\nu)]] +(\mu \nu | 00) - \frac{1}{2}(\mu 0|0 \nu) + (\mu \nu | 0'0') - \frac{1}{2}(\mu 0'|0' \nu) \\
\implies F_{\mu \nu} & = h_{\mu \nu} + \sum_{l=1}^k[2(\mu \nu| ll) - (\mu l|l\nu)]] + \frac{1}{2}[2(\mu \nu | 00) - (\mu 0|0 \nu)] + \frac{1}{2}[2(\mu \nu | 0'0') - (\mu 0'|0' \nu)]
\end{split}
\end{equation*}

\subsubsection*{B - Matrix Elements for $\hat{H}$ and $\hat{\mu}$}

\begin{table}[H]
\caption{FOFEM CSFs obtained from the $\hat{S}^2$ matrix and their transition dipole moments with the open-shell singlet and triplet CSFs.}
    \centering
    \renewcommand{\arraystretch}{1.5} 
\begin{tabularx}{\textwidth}{|p{8cm}|X|}
    \hline
         \textbf{CSF} & \textbf{Transition dipole moment} \\ \hline
         \multirow{2}{8cm}{$\ket{^5\Psi_1^{1'}} = \frac{1}{\sqrt{6}}
          (\ket{\Psi_{\bar{1} ~\bar{0}}^{\bar{1'} \bar{0'}}} -
          \ket{\Psi_{1 ~\bar{0}}^{1' \bar{0'}}} +
          \ket{\Psi_{\bar{1} ~0}^{\bar{0'} 1'}}  +
          \ket{\Psi_{1 ~0}^{1' 0'}} +
          \ket{\Psi_{1 ~\bar{0}}^{0' \bar{1'}}} -
          \ket{\Psi_{\bar{1} ~0}^{\bar{1'} 0'}})$} & $\bra{^3\text{OS}}\mu \ket{^5\Psi_1^{1'}} = 0$ \\ 
          & $\bra{^1\text{OS}}\mu \ket{^5\Psi_1^{1'}} = 0$ \\ \hline
         \multirow{2}{8cm}{$\ket{^{3}\Psi_1^{1'}} = \frac{1}{\sqrt{2}}
            (\ket{\Psi_{\bar{1} ~\bar{0}}^{\bar{1'} \bar{0'}}} -
            \ket{\Psi_{1 ~0}^{1' 0'}})$} & $\bra{^3\text{OS}}\mu \ket{^{3}\Psi_1^{1'}} = \mu_{1,1^{'}}$ \\
          & $\bra{^1\text{OS}}\mu \ket{^3\Psi_1^{1'}} = 0$ \\ \hline
          \multirow{2}{8cm}{$\ket{^{3^{'}}\Psi_1^{1'}} = \frac{1}{\sqrt{2}}
            (\ket{\Psi_{1 ~\bar{0}}^{1' \bar{0'}}} -
            \ket{\Psi_{\bar{1} ~0}^{\bar{1'} 0'}})$} & $\bra{^3\text{OS}}\mu \ket{^{3'}\Psi_1^{1'}} = \mu_{1,1^{'}}$ \\ 
          & $\bra{^1\text{OS}}\mu \ket{^{3'}\Psi_1^{1'}} = 0$ \\ \hline
          \multirow{2}{8cm}{$\ket{^{3^{''}}\Psi_1^{1'}} = \frac{1}{\sqrt{2}}
            (\ket{\Psi_{\bar{1} ~0}^{\bar{0'} 1'}} -
            \ket{\Psi_{1 ~\bar{0}}^{0' \bar{1'}}})$} & $\bra{^3\text{OS}}\mu \ket{^{3''}\Psi_1^{1'}} = 0$ \\
\hline
    \end{tabularx}
\end{table}

{\centering
\renewcommand{\arraystretch}{1.2} 
\begin{longtable}{|m{0.22\textwidth}|m{0.7\textwidth}|}
\caption{Diagonal elements of the XCIS-D matrix, which represent the energies of each CSF. Indices $i,j,k,l$ represent core MOs, while indices $p,q,r,s$ represent virtual MOs. Energies are written relative to $E_0 = \sum_i (h_{ii} + F^\text{d}_{ii}) + \frac{1}{2}(h_{00} + h_{0'0'} + F^\text{d}_{00} + F^\text{d}_{0'0'})$ to allow more convenient comparison.} \\
\hline
     \makecell{\textbf{Diagonal} \\ \textbf{Matrix Element}} & \makecell{\textbf{Expression in terms of $E_0$}} \\ 
     \hline
     \endfirsthead
     \hline
     \makecell{\textbf{Diagonal} \\ \textbf{Matrix Element}} & \makecell{\textbf{Expression in terms of $E_0$}} \\ 
     \hline
     \endhead

     \hline
     \multicolumn{2}{c}{\textit{Continued on next page...}} \\
     \endfoot

     \hline
     \endlastfoot

     $\bra{^1\text{OS}}\hat{H} \ket{^1\text{OS}}$ & $E_0 - \frac{1}{4}(J_{00} + J_{0'0'}) + \frac{3}{2}K_{00'}$ \\
     $\bra{^1\text{ZW}^-}\hat{H} \ket{^1\text{ZW}^-}$ & $E_0 + \frac{1}{4}(J_{00} + J_{0'0'}) - J_{00'} - \frac{1}{2}K_{00'}$ \\
     $\bra{^1\text{ZW}^+}\hat{H} \ket{^1\text{ZW}^+}$ & $E_0 + \frac{1}{4}(J_{00} + J_{0'0'}) - J_{00'} + \frac{3}{2}K_{00'}$ \\
     $\bra{^{1}\text{CS}_i^0}\hat{H} \ket{^{1}\text{CS}_i^0}$ & $E_0 + F_{00}^\text{d} - F_{ii}^\text{d} - J_{i0} + \frac{1}{4}(J_{00} - J_{0'0'}) + \frac{3}{2}K_{i0'} + \frac{1}{2}K_{i0}$ \\
     $\bra{^{1}\text{CS}_i^{0'}}\hat{H} \ket{^{1}\text{CS}_i^{0'}}$ & $E_0 + F_{0'0'}^\text{d} - F_{ii}^\text{d} - J_{i0'} + \frac{1}{4}(J_{0'0'} - J_{00}) + \frac{3}{2}K_{i0} + \frac{1}{2}K_{i0'}$ \\
     $\bra{^{1}\text{SV}_0^p}\hat{H} \ket{^{1}\text{SV}_0^p}$ & $E_0 + F_{pp}^\text{d} - F_{00}^\text{d} - J_{0p} + \frac{1}{4}(J_{00} - J_{0'0'}) + \frac{1}{2}K_{0p} + \frac{3}{2}K_{0'p}$ \\
     $\bra{^{1}\text{SV}_{0'}^p}\hat{H} \ket{^{1}\text{SV}_{0'}^p}$ & $E_0 + F_{pp}^\text{d} - F_{0'0'}^\text{d} - J_{0'p} + \frac{1}{4}(J_{0'0'} - J_{00}) + \frac{1}{2}K_{0'p} + \frac{3}{2}K_{0p}$ \\
     $\bra{^{1S}\text{CV}_{i}^p}\hat{H} \ket{^{1S}\text{CV}_{i}^p}$ &  $E_0 + F_{pp}^\text{d} - F_{ii}^\text{d} - J_{ip} + 2K_{ip} - \frac{1}{4}(J_{00} + J_{0'0'}) + \frac{3}{2}K_{00'}$ \\
\end{longtable}
}

{\centering
\renewcommand{\arraystretch}{1.2} 
\begin{longtable}{|m{0.22\textwidth}|m{0.7\textwidth}|}
\caption{Off-diagonal elements of the singlet block of the XCIS-D matrix, which represent the energetic coupling between each CSF. Indices $i,j,k,l$ represent core MOs, while indices $p,q,r,s$ represent virtual MOs. The $\delta$ symbols used is the Kronecker delta, used to show when CSFs how couplings change when CSFs have common core or virtual orbitals. Note that the XCIS-D matrix is symmetric meaning only the upper diagonal elements are shown.} \\
    \hline
     \makecell{\textbf{Off-Diagonal} \\ \textbf{$\hat{H}$ Matrix Element}} & \makecell{\textbf{Expression in terms of $F^\text{d}$ and two electron integrals}} \\ 
     \hline
     \endfirsthead
     \hline
     \makecell{\textbf{Off-Diagonal} \\ \textbf{$\hat{H}$ Matrix Element}} & \makecell{\textbf{Expression in terms of $F^\text{d}$ and/or two electron integrals}} \\ 
     \hline
     \endhead

     \hline
     \multicolumn{2}{c}{\textit{Continued on next page...}} \\
     \endfoot

     \hline
     \endlastfoot
    
     $\bra{^{1}\text{OS}} \hat{H} \ket{^{1}\text{ZW}^-}$ & $(00'|00) - (00'|0'0')$ \\
     $\bra{^{1}\text{OS}} \hat{H} \ket{^{1}\text{ZW}^+}$ & $2F_{00'}^\text{d}$ \\
     $\bra{^{1}\text{OS}} \hat{H} \ket{^{1}\text{CS}_i^0}$ & $-F_{i 0}^{\text{d}} + \frac{3}{2}(i0'|0'0) - \frac{1}{2}(i0|00)$ \\
     $\bra{^{1}\text{OS}} \hat{H} \ket{^{1}\text{CS}_i^{0'}}$ & $F_{i 0'}^{\text{d}} + \frac{1}{2}(i0'|0'0')  - \frac{3}{2}(i0|00')$ \\
     $\bra{^{1}\text{OS}} \hat{H} \ket{^{1}\text{SV}_0^p}$ & $F_{0p}^{\text{d}} + \frac{3}{2}(p0'|0'0)  - \frac{1}{2}(p0|00)]$ \\
     $\bra{^{1}\text{OS}} \hat{H} \ket{^{1}\text{SV}_{0'}^p}$ & $-F_{0'p}^{\text{d}} + \frac{1}{2}(p0'|0'0') - \frac{3}{2}(p0|00')$ \\
     $\bra{^{1}\text{OS}} \hat{H} \ket{^{1S}\text{CV}_{i}^p}$ & $\sqrt{2}F_{ip}^\text{d}$ \\
     $\bra{^{1}\text{OS}} \hat{H} \ket{^{1T}\text{CV}_{i}^p}$ & $\frac{\sqrt{3}}{\sqrt{2}}[(i0'|0'p) - (i0|0p)]$ \\
\hline
     
\end{longtable}
}

{\centering
\renewcommand{\arraystretch}{1.2} 
\begin{longtable}{|m{0.22\textwidth}|m{0.7\textwidth}|}
\caption{Off-diagonal elements of the triplet block of the XCIS-D matrix, which represent the energetic coupling between each CSF. Indices $i,j,k,l$ represent core MOs, while indices $p,q,r,s$ represent virtual MOs. The $\delta$ symbols used is the Kronecker delta, used to show when CSFs how couplings change when CSFs have common core or virtual orbitals. Note that the XCIS-D matrix is symmetric meaning only the upper diagonal elements are shown.} \\
    \hline
     \makecell{\textbf{Off-Diagonal} \\ \textbf{$\hat{H}$ Matrix Element}} & \makecell{\textbf{Expression in terms of $F^\text{d}$ or two electron integrals}} \\ 
     \hline
     \endfirsthead
     \hline
     \makecell{\textbf{Off-Diagonal} \\ \textbf{$\hat{H}$ Matrix Element}} & \makecell{\textbf{Expression in terms of $F^\text{d}$ or two electron integrals}} \\ 
     \hline
     \endhead

     \hline
     \multicolumn{2}{c}{\textit{Continued on next page...}} \\
     \endfoot

     \hline
     \endlastfoot
    
     $\bra{^{3}\text{OS}}\hat{H} \ket{^3\text{CS}_i^0}$ &  $F_{i0}^\text{d} + \frac{1}{2}(i0|00) + \frac{1}{2}(i0'|0'0)$ \\
         $\bra{^{3}\text{OS}}\hat{H} \ket{^3\text{CS}_i^{0'}}$ &  $F_{i0'}^\text{d} + \frac{1}{2}(i0'|0'0') + \frac{1}{2}(i0|00')$ \\
         $\bra{^{3}\text{OS}}\hat{H} \ket{^3\text{SV}_0^p}$ &  $F_{0p}^\text{d} - \frac{1}{2}(0p|00) - \frac{1}{2}(00'|0'p)$ \\
         $\bra{^{3}\text{OS}}\hat{H} \ket{^3\text{SV}_{0'}^p}$ &  $-F_{0'p}^\text{d} + \frac{1}{2}(0'p|0'0') + \frac{1}{2}(0'0|0p)$ \\
         $\bra{^{3}\text{OS}}\hat{H} \ket{^{3T}\text{CV}_i^p}$ &  $\sqrt{2}F_{ip}^\text{d}$ \\
         $\bra{^{3}\text{OS}}\hat{H} \ket{^{3S}\text{CV}_i^p}$ &  $\frac{\sqrt{2}}{2}[(i0|0p) - (i0'|0'p)]$ \\
         $\bra{^{3}\text{OS}}\hat{H} \ket{^{3X}\text{CV}_i^p}$ &  $(i0|0p) + (i0'|0'p)$ \\
         $\bra{^{3}\text{OS}}\hat{H} \ket{^{3}\text{CS}_{i~~\bar{j}}^{0'\bar{0'}}}$ &  $(i0|0'j) - (i0'|0j)$ \\
\hline
     
\end{longtable}
}

\begin{table}[h]
\caption{Hamiltonian Interactions for the Quintet core ($i, j$) $\rightarrow$ virtual ($p, q$) CSF. The $\delta$ symbols used is the Kronecker delta, used to show when CSFs how couplings change when CSFs have common core or virtual orbitals.}
    \centering
    \renewcommand{\arraystretch}{1.5} 
    \begin{tabular}{|p{3.8cm}|p{13cm}|}
    \hline
         $\hat{H}$ Matrix element & =  \\ \hline
         $\bra{^{5}\text{CV}_i^p}\hat{H} \ket{^{5}\text{CV}_j^q}$  & $\delta_{ij}\delta_{pq}E_Q - \delta_{pq}(1 - \delta_{ij})[F_{ij}^\text{d} + \frac{1}{2}(i0|0j)  + \frac{1}{2}(i0'|0'j) + (ij|pp)] + \delta_{ij}(1 - \delta_{pq})[F_{pq}^\text{d} - (pq|ii) - \frac{1}{2}(p0|0q) - \frac{1}{2}(p0'|0'q)] - (1 - \delta_{ij})(1 - \delta_{pq})(ij|pq)]$ \\ \hline
         
    \end{tabular}
\end{table}

{\centering
\renewcommand{\arraystretch}{1.2} 
\begin{longtable}{|m{0.2\textwidth}|m{0.5\textwidth}|}
\caption{Analytical expressions for the singlet block of the transition dipole matrix in the XCIS-D basis. Indices $i,j,k,l$ represent core MOs, while indices $p,q,r,s$ represent virtual MOs. $\sum_o$ represents a sum over all occupied orbitals. The $\delta$ symbols used is the Kronecker delta, used to show when CSFs how couplings change when CSFs have common core or virtual orbitals. Note that the matrix is symmetric meaning only the upper diagonal elements are shown.} \\
    \hline
     \makecell{\textbf{Off-Diagonal} \\ \textbf{$\hat{\mu}$ Matrix Element}} & \makecell{\textbf{Expression in terms of $F^\text{d}$ or two electron integrals}} \\ 
     \hline
     \endfirsthead
     \hline
     \makecell{\textbf{Off-Diagonal} \\ \textbf{$\hat{\mu}$ Matrix Element}} & \makecell{\textbf{Expression in terms of $F^\text{d}$ or two electron integrals}} \\ 
     \hline
     \endhead

     \hline
     \multicolumn{2}{c}{\textit{Continued on next page...}} \\
     \endfoot

     \hline
     \endlastfoot
    
     $\bra{^{1}\text{OS}} \hat{\mu} \ket{^{1}\text{ZW}^-}$ & $0$ \\
         $\bra{^{1}\text{OS}} \hat{\mu} \ket{^{1}\text{ZW}^+}$ & $2\mu_{00'}$ \\
         $\bra{^{1}\text{OS}} \hat{\mu} \ket{^{1}\text{CS}_i^0}$ & $-\mu_{i0}$ \\
         $\bra{^{1}\text{OS}} \hat{\mu} \ket{^{1}\text{CS}_i^{0'}}$ & $\mu_{i0'}$ \\
         $\bra{^{1}\text{OS}} \hat{\mu} \ket{^{1}\text{SV}_0^p}$ & $\mu_{0j}$ \\
         $\bra{^{1}\text{OS}} \hat{\mu} \ket{^{1}\text{SV}_{0'}^p}$ & $-\mu_{0'j}$ \\
         $\bra{^{1}\text{OS}} \hat{\mu} \ket{^{1S}\text{CV}_{i}^p}$ & $\sqrt{2}\mu_{ip}$ \\
         $\bra{^{1}\text{OS}} \hat{\mu} \ket{^{1T}\text{CV}_{i}^p}$ & $0$ \\
\end{longtable}
}

{\centering
\renewcommand{\arraystretch}{1.2} 
\begin{longtable}{|m{0.2\textwidth}|m{0.5\textwidth}|}
\caption{Analytical expressions for the triplet block of the transition dipole matrix in the XCIS-D basis. Indices $i,j,k,l$ represent core MOs, while indices $p,q,r,s$ represent virtual MOs. $\sum_o$ represents a sum over all occupied orbitals. The $\delta$ symbols used is the Kronecker delta, used to show when CSFs how couplings change when CSFs have common core or virtual orbitals. Note that the matrix is symmetric meaning only the upper diagonal elements are shown.} \\
    \hline
     \makecell{\textbf{Off-Diagonal} \\ \textbf{$\hat{\mu}$ Matrix Element}} & \makecell{\textbf{Expression in terms of $F^\text{d}$ or two electron integrals}} \\ 
     \hline
     \endfirsthead
     \hline
     \makecell{\textbf{Off-Diagonal} \\ \textbf{$\hat{\mu}$ Matrix Element}} & \makecell{\textbf{Expression in terms of $F^\text{d}$ or two electron integrals}} \\ 
     \hline
     \endhead

     \hline
     \multicolumn{2}{c}{\textit{Continued on next page...}} \\
     \endfoot

     \hline
     \endlastfoot
    
     $\bra{^{3}\text{OS}}\hat{\mu} \ket{^3\text{CS}_i^0}$ &  $\mu_{i0}$ \\
         $\bra{^{3}\text{OS}}\hat{\mu} \ket{^3\text{CS}_i^{0'}}$ &  $\mu_{i0'}$ \\
         $\bra{^{3}\text{OS}}\hat{\mu} \ket{^3\text{SV}_0^p}$ &  $\mu_{0p}$ \\
         $\bra{^{3}\text{OS}}\hat{\mu} \ket{^3\text{SV}_{0'}^p}$ & $-\mu_{0'p}$ \\
         $\bra{^{3}\text{OS}}\hat{\mu} \ket{^{3T}\text{CV}_i^p}$ &  $\sqrt{2}\mu_{ip}$ \\
         $\bra{^{3}\text{OS}}\hat{\mu} \ket{^{3S}\text{CV}_i^p}$ &  $0$ \\
         $\bra{^{3}\text{OS}}\hat{\mu} \ket{^{3X}\text{CV}_i^p}$ &  $0$ \\
         $\bra{^{3}\text{OS}}\hat{\mu} \ket{^{3}\text{CS}_{i~~\bar{j}}^{0'\bar{0'}}}$ &  $0$ \\
\end{longtable}
}

\begin{table}[h]
\caption{Transition Dipole Moments for the Quintet core ($i, j$) $\rightarrow$ virtual ($p, q$) CSF.  $\sum_o$ represents a sum over all occupied orbitals. The $\delta$ symbols used is the Kronecker delta, used to show when CSFs how couplings change when CSFs have common core or virtual orbitals.}
    \centering
    \renewcommand{\arraystretch}{1.5} 
    \begin{tabular}{|p{4cm}|p{8.5cm}|}
    \hline
    $\hat{H}$ Matrix element & =  \\ \hline
     $\bra{^{5}\text{CV}_i^p}\hat{\mu} \ket{^{5}\text{CV}_j^q}$ &  $\delta_{ij}\delta_{pq}\sum_o \mu_{oo} - \delta_{pq}(1 - \delta_{ij})\mu_{ij} + \delta_{ij}(1 - \delta_{pq})\mu_{pq}$ \\ \hline
    \end{tabular}
\end{table}

\newpage

\subsection*{III - Test calculations on small/medium diradicals}

\subsubsection*{A - Computational Details}

For each of the molecules studied, geometries were first optimised with the GAMESS quantum chemistry package \cite{Schmidt1993, Barca2020}, using unrestricted density functional theory with the PBE functional. The 6-31+G$^{**}$ basis set was used for all \textit{ab-initio} calculations described. The multiplicity was set to 3 to ensure that the calculations converged towards the open-shell triplet geometry, and the optimisations were constrained to preserve the respective point-group symmetry of each molecule. For cyclobutadiene, D$_4$h symmetry was enforced during DFT optimisation to prevent a Jahn-Teller distortion to a closed-shell geometry \cite{Borden1996, Reeves1969, Monino2022}. For trimethylenemethane and \textit{meta-}xylylene, point group symmetries of D$_3$h and C$_2$v respectively were enforce. Optimised geometries were then passed into the D-ExROPPP program, which calculates molecular orbitals and then excited states. A convergence tolerance of $2.5 \times 10 ^{-15}$ was used to check whether the density matrix was appropriately converged.

For the multi-reference methods, this process had to be done in two stages, with molecular orbitals first obtained with single-point energy calculations at the restricted open-shell Hartree-Fock (ROHF) level of theory. These MOs were then used to run state-averaged complete active space self-consistent field (SA-CASSCF) calculations with a ($n$,$n$) active space, where n is the number of $\pi$-orbitals in each molecule. For cyclobutadiene and trimethylenemethane, a (4,4) active space was used while for \textit{meta-}xylylene a (8,8) active space was used. The state averaging procedure means that the molecular orbitals and CI coefficients are self-consistently optimised with respect to the average energy of a given set of electronic states, rather than solely the ground state. For Cyclobutadiene, this was selected to be the 5 lowest energy singlet states, for trimethylenemethane the energy was averaged over the 5 lowest energy triplet states and for \textit{meta-}xylylene the 8 lowest energy triplets were selected. Alongside SA-CASSCF, results were also obtained using the generalised multiconfigurational quasi-degenerate perturbation theory method (GMC-QDPT). For GMC-QDPT calculations, a level shift of 0.02 hartrees was used to prevent intruder states.

\subsubsection*{B - Results for Ground and Excited States}

\begin{table}[H]
\caption{Summary of the 7 lowest energy electronic states of the mXYL radical obtained using various methods. Energies are given relative to that of their respective ground state. Configurations shown are those with the highest weighting in the overall excited state, as shown by the coefficients in brackets. All results are fully spin pure, meaning $\langle\hat{S^2}\rangle = 0$ for singlet states and $\langle\hat{S^2}\rangle = 2$ for triplets.}
    \centering
    \begin{tabular}{|c|l|c|l|c|l|}
    \hline
         \makecell{\textbf{D-ExROPPP} \\ \textbf{Energy}} & \makecell{\textbf{D-ExROPPP} \\ \textbf{Configurations}} & \makecell{\textbf{CASSCF} \\ \textbf{Energy}} & \makecell{\textbf{CASSCF} \\ \textbf{Configurations}} & \makecell{\textbf{GMC-QDPT} \\ \textbf{Energy}} & \makecell{\textbf{GMC-QDPT} \\ \textbf{Configurations}} \\ \hline
         
         0 & \makecell{$\ket{^3\text{OS}}$ $(0.95)$} & 0 & \makecell{$\ket{^3\text{OS}}$ $(0.97)$} & 0 & \makecell{$\ket{^3\text{OS}}$ $(0.90)$} \\ \hline
         
         0.485 & \makecell{$\ket{^1\text{ZW}^-}$ $(0.89)$} & 0.338 & \makecell{$\ket{^1\text{ZW}^-}$ $(0.98)$} & 0.502 & \makecell{$\ket{^1\text{ZW}^-}$ $(0.81)$} \\ \hline
         
         1.116 & $\begin{aligned}
         \rule{0pt}{4ex} &\ket{^1\text{OS}} (0.86), \\ & \ket{^1\text{CS}_2^{0'}} (-0.28) \rule[-2ex]{0pt}{0pt}
         \end{aligned}$ & 2.437 & $\begin{aligned}
         \rule{0pt}{4ex} &\ket{^1\text{OS}} (0.62), \\ & \ket{^1\text{CS}_1^{0'}} (-0.39) \rule[-2ex]{0pt}{0pt}
         \end{aligned}$  & 1.235 & $\begin{aligned}
         \rule{0pt}{4ex} &\ket{^1\text{OS}} (0.79), \\ & \ket{^1\text{CS}_1^{0'}} (-0.30) \rule[-2ex]{0pt}{0pt}
         \end{aligned}$ \\ \hline
          
         2.750 & $\begin{aligned}
         \rule{0pt}{4ex} &\ket{^1\text{SV}_{0'}^{2'}} (0.53), \\ & \ket{^1\text{CS}_{2}^{0}} (0.52) \rule[-2ex]{0pt}{0pt} \end{aligned}$ & 3.830 & $\begin{aligned}
         \rule{0pt}{4ex} &\ket{^1\text{SV}_{0'}^{2'}} (0.56), \\ & \ket{^1\text{CS}_{2}^{0}} (-0.41) \rule[-2ex]{0pt}{0pt} \end{aligned}$  & 3.047 & $\begin{aligned}
         \rule{0pt}{4ex} &\ket{^1\text{CS}_{2}^{0}} (0.60), \\ & \ket{^1\text{SV}_{0'}^{2'}} (0.54) \rule[-2ex]{0pt}{0pt} \end{aligned}$ \\ \hline
         
         2.813 & \makecell{$\ket{^1\text{ZW}^+}$ $(0.93)$} & 3.754 & \makecell{$\ket{^1\text{ZW}^+}$ $(0.61)$} & 2.352 & \makecell{$\ket{^1\text{ZW}^+}$ $(0.89)$} \\ \hline

         2.857 & $\begin{aligned}
         \rule{0pt}{4ex} &\ket{^3\text{SV}_{0}^{1'}} (0.65), \\ & \ket{^3\text{CS}_{1}^{0}} (-0.63) \rule[-2ex]{0pt}{0pt} \end{aligned}$ & 4.008 & $\begin{aligned}
         \rule{0pt}{4ex} &\ket{^3\text{CS}_{1}^{0'}} (0.45), \\ & \ket{^3\text{CS}_{2}^{0}} (-0.39) \rule[-2ex]{0pt}{0pt} \end{aligned}$ & 3.279 & $\begin{aligned}
         \rule{0pt}{4ex} &\ket{^3\text{CS}_{2}^{0}} (0.49), \\ & \ket{^3\text{SV}_{0}^{1'}} (0.41) \rule[-2ex]{0pt}{0pt} \end{aligned}$  \\ \hline

         2.986 & $\begin{aligned}
         \rule{0pt}{4ex} &\ket{^1\text{CS}_{1}^{0'}} (0.54), \\ & \ket{^1\text{CS}_{2}^{0}} (-0.44) \rule[-2ex]{0pt}{0pt} \end{aligned}$ & 4.129 & $\begin{aligned}
         \rule{0pt}{4ex} &\ket{^1\text{SV}_{0}^{2'}} (0.44), \\ & \ket{^1\text{CS}_{2}^{0}} (-0.43) \rule[-2ex]{0pt}{0pt} \end{aligned}$  & 3.286 & $\begin{aligned}
         \rule{0pt}{4ex} &\ket{^1\text{CS}_{2}^{0}} (0.49), \\ & \ket{^1\text{SV}_{0}^{3'}} (-0.42) \rule[-2ex]{0pt}{0pt} \end{aligned}$  \\ \hline

         \hline
    \end{tabular}
\end{table}

\begin{table}[H]
\caption{Summary of the electronic states of the CBD radical obtained using the D-ExROPPP method but with different SOMOs and configuration spaces. The localised and delocalised SOMOs used are shown in Fig. 2. As described in the methodology section, XCIS results are obtained from a configuration interaction treatment that only includes CSFs describing single excitations, wheras XCIS-D means that specific double excitations are also included. The configuration coefficients shown in brackets are those of the XCIS-D excited states, although the XCIS coefficients almost within $\pm$ 0.02 of these values.}
    \centering
    \small{
    \begin{tabular}{|c|w{c}{1.8cm}|w{c}{1.8cm}|c|w{c}{1.8cm}|w{c}{1.8cm}|}
    \hline
         \textbf{'Localised'} & \multicolumn{2}{c|}{\textbf{'Localised' Energy (eV)}} & \textbf{'Delocalised'} & \multicolumn{2}{c|}{\textbf{'Delocalised' Energy (eV)}} \\ \cline{2-3} \cline{5-6}
         \textbf{Configurations} & \textbf{XCIS} & \textbf{XCIS-D} & \textbf{Configurations} & \textbf{XCIS} & \textbf{XCIS-D} \\ \hline

         \makecell{$\ket{^1\text{OS}}$ $(0.97)$} & 0 & 0 & \makecell{$\ket{^1\text{ZW}^-}$ $(0.97)$} & 0 & 0 \\ \hline

         $\begin{aligned}
         \rule{0pt}{2ex} &\ket{^3\text{OS}} (0.97), \\ & \ket{^{3S}\text{CV}_1^{1'}} (0.24) \rule[-1ex]{0pt}{0pt}
         \end{aligned}$ & 0.355 & 0.355 & \makecell{$\ket{^3\text{OS}}$ $(1.00)$} & 0.811 & 0.811 \\ \hline

         \makecell{$\ket{^1\text{ZW}^+}$ $(0.95)$} & 2.770 & 1.938 & $\begin{aligned}
         \rule{0pt}{2ex} &\ket{^1\text{ZW}^+} (0.91), \\ & \ket{^{1T}\text{CV}_1^{1'}} (0.21) \rule[-1ex]{0pt}{0pt}
         \end{aligned}$ & 2.682 & 1.510 \\ \hline

         \makecell{$\ket{^1\text{ZW}^-}$ $(1.00)$} & 2.770 & 2.770 & \makecell{$\ket{^1\text{OS}}$ $(1.00)$} & 2.770 & 2.770 \\ \hline
         
         $\begin{aligned}
         \rule{0pt}{4ex} &\ket{^3\text{CS}_{1}^{0}} (0.71), \\ &\ket{^3\text{SV}_{0}^{1}} (0.71) 
         \end{aligned}$ & 3.462 & 3.462 & $\begin{aligned}
         \rule{0pt}{2ex} &\ket{^3\text{CS}_{1}^{0}} (0.53), \\ &\ket{^3\text{SV}_{0}^{1}} (0.47), \\ &\ket{^3\text{CS}_{1}^{0'}} (0.47), \\ &\ket{^3\text{SV}_{0'}^{1}} (-0.53) \rule[-1ex]{0pt}{0pt}
         \end{aligned}$ & 3.462 & 3.462 \\

         \hline
    \end{tabular}
    }
\end{table}

\begin{table}[H]
\caption{Summary of the electronic states of the TMM radical obtained using the D-ExROPPP method but with different SOMOs and configuration spaces. As described in the methodology section XCIS results are obtained from a configuration interaction treatment that only includes CSFs describing single excitations, wheras XCIS-D means that specific double excitations are also included. The configuration coefficients shown in brackets are those of the XCIS-D excited states, although the XCIS coefficients almost within $\pm$ 0.02 of these values.}
    \centering
    \begin{tabular}{|c|w{c}{1.8cm}|w{c}{1.8cm}|c|w{c}{1.8cm}|w{c}{1.8cm}|}
    \hline
         \textbf{'Standard'} & \multicolumn{2}{c|}{\textbf{'Standard' Energy (eV)}} & \textbf{'Rotated'} & \multicolumn{2}{c|}{\textbf{'Rotated' Energy (eV)}} \\ \cline{2-3} \cline{5-6}
         \textbf{Configurations} & \textbf{XCIS} & \textbf{XCIS-D} & \textbf{Configurations} & \textbf{XCIS} & \textbf{XCIS-D} \\ \hline

         $\begin{aligned}
         \rule{0pt}{4ex} &\ket{^3\text{OS}} (0.97), \\ & \ket{^{3S}\text{CV}_1^{1'}} (-0.23) \rule[-2ex]{0pt}{0pt}
         \end{aligned}$ & 0 & 0 & $\begin{aligned}
         \rule{0pt}{4ex} &\ket{^3\text{OS}} (0.97), \\ & \ket{^{3S}\text{CV}_1^{1'}} (-0.23) \rule[-2ex]{0pt}{0pt}
         \end{aligned}$ & 0 & 0 \\ \hline

         \makecell{$\ket{^1\text{OS}} (0.88)$} & 0.893 & 0.893 & \makecell{$\ket{^1\text{ZW}^-} (0.88)$} & 0.893 & 0.893 \\ \hline

         \makecell{$\ket{^1\text{ZW}^-} (0.88)$} & 0.893 & 0.893 & \makecell{$\ket{^1\text{OS}} (0.88)$} & 0.893 & 0.893 \\ \hline

         $\begin{aligned}
         \rule{0pt}{4ex} &\ket{^3\text{CS}_{1}^{0'}} (0.66), \\ &\ket{^3\text{SV}_{0'}^{1}} (0.66) 
         \end{aligned}$ & 3.177 & 3.177 & $\begin{aligned}
         \rule{0pt}{6ex} &\ket{^3\text{CS}_{1}^{0}} (0.46), \\ &\ket{^3\text{SV}_{0}^{1}} (-0.46), \\ &\ket{^3\text{CS}_{1}^{0'}} (0.46), \\ &\ket{^3\text{SV}_{0'}^{1}} (0.46) \rule[-3ex]{0pt}{0pt}
         \end{aligned}$ & 3.462 & 3.462 \\ \hline

         \makecell{$\ket{^1\text{ZW}^+}$ $(0.96)$} & 4.257 & 3.736 & \makecell{$\ket{^1\text{ZW}^+}$ $(0.96)$} & 4.257 & 3.736 \\ \hline

         \hline
    \end{tabular}
\end{table}

\begin{table}[H]
\caption{Summary of the electronic states of the mXYL radical obtained using the D-ExROPPP method but with different SOMOs and configuration spaces. The spatial distributions of standard and rotated orbitals are shown in figures 11 and 12. XCIS means that only CSFs describing single excitations are included in the CI matrix, wheras XCIS-D means that specific double excitations are also included as described in the methodology section. The coefficients for the two sets of configurations are taken from the XCIS-D excited states, although the XCIS coefficients are all within 0.01 of these values.}
    \centering
    \begin{tabular}{|c|w{c}{1.8cm}|w{c}{1.8cm}|c|w{c}{1.8cm}|w{c}{1.8cm}|}
    \hline
         Standard & \multicolumn{2}{c|}{Standard Energy (eV)} & Rotated & \multicolumn{2}{c|}{Rotated Energy (eV)} \\ \cline{2-3} \cline{5-6}
         Configurations & XCIS & XCIS-D & Configurations & XCIS & XCIS-D \\ \hline

         \makecell{$\ket{^3\text{OS}}$ $(0.95)$} & 0 & 0 & \makecell{$\ket{^3\text{OS}}$ $(0.95)$} & 0 & 0 \\ \hline
         
         \makecell{$\ket{^1\text{ZW}^-}$ $(0.89)$} & 0.498 & 0.485 & \makecell{$\ket{^1\text{OS}}$ $(0.91)$} & 0.562 & 0.534 \\ \hline

         \makecell{$\ket{^1\text{OS}}$ $(0.86)$} & 1.187 & 1.116 & \makecell{$\ket{^1\text{ZW}^-}$ $(0.84)$} & 1.039 & 1.018 \\ \hline

         $\begin{aligned}
         \rule{0pt}{4ex} &\ket{^3\text{SV}_{0'}^{1'}} (0.68), \\ & \ket{^3\text{CS}_{1}^{0'}} (-0.59) \rule[-2ex]{0pt}{0pt} \end{aligned}$  & 2.723 & 2.726 & $\begin{aligned}
         \rule{0pt}{4ex} &\ket{^3\text{SV}_{0'}^{1'}} (0.58), \\ & \ket{^3\text{SV}_{0'}^{2'}} (-0.50) \rule[-2ex]{0pt}{0pt} \end{aligned}$ & 2.449 & 2.423 \\ \hline

         $\begin{aligned}
         \rule{0pt}{4ex} &\ket{^3\text{SV}_{0}^{2'}} (0.55), \\ & \ket{^3\text{CS}_{2}^{0'}} (0.54) \rule[-2ex]{0pt}{0pt} \end{aligned}$ & 2.854 & 2.782 & $\begin{aligned}
         \rule{0pt}{4ex} &\ket{^3\text{SV}_{0}^{1'}} (0.52), \\ & \ket{^3\text{CS}_{1}^{0}} (-0.49) \rule[-2ex]{0pt}{0pt} \end{aligned}$ & 2.684 & 2.639 \\ \hline

         \makecell{$\ket{^1\text{ZW}^+}$ $(0.92)$} & 3.311 & 2.812 & \makecell{$\ket{^1\text{ZW}^+}$ $(0.93)$} & 3.218 & 2.731 \\ \hline

         \hline
    \end{tabular}
\end{table}

\newpage
\subsection*{IV - Calculations on Emissive Diradicals}

\subsubsection*{A - Computational Details}

Optimised geometry files for each of the large emissive diradicals studied were obtained from the supporting information sections of the papers referenced in Table A13. For D-ExROPPP calculations, a convergence tolerance of $2.5 \times 10 ^{-15}$ was used to check whether the density matrix was appropriately converged by the SCF procedure. Spin-flip time-dependent density functional theory calculations were run using the Orca quantum chemistry package \cite{Neese2022}. A 6-31G$^{**}$ basis set was used for all TD-DFT calculations and 30 electronic states were obtained in each case.

\subsubsection*{C - Results for Ground and Excited States}

\begin{table}[H]
\caption{Singlet triplet gaps ($\Delta E_\text{S-T} = E_\text{T} - E_\text{S}$) predicted by D-ExROPPP alongside those predicted from variable-temperature electron paramagnetic resonance (VT-EPR) measurements. \\ $^*$ VT-EPR measurements were taken but data suggested that $\Delta E_\text{S-T} \approx 0$. \\ $^{**}$ No VT-EPR data is available.}
    \centering
    \begin{tabular}{|c|c|l|c|l|c|}
    \hline
         \makecell{\textbf{Molecule} \\ \textbf{Name}} & \makecell{\textbf{D-ExROPPP} \\ \textbf{GS Multiplicity}} & \makecell{\textbf{D-ExROPPP} \\ $\Delta E_\text{S-T}$ \textit{(eV)}} & \makecell{\textbf{Experimental} \\ \textbf{GS Multiplicity}} & \makecell{\textbf{Experimental} \\ $\Delta E_\text{S-T}$ \textit{(eV)}} & \makecell{\textbf{Ref.}} \\ \hline

        TTM-TTM & Singlet & -0.692 & Singlet & -0.135 $\pm$ 0.002 & \cite{Chang2024} \\ \hline
        \textit{para-}TTM-Ph-TTM & Singlet & -0.083 & Singlet & -0.065 & \cite{Abdurahman2023} \\ \hline 
        \textit{meta-}TTM-Ph-TTM & Singlet & -0.091 & N/A$^{**}$ & N/A$^{**}$ & \cite{Kopp2024} \\ \hline
        DTA & Singlet & -0.087 & Singlet & -0.014 & \cite{Tong2025} \\ \hline
        5-7-ICz-TTM$_2$ & Singlet & -0.028 & Singlet & -0.0005 & \cite{Arnold2025} \\ \hline
        5-8-ICz-TTM$_2$ & Singlet & -0.073 & Degenerate$^{*}$ & $\approx0^{*}$ & \cite{Arnold2025} \\ \hline
        5-11-ICz-TTM$_2$ & Singlet & -0.085 & Degenerate$^{*}$ & $\approx0^{*}$ & \cite{Arnold2025} \\ \hline
        pseudo-\textit{o}-PCP-TTM$_2$ & Triplet & 0.011 & $\approx$ Degenerate$^{*}$ & $\approx 0^{*}$ & \cite{Schneburg2026} \\ \hline
        pseudo-\textit{p}-PCP-TTM$_2$ & Triplet & 0.015 & Degenerate$^{*}$ & $\approx0^{*}$ & \cite{Schneburg2026} \\ \hline
        PyBTM-Ph-PyBTM & Singlet & -0.087 & Singlet & -0.044 & \cite{Hattori2024} \\ \hline
        PyBTM-Hex$_2$-PyBTM & Singlet & -0.087 & Singlet & -0.031 & \cite{Hattori2024} \\ \hline
        PCz-(PyBTM')$_2$ & Singlet & -0.039 & Singlet & -0.002 & \cite{Mizuno2024} \\ \hline
        THDBA-(PyBTM')$_2$ & Triplet & 0.024 & Triplet & 0.0007 $\pm$ 0.0005 & \cite{Matsuoka2023} \\ \hline
        TPA(Me)-(PyBTM'')$_2$ & Singlet & -0.050 & Singlet & -0.0015 & \cite{Hattori2019} \\ \hline
         
         \hline
    \end{tabular}
\end{table}

\begin{table}[H]
\caption{SF-TD-DFT ground state multiplicities and singlet-triplet energy gaps ($\Delta E_\text{S-T}$) for all diradical molecules computed with B3LYP, CAM-B3LYP, and PBE0 functionals. $\langle S^2 \rangle$ is the expectation value of the total spin operator, where $\langle S^2 \rangle < 1$ indicates a ground state of mostly singlet character and $\langle S^2 \rangle >1$ indicates a ground state of mostly triplet character.}
    \centering
    \renewcommand{\arraystretch}{1.5}
    \begin{tabular}{|c|c|l|c|l|c|l|}
    \hline
        \makecell{\textbf{Molecule} \\ \textbf{Name}} &
        \makecell{\textbf{B3LYP} \\ \textbf{GS $\langle \hat{S^2}\rangle$}} &
        \makecell{\textbf{B3LYP} \\ $\Delta E_\text{S-T}$ \textit{(eV)}} &
        \makecell{\textbf{CAM-B3LYP} \\ \textbf{GS  $\langle \hat{S^2}\rangle$}} &
        \makecell{\textbf{CAM-B3LYP} \\ $\Delta E_\text{S-T}$ \textit{(eV)}} &
        \makecell{\textbf{PBE0} \\ \textbf{GS  $\langle \hat{S^2}\rangle$}} &
        \makecell{\textbf{PBE0} \\ $\Delta E_\text{S-T}$ \textit{(eV)}} \\ \hline
         
        TTM-TTM &
         0.08 & -0.135 &
         0.21 & -0.309 &
         0.11 & -0.155 \\ \hline
         
        \textit{para}-TTM-Ph-TTM &
         0.08 & -0.0068 &
         0.10 & -0.0028 &
         0.10 & -0.0058 \\ \hline
         
        \textit{meta}-TTM-Ph-TTM &
         2.06 & 0.00099 &
         2.18 & 0.00090 &
         2.10 & 0.0012 \\ \hline
         
        DTA &
         0.11 & -0.070 &
         0.13 & -0.028 &
         0.13 & -0.060 \\ \hline
         
        5-7-ICz-TTM$_2$ &
         1.02 & -0.025 &
         1.15 & -0.020 &
         1.07 & -0.027 \\ \hline
         
        5-8-ICz-TTM$_2$ &
         0.54 & -0.0046 &
         1.00 & 0.0027 &
         0.72 & -0.0044 \\ \hline
         
        5-11-ICz-TTM$_2$ &
         0.13 & -0.0048 &
         0.60 & -0.0014 &
         0.20 & -0.0035 \\ \hline
         
        \textit{pseudo}-\textit{o}-PCP-TTM$_2$ &
         1.29 & 0.00065 &
         1.65 & 0.00012 &
         1.35 & 0.00059 \\ \hline
         
        \textit{pseudo}-\textit{p}-PCP-TTM$_2$ &
         0.19 & -0.00098 &
         0.51 & -0.035 &
         0.30 & -0.033 \\ \hline
         
        PyBTM-Ph-PyBTM &
         0.08 & -0.013 &
         0.12 & -0.0052 &
         0.11 & -0.011 \\ \hline
         
        PyBTM-Hex$_2$-PyBTM &
         0.08 & -0.0016 &
         0.12 & -0.0006 &
         0.10 & -0.0013 \\ \hline
         
        PCz-(PyBTM$'$)$_2$ &
         0.03 & -3.34 &
         0.08 & -3.68 &
         0.04 & -3.42 \\ \hline
         
        THDBA-(PyBTM$'$)$_2$ &
         1.05 & 0.0063 &
         1.18 & 0.0033 &
        1.10 & 0.0070 \\ \hline
         
        TPA(Me)-(PyBTM$''$)$_2$ &
         0.12 & -0.0090 &
         0.17 & -0.0017 &
        0.15 & -0.0070 \\ \hline
     
    \end{tabular}

\end{table}

\begin{figure}[H]
\centering
\includegraphics[width=1\linewidth, height=0.69\linewidth]{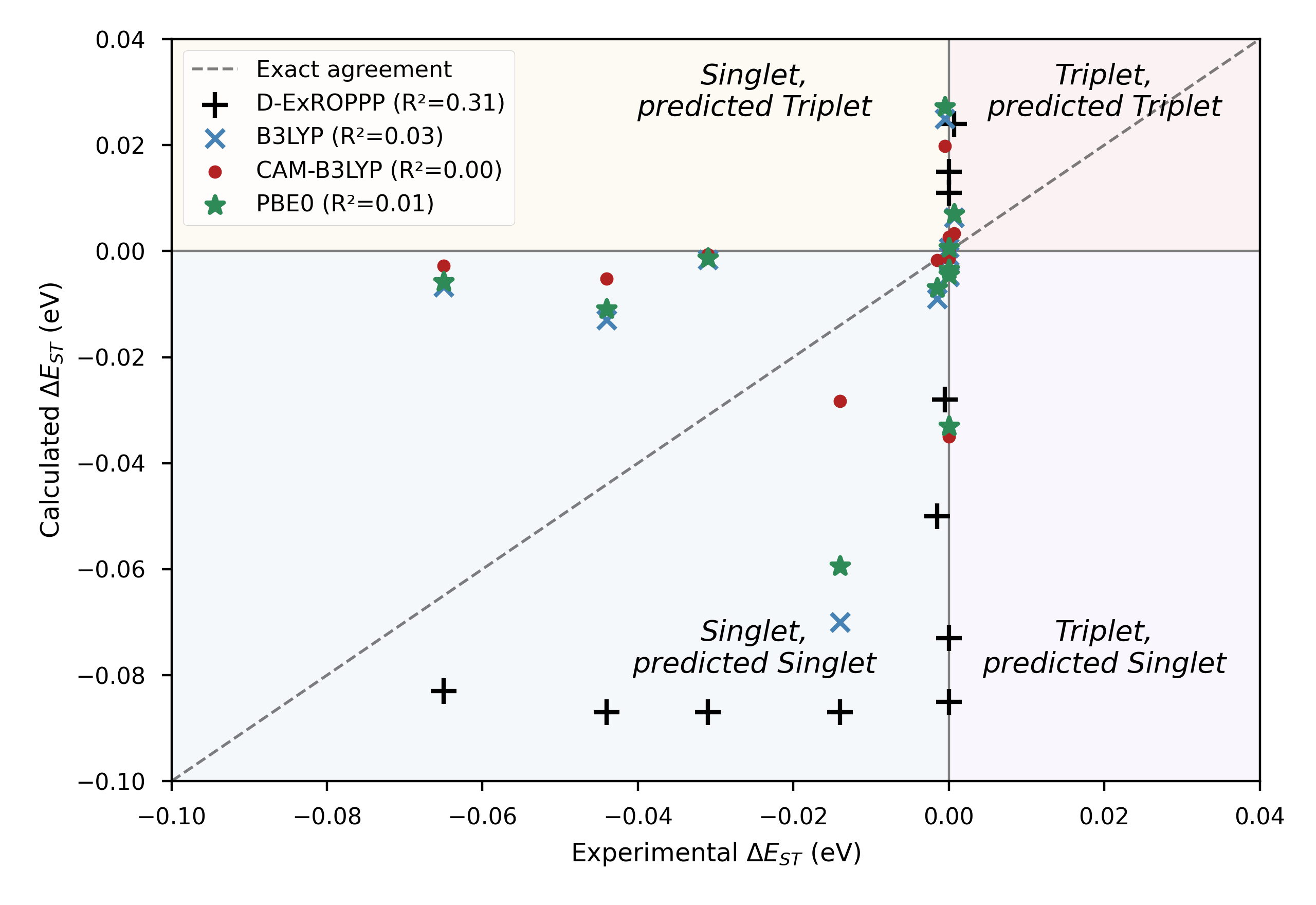}
\caption{Comparison of $\Delta E_\text{S-T}$ predictions obtained by different computational methods for the 13 diradicals studied with VT-EPR data available. Data points are found in tables A13 and A14.}
\end{figure}

\begin{table}[H]
\caption{Experimental and calculated first peak transition energies for the diradical molecule set. D-ExROPPP and TD-DFT (with B3LYP functional) values are obtained by inspected the oscillator strengths ($f$) of the first 29 excited states and selecting the first state in the Visible/Near-IR range with a non-negligible value of $f$. Experimental energies are obtained by predicting the positions of the lowest energy Visible/Near-IR in absorption spectra and converting to eV.}
    \centering
    \begin{tabular}{|c|l|l|l|l|l|}
    \hline
         \makecell{\textbf{Molecule} \\ \textbf{Name}} &
            \makecell{\textbf{Exp.} \\ \textbf{Energy} (eV)} &
            \makecell{\textbf{D-ExROPPP} \\ \textbf{Energy} (eV)} & \makecell{\textbf{D-ExROPPP} \\ $\langle\hat{S^2}\rangle$} &
            \makecell{\textbf{SF-TD-DFT} \\ \textbf{Energy} (eV)} & \makecell{\textbf{SF-TD-DFT} \\ $\langle\hat{S^2}\rangle$} \\ \hline

            TTM-TTM                                         & 2.049 & 1.904 & 0.00 & 1.443 & 1.15 \\ \hline
            \textit{para}-TTM-Ph-TTM                                      & 2.214 & 2.638 & 0.00 & 2.292 & 1.11 \\ \hline
            \textit{meta}-TTM-Ph-TTM                              & 2.167 & 2.485 & 0.00 & 1.694 & 1.09 \\ \hline
            DTA                                             & 1.471 & 1.585 & 2.00 & 1.589 & 0.65 \\ \hline
            5-7-ICz-TTM$_2$                                 & 1.934 & 2.250 & 2.00 & 1.805 & 1.07 \\ \hline
            5-8-ICz-TTM$_2$                                 & 1.876 & 2.326 & 2.00 & 1.831 & 1.07 \\ \hline
            5-11-ICz-TTM$_2$                                & 1.839 & 2.353 & 2.00 & 1.842 & 1.00 \\ \hline
            \textit{pseudo}-\textit{o}-PCP-TTM$_2$          & 1.853 & 1.907 & 2.00 & 1.968 & 1.30 \\ \hline
            \textit{pseudo}-\textit{p}-PCP-TTM$_2$          & 1.848 & 1.919 & 2.00 & 1.996 & 1.24\\ \hline
            PyBTM-Ph-PyBTM                                  & 2.183 & 2.500 & 0.00 & 1.655 & 1.02 \\ \hline
            PyBTM-Hex-PyBTM                                 & 2.283 & 2.627 & 0.00 & 1.640 & 1.03 \\ \hline
            PCz-(PyBTM)$_2$                               & 1.899 & 2.465 & 2.00 & 3.73 & 1.88 \\ \hline
            THDBA-(PyBTM)$_2$                               & 1.968 & 1.959 & 2.00 & 1.631 & 1.09 \\ \hline
            TPA(Me)-(PyBTM$''$)$_2$                         & 1.579 & 1.388 & 2.00 & 1.296 & 0.42 \\ \hline

    \end{tabular}
\end{table}

\begin{table}[H]
\caption{Comparison of the wall-clock time taken by various computational methods to calculate the first 30 electronic states of large emissive diradicals.\\ $^{*}$ Estimated single-core wall clock time calculated using eq. 17 with CPU-Utilisation estimated at $75\%$. \\ $^{**}$ Approximate increase in computational speed when running D-ExROPPP with an XCIS-D basis compared to SF-TD-DFT/B3LYP. $\text{Speed-up Factor} = \frac{t_\text{SF-TD-DFT}}{t_\text{D-ExROPPP}}$}
    \centering
    \begin{tabular}{|c|l|l|l|}
    \hline
         \makecell{\textbf{Molecule} \\ \textbf{Name}} & \makecell{$t_\text{D-ExROPPP}$$^*$ \\ / s} & \makecell{$t_\text{SF-TD-DFT}$$^{**}$ \\ / s} & \makecell{\textbf{Speed-up} \\ \textbf{Factor}$^{***}$} \\ \hline
         
         TTM-TTM & 17 & 16434 & 966  \\ \hline
        \textit{para-}TTM-Ph-TTM & 22 & 3438 & 156  \\ \hline 
        \textit{meta-}TTM-Ph-TTM & 23 & 3132 & 137 \\ \hline
        DTA & 34 & 3600 & 106   \\ \hline
        5-7-ICz-TTM$_2$ & 86 & 5202 & 60  \\ \hline
        5-8-ICz-TTM$_2$ & 85 & 4824 & 57 \\ \hline
        5-11-ICz-TTM$_2$ & 88 & 5112 & 58 \\ \hline
        pseudo-\textit{o}-PCP-TTM$_2$ & 86 & 5166 & 60  \\ \hline
        pseudo-\textit{p}-PCP-TTM$_2$ & 81 & 5202 & 62  \\ \hline
        PyBTM-Ph-PyBTM & 20 & 2898 & 145  \\ \hline
        PyBTM-Hex$_2$-PyBTM & 55 & 4950 & 90  \\ \hline
        PCz-(PyBTM')$_2$ & 61 & 4896 & 82 \\ \hline
        THDBA-(PyBTM')$_2$ & 89 & 5184 & 58  \\ \hline
        TPA(Me)-(PyBTM'')$_2$ & 106 & 4932 & 47  \\ \hline

         \hline
    \end{tabular}
\end{table}

\end{document}